\documentclass[a4paper,11pt]{article}
\pdfoutput=1
\usepackage{jheppub}

\usepackage{amsmath,amssymb,amsfonts,cases}
\usepackage{amsthm}
\usepackage{enumitem}
\usepackage{adjustbox}
\usepackage{graphicx,caption,subcaption}
\usepackage[T1]{fontenc}
\usepackage{enumitem}
\usepackage{empheq}
\usepackage{dsfont}
\usepackage[mathscr]{euscript}
\usepackage{dcolumn}
\usepackage{bm}
\usepackage{bbm}
\usepackage{slashed}
\usepackage{wrapfig}
\usepackage{mathtools}
\usepackage{tikz}
\usepackage{slashbox}
\usepackage{tikz-cd}
\usepackage{hyperref}

\newcommand{\beq}{\begin{equation}}
\newcommand{\eeq}{\end{equation}}
\newcommand{\ba}{\begin{align}}
\newcommand{\ea}{\end{align}}
\newcommand{\bea}{\begin{eqnarray}}
\newcommand{\eea}{\end{eqnarray}}
\newcommand{\bi}{\begin{itemize}}
\newcommand{\ei}{\end{itemize}}
\newcommand{\ben}{\begin{enumerate}}
\newcommand{\een}{\end{enumerate}}

\newcommand{\lsp}{\hspace{0.5pt}}

\def\Tr{{\rm Tr}}

\renewcommand{\d}{\partial}

\newcommand{\ran}{\rangle}

\newcommand{\ket}[1]{| #1 \ran}

\newcommand{\lrpr}{\raise 1ex\hbox{$^\leftrightarrow$} \hspace{-9pt} \slashed{\partial}}
\newcommand{\lpr}{\raise 1ex\hbox{$^\leftarrow$} \hspace{-9pt}\slashed{\partial}}
\newcommand{\rpr}{\raise 1ex\hbox{$^\rightarrow$} \hspace{-9pt}\slashed{\partial}}

\renewcommand{\a}{\alpha}
\renewcommand{\b}{\beta}
\renewcommand{\d}{\delta}

\newcommand{\G}{\Gamma}
\newcommand{\g}{\gamma}
\renewcommand{\k}{\kappa}

\newcommand{\la}{\lambda}
\newcommand{\m}{\mu}
\newcommand{\n}{\nu}

\renewcommand{\r}{\rho}

\def\a{\alpha}
\def\b{\beta}
\def\d{\delta}

\def\k{\kappa}

\def\m{\mu}
\def\n{\nu}

\def\r{\rho}

\def\G{\Gamma}

\def\O{\Omega}

\newcommand{\pd}{\partial}

\newcommand\eq[1]{eq.~(\ref{eq:#1})}
\newcommand\tbl[1]{Table~\ref{tab:#1}}
\newcommand{\sn}[1]{section~\ref{sec:#1}}
\newcommand{\fig}[1]{Figure~\ref{fig:#1}}

\newcommand\mf[1]{{\mathfrak{#1}}}
\newcommand\mc[1]{{\mathcal{#1}}}

\newcommand\CD{{\mc{D}}}

\newcommand\CM{{\mc{M}}}

\newcommand\CO{{\mc{O}}}

\newcommand\CZ{{\mc{Z}}}

\definecolor{cardinal}{rgb}{0.6,0,0}
\definecolor{darkgreen}{rgb}{0,0.5,0}
\definecolor{golden}{rgb}{0.92, 0.7, 0}
\definecolor{midnight}{rgb}{0, 0, 0.5}
\definecolor{darkblue}{rgb}{0.2, 0, 0.8}

\allowdisplaybreaks[4]

\preprint{UUITP-53/22}

\title{\LARGE Boundaries in Free Higher Derivative Conformal Field Theories}

\author[a,b]{Adam Chalabi,}
\author[c]{Christopher P.\ Herzog,}
\author[d]{Krishnendu Ray,}
\author[e,f]{Brandon Robinson,}
\author[g]{Jacopo Sisti,}
\author[c]{and Andreas Stergiou}

\affiliation[a]{Niels Bohr Institute, University of Copenhagen, Blegdamsvej 17, 2100 Copenhagen, Denmark}
\affiliation[b]{STAG Research Centre, Physics and Astronomy, University of Southampton, Highfield, Southampton, SO17 1BJ, UK}
\affiliation[c]{Department of Mathematics, King's College London, Strand, London, WC2R 2LS, UK}
\affiliation[d]{Rudolf Peierls Centre for Theoretical Physics, University of Oxford, Parks Road, Oxford, OX1 3PU, UK}
\affiliation[e]{Instituut voor Theoretische Fysica, K.U. Leuven, Celestijnenlaan 200D, BE-3001 Leuven, Belgium}
\affiliation[f]{INFN, sezione di Milano-Bicocca,
Piazza della Scienza 3, I-20126 Milano, Italy}
\affiliation[g]{Department of Physics and Astronomy, Uppsala University, Box 516, SE-75120 Uppsala, Sweden}

\emailAdd{adam.chalabi@nbi.ku.dk}
\emailAdd{christopher.herzog@kcl.ac.uk}
\emailAdd{krishnendu.ray@sjc.ox.ac.uk}
\emailAdd{brandon.robinson@mib.infn.it}
\emailAdd{jacopo.sisti@physics.uu.se}
\emailAdd{andreas.stergiou@kcl.ac.uk}

\abstract{We consider free higher derivative theories of scalars and Dirac fermions in the presence of a boundary in general dimension.  We establish a method for finding consistent conformal boundary conditions in these theories by removing certain boundary primaries from the spectrum.
A rich set of renormalization group flows between various conformal boundary conditions
is revealed, triggered by deformations quadratic in the boundary primaries.  We compute the free energy of these theories on a hemisphere, and show that the boundary $a$-theorem is generally violated along boundary flows
as a consequence of bulk non-unitarity.
We further characterize the boundary theory by computing the two-point function of the displacement operator.
}

\begin{document}
\maketitle
\flushbottom

\section{Introduction}\label{sec:introduction}

The higher derivative scalar field theory with action
\begin{equation}
\label{Boxsquaredtheory}
 \int d^d x \, \phi\lsp \Box^2 \phi
\end{equation}
has a dubious reputation among relativistic quantum field theorists.
For a metric with Lorentzian signature, a $\partial_t^4$ time derivative
in the Lagrangian indicates the presence of ghosts and the Ostrogradski instability.
If we think of $\Box^2 \phi = 0$ governing just the spatial degrees of freedom, then we are forced to treat time differently and sacrifice the sacred cow of Lorentz invariance.
Even in a purely Euclidean context,
if we insist on conformal symmetry, then the scaling dimension of $\phi$ is below the unitarity bound, making the theory
non-unitary.

Yet the equation of motion $\Box^2 \phi = 0$ in a Euclidean context is of enormous practical interest.  It was shown about two hundred years ago that this equation for $d=2$ governs small deformations of thin plates \cite{germain1821recherches,lagrange1828note,poisson1829memoire}. Since then it has played an important role in the engineering community where to this day it is discussed in textbooks \cite{timoshenko1951theory,slaughter2012linearized,landau1986theory}.  (The $d=1$ case governs the deformation of beams.)\footnote{ The flow of a viscous incompressible fluid
 on a plane is described by the biharmonic equation \cite{rayleigh1893xxxviii}.
 We should also mention the semi-flexible polymers which are governed by a one dimensional biharmonic equation of motion \cite{burkhardt2010fluctuations}.
 There is further extensive literature on Lifshitz theories, which involve a $\Box^2$ kinetic term in some directions and a $\Box$ kinetic term in others (see e.g.\ refs.~\cite{Diehl:2003kv,Diehl:2004bjl,diehl2006boundary} for a discussion of Lifshitz theories with a boundary).
}

Our interest in the theory (\ref{Boxsquaredtheory}) is more formal.  It and its bosonic and fermionic cousins,
\begin{equation}
 \int d^d x \, \phi\lsp \Box^k \phi \quad \mbox{and} \quad   \int d^d x \, \overline \psi\lsp \slashed{\partial}^{2k+1} \psi \ ,
\end{equation}
for $k$ a positive integer
provide nontrivial yet tractable examples of Conformal Field Theories (CFTs) where we can try to classify the set of conformal boundary
conditions and the Renormalization Group (RG) flows between them.
Issues with ghosts, instabilities, and non-unitarity we are able for the most part to ignore.
That the $\Box^2 \phi = 0$ theory has an engineering application
will give our results in this case an interesting physical interpretation, as we will see.

As a warm-up, let us try to establish under what boundary conditions the operator $\Box^2 $ is symmetric with respect to the inner product (\ref{Boxsquaredtheory}).  For an elliptic problem, we should try to enforce two boundary conditions along each edge.
For simplicity, we will consider here a single, flat edge along the plane $x^n = 0$.  Two obvious possibilities are to set $\Phi^{(0)} \equiv \phi|_{\rm bry} = 0$
and $\Phi^{(1)} \equiv \partial_n \phi|_{\rm bry} = 0$ along the edge.  The boundary conditions for higher powers of the normal derivative are more ambiguous
however.  We are free to consider the homogeneously scaling combinations
\begin{eqnarray*}
\Phi^{(2)} \equiv \partial_n^2 \phi + \nu\lsp \Box_\parallel \phi \biggr|_{\rm bry}  \quad \mbox{and} \quad \Phi^{(3)}\equiv  \partial_n^3 \phi + \mu\lsp \partial_n \Box_\parallel \phi \biggr|_{\rm bry}
\end{eqnarray*}
with arbitrary coefficients $\mu$ and $\nu$ and still have boundary conditions compatible with the residual rotational and translational invariance of the problem.

If we integrate by parts several times, we find that
\begin{align}
\int d^d x \, \phi' \Box^2  \phi = \int d^d x \, (\Box^2 \phi') \phi + \int &d^{d-1} x \biggr(\Phi'^{(0)}  \Phi^{(3)} - \Phi'^{(3)}  \Phi^{(0)} - \Phi'^{(1)}   \Phi^{(2)}+ \Phi'^{(2)} \Phi^{(1)} \nonumber \\
& + (2-\mu-\nu) ( \Phi'^{(0)}\Box_\parallel \Phi^{(1)} - \Phi'^{(1)} \Box_\parallel  \Phi^{(0)}  ) \biggr) \ .
\end{align}
There are three pairs of boundary conditions which will eliminate the  boundary terms,
\begin{align}
DD :& \; \Phi^{(0)} = 0 \ , \; \; \Phi^{(1)} = 0\ , \\
DN :& \; \Phi^{(0)} = 0 \ , \; \; \Phi^{(2)} = 0\ , \\
ND :& \; \Phi^{(1)} = 0 \ , \; \; \; \Phi^{(3)} = 0\ ,
\end{align}
and a fourth which works only if $\mu+\nu = 2$,
\begin{equation}\label{eq:NN_intro}
NN : \Phi^{(2)} = 0 \ , \; \; \; \Phi^{(3)} = 0 \ .
\end{equation}

The $DD$, $DN$, and $NN$ boundary conditions were identified over a hundred years ago by Kirchoff and others
as important in the description of the equilibrium
configuration of a thin plate \cite{love1888xvi,kirchoff1876vorlesungen,kirchoff1850balance}.\footnote{ The labeling convention $DD$, $DN$, etc.\ is our own.  We will explain the reason for our choice of the letters $N$ and $D$ in section \ref{sec:primaries}.
}
The condition $DD$ describes a clamped edge, i.e.\ a plate embedded in a wall such that not only its position is fixed but the horizontal angle at which it protrudes as well.  An edge that is free and simply supported, for example by a wedge support underneath, is described by $DN$.
Finally the boundary condition $NN$ (along with the constraint $\mu+\nu = 2$) describes a free and unsupported edge.
The textbook derivations of these boundary conditions often rely on a free body diagram of the edge along with corresponding forces and torques, and not on abstract properties of the operator $\Box^2$.

 The coefficient $\nu$ for the  plate is Poisson's ratio:
the distance a material is compressed divided by the distance it expands laterally.
In an isotropic, two parameter model of elasticity,
it is determined by the bulk modulus $K$ and shear modulus $\sigma$ \cite{landau1986theory}:
\begin{align}
\label{Poisson}
\nu = \frac{1}{2} \frac{3K - 2\sigma}{3K+\sigma} \ .
\end{align}
The bulk modulus $K$ determines the energy change of a material under compression while $\sigma$ gives the same for shear deformations.  For thermodynamic stability, both must be positive semi-definite, which in turn means Poisson's ratio is bounded between $-1$ (for $K=0$) and $\frac{1}{2}$ (for $\sigma = 0$).  Liquids, which are incompressible
and offer little if any resistance to shear forces,
have $\nu = \frac{1}{2}$.
Materials with negative Poisson's ratio are sometimes called auxetic materials.
We will see later that the theory is scale invariant for any $\nu$, but
the requirement of conformal invariance forces us to choose $\nu = -1$ for our boundary conditions.
Such materials are shape invariant under compression and have a stress-tensor with a vanishing trace.\footnote{ For a history of Poisson's ratio, see ref.\  \cite{greaves2013poisson}.
}

It is interesting that we land on the same boundary conditions through the requirement that $\Box^2$ be a symmetric operator.  In the text, we will impose a stronger requirement, that
these $\Phi^{(i)}$ be primary operators with respect to the boundary conformal group.  This requirement will allow us to
systematically analyze not just the $\Box^2 \phi$ theory but all of the  $\Box^k \phi$ theories as well as their fermionic
cousins $\slashed{\partial}^{2k+1} \psi$.  Moreover, the requirement of conformal invariance will fix the coefficients
in the boundary conditions uniquely, choosing $\nu = -1$ in the $\Box^2 \phi$ case.

In addition to material science, our work has a connection to results in conformal geometry \cite{case2018boundary, CaseLuo}.  A natural curved space generalization of $\Box^2$ is the Paneitz operator \cite{Fradkin:1982xc,Fradkin:1981jc}.  The Paneitz operator $\triangle_4$ is a fourth order differential operator which has the property that, under Weyl transformations of the metric, $g_{\mu\nu} \to \Omega^2 g_{\mu\nu}$, it transforms covariantly
\begin{equation}
\label{Paneitzrule}
\Omega^{-\frac{d}{2}-2} \triangle_4(g) \phi = \triangle_4(\Omega^2 g) \Omega^{-\frac{d}{2}+2} \phi \ .
\end{equation}
More generally, $\Box^k$ is promoted in curved space to a GJMS operator $\triangle_{2k}$
\cite{graham1992conformally}, and there is a similar
promotion of $\slashed{\pd}^{2k+1}$ to the spinor GJMS operator $\slashed{\triangle}_{2k+1}$~\cite{fischmann2013conformal,holland2001conformally,deBerredo-Peixoto:2001uob}. The case $\triangle_2 = \Box +  \frac{d-2}{4 (d-1)} R$ gives the familiar equation of motion for the conformally coupled scalar.
The rule (\ref{Paneitzrule}) guarantees that the curved space generalization of (\ref{Boxsquaredtheory}) is Weyl
invariant.  In flat space, we can then expect a residual conformal invariance -- the set of diffeomorphisms that leave the
metric invariant up to Weyl rescaling.

The relation between our theories and the GJMS operators leads to an interesting discussion about the stress-energy tensor
 \cite{Stergiou:2022qqj,Osborn:2016bev}.  Being careful about the higher derivative terms, the usual
 Noether procedure for translational symmetry can be used on our
 $\Box^k \phi$ and $\slashed{\partial}^{2k+1} \psi$ theories to derive a stress tensor.  However, this stress tensor is
 not traceless nor in fact a primary with respect to the conformal group.  However, it can be
 ``improved'' in a unique way to have these properties.\footnote{An extension of the usual Noether procedure that yields the improved stress tensor is described in~\cite{OsbornLectures}.}
 There is a corresponding stress tensor that can be derived by varying the curved space action, involving the GJMS
 operators, with respect to the metric.  A conjecture of  \cite{Stergiou:2022qqj,Osborn:2016bev} is that the ``improved'' stress tensor
 and the stress tensor that comes from the curved space action are generically the same.\footnote{  In fact, there is an obstruction to the construction of this stress tensor for the scalar theory
  which can occur in $d=2k$, $2k-2$, $2k-4$, \ldots.
 The stress tensor for the $\Box^2 \phi$ theory in $d=2$ cannot be made traceless, which has given rise to a lengthy discussion in the literature
 about scale vs.\ conformally invariant theories of elasticity \cite{Ferrari:1995gc,Karananas:2015ioa,Nakayama:2016cyh,Riva:2005gd}.
  }

As we have already stated, our interest here is in these conformal theories in the presence of a boundary.
 In studying the conditions under which $\triangle_4$ and $\triangle_6$ are symmetric operators, Case and Luo \cite{case2018boundary,CaseLuo} characterized boundary differential operators $B^3_i$, $i=0,1,2,3$ and $B^5_i$, $i=0, 1, \ldots, 5$, respectively.
 The $B^3_i$ in the flat space limit become perhaps not
 surprisingly precisely our $\Phi^{(i)}$.
 The $B^5_i$ also match our results for the $\Box^3 \phi$ theory, as we will see later.
 To our knowledge, explicit formulae for the
 higher order $\triangle_k$ are not yet available but see refs.~\cite{branson2001conformally,gover2018conformal}.

 \subsection*{Results}

 The work is dense with many new results about these $\Box^k\phi$ and $\slashed{\partial}^{2k+1}\psi$
 theories, and it is useful to tabulate the highlights:

 \begin{itemize}

 \item
 For the $\Box^k\phi$ theory, there are $2k$ boundary primary operators that one can construct from a single $\phi$ and derivatives.  We label them $\Phi^{(k,q)}$ where $q = 0, 1, \ldots, 2k-1$.  There is a natural pairing among the boundary primaries, $(\Phi^{(k,q)},  \Phi^{(k,2k-q-1)})$.  Conformal boundary conditions consist of choosing one member of each pair to set to zero.

 \item
 For the $\slashed{\partial}^{2k+1}\psi$ theory, there are $4k+2$ boundary primaries that one can construct from a single $\psi$ field, derivatives and gamma matrices.  We label them $\Psi^{(k,q)}_\pm$ where now $q = 0, 1, \ldots, 2k$.
 There is a similar natural pairing
 that now also involves the chirality with respect to the gamma matrix in the normal direction $\gamma_n$:
 $$(\Psi^{(k,q)}_+ , \Psi^{(k,2k-q)}_-) \ .$$  In this case, conformal boundary conditions consist of setting either $\Psi^{(k,q)}_+$ or $\Psi^{(k,2k-q)}_-$ to zero for each pair.

\item
In the case of the $\Box^2\phi$ theory, we explicitly show that the generators
of the conformal symmetry group are time independent given the boundary conditions.
For the $DD$, $ND$, and $DN$ cases, we find the usual Cardy boundary condition
that the normal-tangential component of the stress tensor must vanish,
$T^{na}|_{\rm bry}=0$ \cite{Cardy:1984bb}.  The $NN$ case is more subtle, as it requires a boundary contribution to the
stress tensor in order to maintain conformal symmetry.

\item
We characterize flows between the different boundary conditions.  These flows are caused by adding a term quadratic
in the (nonzero) boundary primaries to the boundary action.  In the case that such a term has scaling dimension less than $d-1$
(i.e.\ relevant in the language of the RG) both fields are set to zero at the infrared (IR) fixed point of the flow.

\item
We compute the $\langle \phi \phi \rangle$
and $\langle \psi \overline \psi \rangle$
two-point functions for the $\Box^k\phi$
and $\slashed{\partial}^{2k+1}\psi$ theories, respectively.

\item
Boundary Conformal Field Theories (BCFTs) come with a protected scalar operator of dimension $d$, the displacement operator, which
is sourced by deformations of the location of the boundary.  We compute the displacement two point function for  the
$\Box^2\phi$ theory, the $\Box^3\phi$ theory and the $\slashed{\partial}^3\psi$ theory with all possible conformal boundary conditions.

The strength of this two point function is related to an anomalous contribution to the trace of the stress tensor. The cases $d=3$ \cite{Herzog:2017kkj, Bianchi:2015liz}, 4 \cite{Herzog:2017xha} and 5 \cite{Chalabi:2021jud,FarajiAstaneh:2021foi} have been worked out in full details, while in \cite{Chalabi:2021jud} such a relation has been discussed perturbatively for any even-dimensional boundary or defect.

\item
We compute the free energy on a hemisphere for our theories.  The curved space requires promoting the $\Box^k$ and
$\slashed{\partial}^{2k+1}$ operators to their GJMS generalizations, but the GJMS operators on the sphere are
sufficiently simple \cite{branson1995sharp} that we are able
to extract the free energy.  While for unitary theories this quantity is expected to be monotonic under RG flow
\cite{Gaiotto:2014gha, Jensen:2015swa, Wang:2021mdq, Kobayashi:2018lil}, our theories are non-unitary and are not ordered by this quantity, as we see explicitly.

\item
We discuss a duality that can be constructed from a pair of our $\Box^k\phi$
theories with conjugate boundary conditions.
By adding the boundary marginal operator
\[
g\lsp \Phi_R^{(k,q)} \Phi_L^{(k,2k-q-1)}
\]
with a large coefficient $g$, we can tune to a related theory where the boundary conditions of these two theories, for
this particular pair of primaries, are swapped.  The theory is dual to the theory with swapped boundary conditions
and deformation $\frac{1}{g} \Phi_L^{(k,q)} \Phi_R^{(k,2k-q-1)}$.
The duality is a generalization of an effect discussed by Witten \cite{Witten:2001ua} in an AdS/CFT context.

 \end{itemize}

An outline of this work is as follows.
Section \ref{sec:background} sets up some conventions and provides further details about the definition of our
$\Box^k\phi$ and $\slashed{\partial}^{2k+1}\psi$ theories.
Section \ref{sec:primaries} is the heart of the work, providing a derivation of the boundary primaries, the corresponding
set of conformal boundary conditions, and a discussion of the flows between them.
The compatibility of our boundary conditions with the Cardy condition $T^{na} = 0$ is also discussed.
Then in section \ref{sec:apps},
we compute  the hemisphere free energy and the displacement two-point function (along with
a brief derivation of the $\langle \phi \phi \rangle$ and $\langle \psi \overline \psi \rangle$ two-point functions in these theories).  Section
\ref{sec:duality} is a brief discussion of a curious duality involving a pair of our boundary theories with conjugate boundary conditions.
A concluding section
\ref{sec:discussion}
mentions some future directions, in particular the problems of quantum gravity, interactions and low dimensions.

There are two appendices.
Appendix \ref{app:scalar-appendix} contains additional details about the higher derivative scalar theories. Appendix~\ref{app:scalar-propagator-appendix} derives the form of the $\Box^k\phi$ propagator without assuming conformal symmetry.
The constraints imposed on this propagator by demanding boundary conformal symmetry are explored in appendix~\ref{app:conformal-constraints-appendix}. Finally, the effect of a relevant boundary deformation on the form of the two-point function is computed in appendix~\ref{app:RGflow}.  Appendix \ref{app:constantterm} discusses the regularization scheme independent constant term in the hemisphere free energy in the odd dimensional cases.

\section{Background}

\label{sec:background}

\subsection{Boundaries in CFTs}

In this subsection, we review some of the basic aspects of boundaries in CFTs that will be useful in the following sections.  We begin by introducing the notation and conventions that we will use throughout unless otherwise noted in exceptional cases.  The background geometry $\CM$ on which we define our theories will typically be a $d$-dimensional Minkowski spacetime $\CM= \mathbb{R}^{d-1,1}$. We choose coordinates $x^\mu$ and mostly plus signature metric $\eta = \text{diag}(-1,+1,\ldots,+1)$ such that the line element is
\begin{align}
ds^2 = \eta_{\mu\nu}dx^\mu dx^\nu,
\end{align}
where the spacetime indices $\mu,\,\nu$ run from $0,\ldots,d-1$ and $x^0\equiv t$.

As written, the space has no boundary, but one can be suitably introduced by the specification and embedding of a boundary hypersurface. The boundary hypersurface takes a single spatial coordinate $x^n$ to be valued on the half-line $\mathbb{R}_{\geq0}$ such that the induced geometry on the boundary manifold $\mathbb{R}^{d-2,1}$ is simply the pullback of the bulk geometry to the boundary by the hypersurface embedding $x^n=0$ .  That is, we take as a coordinatization of $\pd\CM = \mathbb{R}^{d-2,1}$ to be $x^a$ where the boundary spacetime indices $a,\, b$ run from $0,\ldots, d-2$. The boundary metric and line element are $\eta_{ab}$ and
\begin{align}
ds_\parallel^2 = \eta_{ab} \lsp dx^a dx^b \ .
\end{align}
In order to distinguish bulk and boundary, we will denote quantities tangential to the boundary by a ${}_\parallel$ subscript, e.g.\ $f(x_\parallel)$ is a function solely of $x^a$ coordinates.

\subsection{Higher derivative free CFTs}

We start with a manifold ${\mathcal M}$  in $d$ dimensions, a metric $g_{\mu\nu}$ on ${\mathcal M}$,
and a scalar field $\phi(x)$.
We envision a quadratic action for $\phi(x)$ that is invariant both under diffeomorphisms and Weyl transformations.  Moreover, we assume that the Weyl
transformations act on both the metric,
$g_{\mu\nu} \to \Omega(x)^2 g_{\mu\nu}$,
and  on the scalar field,
$\phi(x) \to \Omega^{-\frac{d}{2}+k} \phi(x)$,
where $k \in {\mathbb N}$ and $\Omega(x)$ is a real function:
\begin{align}\label{eq:scalar-GJMS-action}
S^{(k)}_{\phi,\triangle} = -\frac{1}{2}\int d^dx\sqrt{|g|} \phi\triangle_{2k}\phi\,.
\end{align}
This action requires the existence of a special scalar differential operator
$\triangle_{2k}$ with the property that $\triangle_{2k}
\to \Omega^{-\frac{d}{2} - k } \triangle_{2k}\Omega^{\frac{d}{2} - k }$
under Weyl transformations.   These operators are the GJMS operators \cite{graham1992conformally}.

While the precise form of these operators, written out in  terms of curvatures and covariant derivatives, can be complicated, special cases can be
very simple to write down.  The special case that will occupy most of this paper is flat space, where the GJMS operator reduces to an
integer power of the Laplacian, $\triangle_{2k}|_{\text{flat space}} = \Box^k$:
\begin{align}\label{eq:scalar-box-action}
S^{(k)}_{\phi,\Box} = \frac{(-1)^k}{2} \int d^dx \sqrt{|g|} \lsp\phi\lsp\Box^k\phi\,.
\end{align}
We have introduced a sign to guarantee that the expression is positive definite.
Up to boundary terms, an alternative formulation of this action makes the positivity clear,
\begin{align}\label{eq:scalar-symmetric-action}
S^{(k)}_{\phi,\text{ sym}} = \frac{1}{2} \int d^dx \,
\pd_{\m_1}\ldots\pd_{\m_k}\phi\, \pd^{\m_1}\ldots\pd^{\m_k}\phi\ ,
\end{align}
where we have further assumed the Minkowski metric $\eta$.\footnote{We stress that in this case one would need to add additional boundary terms to ensure the theory preserves the maximum possible amount of spacetime symmetries in the presence of a flat edge, i.e.\ the subgroup $SO(d-1,2)$ of the bulk conformal group $SO(d,2)$. These additional terms would bring the action \eq{scalar-symmetric-action} back to \eq{scalar-box-action}.  For this reason, in the rest of the paper we find it convenient to work with the action \eq{scalar-box-action}.}

As a consequence of the invariance of eq.\ (\ref{eq:scalar-GJMS-action}) under diffeomorphism and Weyl transformations, the flat space
action
eq.\ (\ref{eq:scalar-box-action}) has a residual invariance under the conformal group, i.e.\ the set of diffeomorphisms that leave the
metric invariant up to Weyl transformations.  As a consequence, this family (\ref{eq:scalar-box-action}) of free scalar field theories forms
a nice set of examples of CFTs.  Unfortunately,
for all cases $k>1$, they are also known to have certain pathologies.  For example, when written
in a Hamiltonian formulation, the Hamiltonian is linear in one or more
conjugate momenta and hence unbounded below \cite{ostrogradsky1850memoire},
as we remind the reader in section \ref{sec:discussion}. This phenomenon is dubbed the Ostrogradsky instability. Furthermore,
the dimension of the scalar field $\Delta = \frac{d}{2} - k$ is less than the unitarity bound $\frac{d}{2}-1$
on scalar representations of the conformal group \cite{Mack:1975je}.  Hence these theories are non-unitary.
While these issues form a backdrop to the discussion here, we offer no solutions, instead focusing on the task
of understanding how to preserve conformal symmetry in the presence of a flat boundary.

A similar construction can be made for fermions using the spinor GJMS operator \cite{fischmann2013conformal,holland2001conformally,deBerredo-Peixoto:2001uob}:
\begin{align}\label{eq:fermion-GJMS-action}
S^{(k)}_{\psi,\triangle} = i\int d^dx |e| \lsp\overline\psi {\triangle\!\!\!\!\slash\,}_{2k+1}\psi~,
\end{align}
where $e$ denotes the determinant of an orthonormal frame and the Dirac adjoint is $\bar{\psi}=\psi^\dagger \gamma_0$.
In flat space in Minkowski coordinates and for a stationary frame, this action can be written in the symmetric form
\begin{align}\label{eq:fermion-symmetric-action}
S^{(k)}_{\psi,\text{ sym}} = \frac{i}{2}\int d^dx \ (\overline\psi\lsp \rpr_{2k+1} \psi  - \overline\psi\lsp \lpr_{2k+1}\psi)\,,
\end{align}
where $\slashed{\pd}_{2k+1}$ contains $2k+1$ powers of the Dirac operator $\slashed{\pd}=\gamma^\mu\partial_\mu$ and $\{\gamma^\mu,\gamma^\nu\}=2\eta^{\mu\nu}$.
Similar issues with unitarity are present in the fermionic theory, which we shall largely ignore, again focusing on the task of understanding conformal boundary conditions
for spinors.

\section{Boundary primaries and conformal boundary conditions}
\label{sec:primaries}

In this section, we identify a simple class of boundary conformal primaries in free BCFTs.
We consider two examples:
higher derivative theories of scalars and fermions.
We then show how these boundary primaries in higher derivative theories can be used
to construct boundary conditions consistent with a conserved boundary conformal symmetry.

\subsection{Boundary conformal primaries}

To begin, we recall the form of the bulk $d$-dimensional conformal algebra for generators of dilatations $D$, translations $P_\mu$, Lorentz rotations $M_{\mu\nu}$, and special conformal transformations $K_\mu$:
\begin{align}\begin{split}\label{eq:conformal-algebra}
&[D,\,P_\mu] = iP_\mu\,,\quad [D,\,K_\mu] = -i K_\mu\,,\quad [P_\m,\,M_{\n\la}] = i\eta_{\m\la}P_\n -i\eta_{\m\n}P_\la \,,\\\
&[P_\mu,\, K_\nu] = 2i ( M_{\m\n}- \eta_{\mu\nu} D )\,,\quad [K_\mu,\,M_{\n\la}] =i \eta_{\m\la}K_\nu -i\eta_{\m\n}K_\la\,,\\
&[M_{\m\n},\,M_{\la\r}] = i\eta_{\m\la}M_{\n\r} -i\eta_{\n\la}M_{\m\r} -i\eta_{\m\r} M_{\n\la} +i\eta_{\n\r}M_{\m\la}\,.
\end{split}\end{align}
Commutators not given are zero. These generators together generate the group $SO(d,2)$ in Lorentzian signature.
For a given bulk primary state $\ket{\CO_I} \equiv \CO_I(0)\ket{\O}$ formed by the action of the bulk primary operator $\CO_I(x)$ on the vacuum state, the action of the conformal generators must satisfy the following properties
\begin{align}
\begin{split}\label{eq:primary-state-properties}
D\ket{\CO_I}&=i\Delta\ket{\CO_I}\,,\\
M_{\m\n}\ket{\CO_I} &= -(M_{\m\n})_I{}^J\ket{\CO_J}\,,\\
K_\mu\ket{\CO_I} &= 0\,.
\end{split}
\end{align}

The introduction of a boundary breaks this group at least to $SO(d-1,2)$, the conformal group of the boundary.  Starting with the bulk generators
(\ref{eq:primary-state-properties}),
we lose $K_n$, $P_n$, and $M_{\mu n}$ from the symmetry algebra.  The conformal multiplet generated by the $P_\mu$ acting on
a primary ${\CO_I}$ will decompose into a tower of boundary multiplets.  From $\CO_I$, we can form new boundary primary operators of the schematic
form $(P_n^j + \ldots) | \CO_I \rangle$, each of which is annihilated by all of the $K_a$ but not necessarily by
$K_n$.
The precise form of the corrections we now explicate in the case of a scalar and spinor $\CO_I$.

A key lemma in the construction of the boundary primaries is found
by repeated use of eqs.~\eqref{eq:conformal-algebra} and \eqref{eq:primary-state-properties}.  One can inductively prove that the action of $K_\mu$ on a bulk descendant of the primary state $\ket{\CO_I}$ takes the form\footnote{We divide by $P_\mu$ in this formula purely as a book keeping device, in order not to remove the corresponding $P_\mu$ from the product.}
\begin{align}\label{eq:K-Pq-general-primary}
K_\m P_{\n_1}\ldots P_{\n_q}\ket{O_I} = 2 \prod_{i=1}^q P_{\n_i} \sum_{j=1}^q\left(\frac{1}{P_{\n_j}}(iM_{\m\n_j}-(\Delta+q-1)\eta_{\m\n_j})+\sum_{l=j+1}^q\frac{\eta_{\n_j\n_l} P_\m}{P_{\n_j}P_{\n_l}}\right)\ket{\CO_I}\,.
\end{align}

As a first example, let us consider building boundary primary states in a bulk theory of free scalars where the bulk primary states $\ket{\phi}$ have $\Delta = \frac{d-2k}{2}$ and obey $M_{\mu\nu}\ket{\phi}=0$ for all $\m,\,\n$. We denote by $\ket{\Phi^{(k,q)}}$ the boundary primary state that is constructed from the level $q$ descendants of $\ket{\phi}$.
By inspection of \eq{K-Pq-general-primary}, we can easily pick out combinations of bulk descendants of scalar primary states that will give boundary primaries at the first few levels
\begin{align}\label{eq:low-level-scalar-boundary-primaries}
\begin{split}
\ket{\Phi^{(k,0)}}&= \ket{\phi}\,,\\
\ket{\Phi^{(k,1)}}&= P_n\ket{\phi}\,,\\
\ket{\Phi^{(k,2)}}&=\left(P_n^2 - \frac{1}{2k-3}P^aP_a\right) \ket{\phi}\,.
\end{split}
\end{align}
It is worth noting that for $k=1$, $\Phi^{(1,2)} = \Box \phi$, which is the equation of motion.  In other words, for the usual free scalar theory,
there are only two nontrivial boundary primaries, $\Phi^{(1,0)}$ and $\Phi^{(1,1)}$.   The operators $\Phi^{(1,q)}$ for $q\geq 2$ are redundant, i.e.\ they vanish on the equations of motion.
We will see momentarily that more generally, there are exactly $2k$ boundary primaries.
Another interesting case is $k=\frac{3}{2}$ which corresponds to a scalar operator of dimension $\frac{d-3}{2}$, which saturates the unitarity bound of
a $d-1$ dimensional theory.\footnote{While we are mainly interested in theories with integer $k$, the argument in this subsection depends purely on representation theory of the conformal group.  
 Hence we are free to set
 $k$ and correspondingly $\Delta$ to any real or indeed complex value.
}
We expect such an operator to be in the kernel of the boundary Laplacian, $P_a P^a$.  It is perhaps not surprising
then that the boundary primary
$\Phi^{(\frac{3}{2},2)}$
corresponds to this null state $P_a P^a \ket{ \phi}$.

Lorentz invariance suggests the following ansatz generalizing the result (\ref{eq:low-level-scalar-boundary-primaries})
\begin{align}\label{eq:boundary-scalar-primary-ansatz}
\ket{\Phi^{(k,q)}} =\sum_{j=0}^{\lfloor \frac{q}{2} \rfloor} \a^{(k,q)}_{j}P_\parallel^{2j}P_n^{q-2j}  \ket{\phi}\,,
\end{align}
where $P_\parallel^2 = P^aP_a$.
Before computing the $\a^{(k,q)}_{j}$, we make some general observations about these candidate boundary primaries.
\begin{itemize}
\item
By orthogonality, we expect to find only one boundary primary for a given $(k,q)$ pair.  Indeed, the primary at level $q$ depends on $\lfloor \frac{q}{2} \rfloor +1$ coefficients
and must have a vanishing two-point function with all the boundary primaries of $P_n$ degree less than $q$.  If the $P_n$ degrees of two boundary primaries have different parity, then their two-point function automatically vanishes on the boundary.\footnote{This can be seen by taking appropriate derivatives of the bulk two-point function $\langle \phi \phi\rangle$ in a near-boundary expansion.}  There will be $\lfloor \frac{q}{2} \rfloor$ boundary primaries
of smaller degree and the same parity. Insisting that our level $q$ primary is orthogonal to all of them, one can fix the coefficients in \eq{boundary-scalar-primary-ansatz} up to an over-all normalization. We are thus left with a single boundary primary for each level.

\item
By using the equation of motion $\Box^{k} \phi = 0$, we can reduce the $P_n$ degree of a boundary primary below $2k$.  Thus we expect all
$\Phi^{(k,q)}$ boundary primaries with $q \geq 2k$ to be redundant.

\end{itemize}
Taken together, these two observations imply that there will be exactly $2k$ boundary primaries of the form (\ref{eq:boundary-scalar-primary-ansatz}), one for each $q = 0, 1, \ldots, 2k-1$.

The goal then is to determine the coefficients $\a^{(k,q)}_j$ such that $K_a \ket{\Phi^{(k,q)}}=0$.
A special case of (\ref{eq:K-Pq-general-primary}) useful in computing the $\a^{(k,q)}_j$ is
\begin{align}
K_a P_\parallel^{2l}P_n^m\ket{\phi} = \left(2l(2k-2m-2l-1)P_n^2 +m(m-1)P_\parallel^{2}\right)P_aP_\parallel^{2(l-1)}P_n^{m-2}\ket{\phi}\,.
\end{align}
Solving $K_a\ket{\Phi^{(k,q)}}=0$ gives a recursion relation for the coefficients
\begin{align}\label{eq:a-kq-j-recursion}
\a^{(k,q)}_j = -\frac{(2j-q-2)(2j-q-1)}{2j(2k+2j-2q-1)}\a^{(k,q)}_{j-1}\,,
\end{align}
with, choosing a convenient overall normalization, initial data $\a^{(k,q)}_0 =1$. The solution to \eq{a-kq-j-recursion} is easily found,
\begin{align}\label{eq:scalar-boundary-primary-coeffs}
\a^{(k,q)}_{j}= \frac{(-1)^j}{2^j}\frac{q!}{j!(q-2j)!}\frac{(2k-2q-1)!!}{(2k+2j-2q-1)!!}\,,
\end{align}
which determines the coefficients in \eq{boundary-scalar-primary-ansatz} such that $\ket{\Phi^{(k,q)}}$ is a boundary primary.
As expected, we find just one solution for each $(k,q)$ pair.

Given the ability to employ the equation of motion,
the special case $\Phi^{(k,2k)}$ is worth a closer look.  In this case, the coefficients $\alpha_j^{(k,q)}$ reduce to the binomial distribution
\begin{align}
\a^{(k,2k)}_j = \binom{k}{j} \, ,
\end{align}
and the corresponding operator $\Phi^{(k,2k)} = \Box^k \phi$ is the equation of motion for the scalar field.
In other words $\Phi^{(k, 2k)}$ is redundant and therefore null.
More generally, all further $\Phi^{(k,q)}$ with $q=2k+i$ for $i \geq 0$ can be written as $\Box^k$ times a polynomial in $P_n$ and $P_\parallel^2$,
\begin{equation}
\ket{\Phi^{(k,2k+i)}}=\left(\sum_{j=0}^{\lfloor \frac{i}{2} \rfloor} \frac{(2j-1)!! (2k-2j+2i-1)!!}{(2k+2i-1)!!} \binom{i}{2j} P_\parallel^{2j}P_n^{i-2j}\right)P^{2k} \ket{\phi}\ ,
\end{equation}
such that these operators
are also null.

We can also consider, as a concrete example, the theory of higher derivative free fermions (\ref{eq:fermion-symmetric-action}), whose primary states $\ket{\psi}$ have dimension $\Delta = \frac{d-1}{2}-k$.  For this theory, the analysis is changed slightly as compared to the bulk higher derivative scalar since the bulk spinor primaries $\ket{\psi}$ live in non-trivial irreducible representations (irreps) of the bulk Spin group and thus are not annihilated by the conformal generator\footnote{We point out an important relative sign.  These generators will give rise to the commutation relations (\ref{eq:conformal-algebra}).
However, in applying these generators to a state,
we need to include the extra minus sign that appears in (\ref{eq:primary-state-properties}).
}
\begin{align}\label{eq:M-munu-spinor}
M_{\mu\nu}=- \frac{i}{2}\g_{\m\n}\,.
\end{align}
Further in the presence of a boundary, care must be taken as the bulk spinor primary does not live in an irrep of the boundary Spin group. Indeed, we can decompose the bulk $2^{\lfloor\frac{d}{2}\rfloor}$-dimensional Dirac spinor into boundary $2^{\lfloor\frac{d}{2}\rfloor -1}$ spinors, which are eigenfunctions of $\g_n$.
In other words, under the action of the boundary projector $\Pi_\pm = \frac{1}{2}(\mathds{1} \pm \g_n)$,
the boundary primary states $\ket{\psi_\pm} = \Pi_\pm \ket{\psi}$ live in irreps of the boundary Spin group.

In starting the search for a general expression for boundary spinor primaries $\ket{\Psi^{(k,q)}}$, a modification to
\eq{boundary-scalar-primary-ansatz}
needs to be made, where we can contract a free index on $P_a$ not just with another $P^a$ but also with $\g^a$.
The first few primaries take the form
\begin{align}
\label{fermionbryprimaries}
\begin{split}
\ket{\Psi^{(k,0)}} &= \ket{\psi}\,, \\
\ket{\Psi^{(k,1)}} &= \left(P_n\g^n -\frac{1}{2k-1}P_a\g^a\right)\ket{\psi}\,,\\
\ket{\Psi^{(k,2)}} &= \left(P_n^2 +\frac{2}{2k-3}P_aP_n\g^a\g^n -\frac{1}{2k-3}P_\parallel^2\right)\ket{\psi}\,,
\end{split}
\end{align}
which then also need to be projected onto irreps of the boundary Spin group by appropriate use of $\Pi_\pm$.
As $\Pi_\pm$ does not commute with $\gamma^a$, the projection requires that the terms in a near boundary expansion of
$\psi$ have alternating chirality with respect to $\gamma_n$.  If the boundary limit of $\psi$ has positive chirality, then
$\partial_n \psi$ must have negative chirality, $\partial_n^2 \psi$ has again positive chirality and so on.

As a consistency check, we can examine these boundary primaries for $k=0$, where we expect only $\Psi^{(0,0)}$ to be nontrivial.
Indeed we find $\Psi^{(0,1)}$ reduces to the Dirac operator acting on $\ket{\psi}$.
Furthermore $\Psi^{(0,2)}$
 becomes $3 P_n^2 - 2 P_a P_n \gamma^a \gamma^n + P_{\parallel}^2$ acting on $\ket{\psi}$ which factors as $(3\slashed{P}_n + \slashed{P}_\parallel)( \slashed{P}_n + \slashed{P}_\parallel)$, with the Dirac operator to the right. Thus it is also null.
 The $k=\frac{1}{2}$ case is also interesting, as the dimension of $\psi$ saturates the unitarity bounds for a theory in $d-1$ dimensions.  Indeed, $\Psi^{(\frac{1}{2},1)}$ becomes the $d-1$ dimensional Dirac operator acting on $\psi$ in this case.

The computation of the most general boundary spinor primary states constructed from level $q$ descendants of bulk spinor primary states follows the logic of the derivation of \eq{scalar-boundary-primary-coeffs} for the case of higher derivative scalars, and so we will quote the result without reproducing the steps:
\begin{align}
\ket{\Psi^{(k,q)}} = \left(\sum_{j=0}^{\lfloor\frac{q}{2}\rfloor} \a^{(k,q)}_{j}P_\parallel^{2j}P_n^{q-2j}+(-1)^q\sum_{i=0}^{\lfloor\frac{q-1}{2}\rfloor} \frac{(q-2i)\a^{(k,q)}_{i}}{2k+2i-2q+1} P_a P_\parallel^{2i} P_n^{q-2i-1}\g^a\g_n\right)\g_n^q \ket{\psi}\,.
\end{align}
Again, we must project this object using $\Pi_\pm$.
Generalizing the discussion above for $k=0$, note that at level $q=2k+1$, the differential operator acting on $\ket{\psi}$ can be rewritten as the kinetic operator,
$\slashed{\pd}_{2k+1}$.
Hence, there exists a redundant state in the boundary spectrum precisely at $q=2k+1$ for all $k$.
 Indeed, analogous to the scalar case, we claim that for all $q \geq 2k+1$, $\Psi^{(k,q)}$ can be written as a polynomial
in $P_n$, $P_a$, $\gamma_n$ and $\gamma_a$ multiplied by $\slashed{\pd}_{2k+1}$ acting on $\psi$
 and is thus redundant.  In other words, only $\Pi_\pm \Psi^{(k,q)}$ for $q = 0, 1, \ldots, 2k$ are
 nontrivial boundary states.

\subsection{Conformal boundary conditions}

The standard lore of conformal boundary conditions in general dimensions is that in order to preserve conformal symmetry at the boundary, the generalized Cardy condition, ${T^{na}|_{\pd\CM}=0}$, necessarily holds.  In this subsection,
we consider the examples of higher derivative free CFTs introduced above and explore a different definition of conformal boundary conditions.
We define conformal boundary conditions in terms of a set of
 boundary primaries $\{ {\mathcal O}_I \}$ that annihilate the vacuum, $ {\mathcal O}_I | \Omega \rangle = 0$,  and satisfy the variational principle.

It turns out that these definitions of conformal boundary condition may be different.
While we do not have
 good expressions for $T^{na}$ in general, in the case of the $\Box^2 \phi$ theory we can check that the condition $T^{na}|_{\pd\CM}=0$ is too restrictive.

\subsubsection*{Higher derivative scalar BCFT}\label{sec:scalar-primaries}
We begin with the higher derivative scalar CFT in \eq{scalar-box-action}.  Varying with respect to $\phi$ generates a boundary term
\begin{align}\label{eq:scalar-box-k-variation}
\delta S^{(k)}_{\phi,\Box} = \ldots + \frac{(-1)^{k}}{2} \sum_{j=0}^{2k-1}\sum_{l=0}^{k-1-\lfloor\frac{j}{2}\rfloor} (-1)^j\frac{k!}{l!(k-l)!}\int d^{d-1}x_\parallel \pd_n^j\delta\phi \Box^l_{\parallel}\pd_n^{2k-2l-j-1}\phi\,,
\end{align}
where the terms in $\ldots$ are bulk variations that automatically vanish on-shell. Moving forward, we will drop these bulk terms that vanish using the equations of motion.

To illustrate how the boundary variation in \eq{scalar-box-k-variation} is built out of boundary conformal primary operators, let us first consider the case of $k=2$ and then return to generic $k$.  The boundary variation takes a fairly simple form
\begin{align}\label{eq:scalar-box-2-variation}
\delta S^{(2)}_{\phi,\Box}=\frac{1}{2}\int d^{d-1}x_\parallel(\phi( \pd_n^3\delta\phi+2\pd_n\Box_\parallel\delta\phi)-(\pd_n^3\phi +2\pd_n\Box_\parallel\phi)\delta\phi +\pd_n^2\phi\pd_n\delta\phi-\pd_n\phi\pd_n^2\delta\phi)\,.
\end{align}
Recalling the form of the boundary conformal primary states in \eq{boundary-scalar-primary-ansatz}, we can write the corresponding boundary conformal primary operator $\Phi^{(k,q)}$, i.e.\ $\ket{\Phi^{(k,q)}}\equiv i^q\Phi^{(k,q)}\ket{\Omega}$, and find that for $k=2$ the boundary conformal primaries up to $q=3$ are
\begin{align}\begin{split}
\Phi^{(2,0)}&=\phi\,,\\
\Phi^{(2,1)}&=\pd_n\phi\,,\\
\Phi^{(2,2)}&=(\pd_n^2 -\Box_\parallel)\phi\,,\\
\Phi^{(2,3)}&= (\pd_n^3+3\pd_n \Box_\parallel)\phi\,.
\end{split}\end{align}
After a bit of rearranging, \eq{scalar-box-2-variation} can be written as
\begin{align}
\delta S^{(2)}_{\phi,\Box} = \frac{1}{2}\int d^{d-1}x_\parallel(\Phi^{(2,0)}\delta\Phi^{(2,3)}-\Phi^{(2,1)}\delta\Phi^{(2,2)}+\Phi^{(2,2)}\delta\Phi^{(2,1)}-\Phi^{(2,3)}\delta\Phi^{(2,0)})\,.
\end{align}
This form is quite suggestive as it is an integral over an alternating sum of the variation of a boundary primary $\Phi^{(2,q)}$ multiplied by $\Phi^{(2,3-q)}$. Note that the boundary primaries $\Phi^{(k,q)}$ and $\Phi^{(2,2k-q-1)}$ have conformal dimensions ${\Delta = \frac{d-2k+2q}{2}}$ and ${\tilde{\Delta} = \frac{d+2k-2q-2}{2}}$, respectively. These obey the relation $\tilde{\Delta}=d-1-\Delta$, just like a primary operator and its shadow operator. Since the conformal symmetry preserved by the flat boundary is $SO(d-1,2)$, we will sometimes refer to $\Phi^{(k,2k-q-1)}$ as the shadow of $\Phi^{(k,q)}$.

By computing the boundary variation for other low values of $k$, which we will not reproduce here, we are then led to the following ansatz for the boundary variation for generic $k$:
\begin{align}\begin{split}\label{eq:scalar-box-k-variation-ansatz}
\delta S^{(k)}_{\phi,\Box} &= \frac{(-1)^{k}}{2} \int d^{d-1}x_\parallel \sum_{j=0}^{2k-1}(-1)^j \Phi^{(k,j)}\delta\Phi^{(k,2k-j-1)} \\
&= \frac{(-1)^{k}}{2}\int d^{d-1}x_\parallel \sum_{j=0}^{k-1}(-1)^j\left( \Phi^{(k,j)} \delta\Phi^{(k,2k-j-1)}-\delta\Phi^{(k,j)}\Phi^{(k,2k-j-1)}\right).
\end{split}\end{align}
To prove that this is indeed the correct expression for the boundary variation, we need to compare to \eq{scalar-box-k-variation}, and so it is helpful to recast \eq{scalar-box-k-variation-ansatz} in terms of derivatives of $\phi$ and $\delta\phi$. Using the form of the primaries in \eq{boundary-scalar-primary-ansatz}, one can expand the product $\Phi^{(k,j)}\Phi^{(k,2k-j-1)}$,
\begin{align}
 \sum_{j=0}^{2k-1}(-1)^j \Phi^{(k,j)}\Phi^{(k,2k-j-1)}  = \sum_{j=0}^{2k-1}\sum_{l=0}^{k-1-\lfloor\frac{j}{2}\rfloor}(-1)^j \b^{(k,j)}_l \pd^j_n \phi \Box^l_\parallel \pd_n^{2k-2l-j-1}\phi\,.
\end{align}
Re-ordering the sums, one can identify $\beta^{(k,j)}_l$ as a quadratic expression in the $\alpha^{(k,j)}_l$ and check that indeed
\begin{align}
\b^{(k,j)}_l = \sum_{i=0}^{k-1-\lfloor \frac{j}{2}\rfloor}\a^{(k,j+2i)}_i\a_{l-i}^{(k,2k-2i-j-1)} = \frac{k!}{l!(k-l)!}\,,
\end{align}
as required in order that \eq{scalar-box-k-variation} agree with \eq{scalar-box-k-variation-ansatz}.
  Thus, for the scalar higher derivative CFT, solving the boundary variational problem is equivalent to a vanishing condition on the boundary action built as a sum of products of relevant boundary primaries and their shadows.

  Given a boundary term in the variation of the action of the form
  \begin{align}\label{eq:bry_var}
  \frac{1}{2}\int d^{d-1} x_\parallel \left(  \Phi^{(k,j)} \delta\Phi^{(k,2k-j-1)}-\delta\Phi^{(k,j)}\Phi^{(k,2k-j-1)} \right) \ ,
  \end{align}
  we can mutate it into a more canonical form by adding the purely marginal boundary correction to the original action
  \begin{align}
  \label{bryctrterm}
S_{\text{bry}} = -\frac{1}{2}\int d^{d-1} x_\parallel \Phi^{(k,j)} \Phi^{(k,2k-j-1)} \, ,
  \end{align}
  which will cancel one of the original terms leaving
  \begin{align}
 -  \int d^{d-1} x_\parallel \delta \Phi^{(k,j)} \Phi^{(k,2k-j-1)} \, .
  \end{align}
 Assuming $j < k$, we can make an analogy between this term and the $\delta \phi \, \partial_n \phi$ boundary term that shows up in the standard
 variational analysis of a free scalar field.  We can apply either Dirichlet $\delta \phi = 0$ boundary conditions or Neumann $\partial_n \phi = 0$ boundary conditions.  By extension, we make the same definition here, Dirichlet for $\delta \Phi^{(k,j)} = 0$ and Neumann for $\Phi^{(k,2k-j-1)} = 0$, assuming
 that $j < k$, so that $\Phi^{(k,2k-j-1)}$ will involve more $\partial_n$ derivatives in its definition than $\Phi^{(k,j)}$.
 The usual Dirichlet condition where $\Phi^{(k,j)}$ is set equal to a generically nonzero constant is not possible, as it introduces a scale.  The only possible scale invariant Dirichlet conditions are $\Phi^{(k,j)} = 0$ and infinity.  We rule out the case $\Phi^{(k,j)} \to \infty$ in order to have finite field configurations.

The variational principle leads us then naturally to consider the following set of conformally invariant boundary conditions.  We have a $k$-tuple of
pairs of boundary primaries
\begin{align}\label{eq:scalar_bc_pairs}
\biggr( (\Phi^{(k,0)}, \Phi^{(k,2k-1)} ), (\Phi^{(k,1)}, \Phi^{(k,2k-2)} ), \ldots , (\Phi^{(k,k-1)}, \Phi^{(k,k)} ) \biggl) \ .
\end{align}
For each pair, we can set either $\Phi^{(k,j)} =0$, which we call Dirichlet ($D$), or its shadow involving a larger number $\partial_n$ derivatives  to zero, $\Phi^{(k,2k-j-1)} =0$, which we call Neumann ($N$).  In the $k=1$ case, we recover the usual $D$ and $N$ cases.  For $k=2$, there are four possibilities,
$DD$, $ND$, $DN$, and $NN$, and so on for larger $k$.

These boundary conditions are manifestly conformally invariant because they are constructed from the boundary primaries, which have the appropriate
transformation laws under the boundary conformal group.  Note that a more general boundary condition constructed from a linear combination of these boundary
primaries would necessarily introduce a scale,  e.g.\ $\Phi^{(k,0)} + \alpha \lsp\Phi^{(k,1)} = 0$ introduces a scale through $\alpha$.

In our discussion of boundary conditions, we added the particular marginal boundary action eq.~(\ref{bryctrterm}) for each pair of operators $(\Phi^{(k,j)},\Phi^{(k,2k-j-1)})$. The particular choice of over-all sign in eq.~(\ref{bryctrterm}) ensured that the first term in \eq{bry_var} cancels in the boundary variation of the action. Adding a boundary action with the opposite over-all sign, i.e.\ $-S_{\text{bry}}$, one instead removes the second term. With the latter choice, a Dirichlet boundary condition for $\Phi^{(k,j)}=0$ with $j<k$ can be obtained as a free boundary condition, where $\delta \Phi^{(k,2k-j-1)}$ is left arbitrary and its coefficient in the variational argument is interpreted as an equation of motion for the field. Thus, by suitably choosing the sign of the boundary terms for each pair in \eq{scalar_bc_pairs}, one can arrange for an arbitrary $X_0 X_1 \cdots X_{k-1}$ type boundary condition, where $X_i \in \{N, D \}$, to emerge from boundary equations of motion. 

As an additional remark, we notice that had we started our discussion with the symmetric action in \eq{scalar-symmetric-action} instead of \eq{scalar-box-action}, we would have found a different set of boundary conditions, some of them not conformally (even though scale) invariant. This is due to the fact that the action \eq{scalar-symmetric-action} does not preserve the full $SO(d-1,2)$. To elucidate this fact, let us consider for simplicity the case $k=2$. The difference between the two actions is a boundary term that may be written as
\beq
S^{(k=2)}_{\phi,\Box} - S^{(k=2)}_{\phi,\text{ sym}} = \frac{1}{2} \int d^{\lsp d-1}x_{||} \, \left[\Phi^{(2,0)}\Phi^{(2,3)} - \Phi^{(2,1)}\Phi^{(2,2)} -2\lsp \Box_{||} \phi \, \Phi^{(2,1)} \right] \, .
\eeq
While the first two terms preserve the boundary conformal group, the last one breaks it unless either $\phi(0,x_{||}) = 0$ or $\partial_n \phi(0,x_{||}) = 0$.

A beautiful aspect of this system is that it allows for a variety of boundary RG flows.  Whenever there is a Neumann boundary condition
$\Phi^{(k,2k-j-1)} = 0$, there is a corresponding relevant boundary operator that can be added to the action, namely $c \int d^{d-1} x_\parallel  (\Phi^{(k,j)})^2$ where $c$ is a coupling constant  with
positive mass dimension.  The variational principle leads to the boundary equation of motion $\Phi^{(k,2k-j-1)} = c\lsp \Phi^{(k,j)} $.  Because of its
positive mass dimension, $c$ effectively becomes infinite at low energies and the system is driven to the corresponding Dirichlet boundary condition
$\Phi^{(k,j)} = 0$ through the flow.

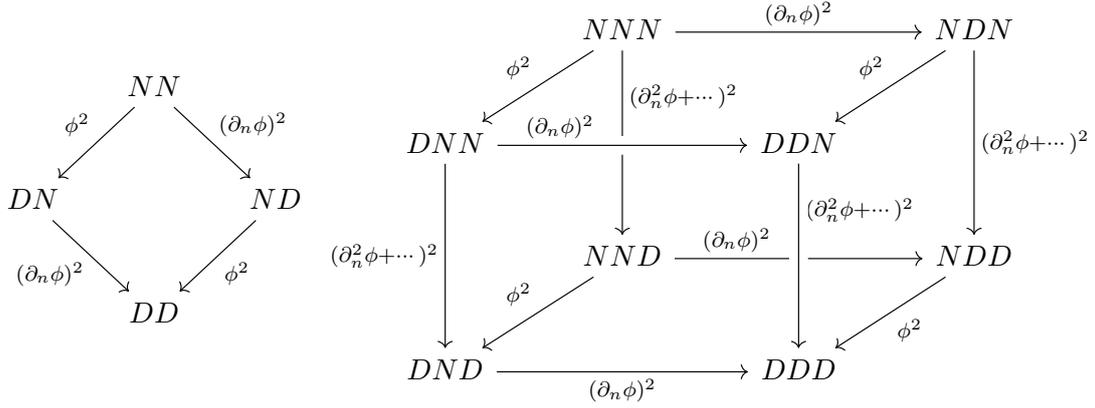
\begin{figure}
\centering
\begin{tikzcd}[row sep=2.5em, column sep=1.5em]
& NN \ar{dl}[swap]{\phi^2} \ar{dr}[]{(\partial_n\phi)^2} \\
DN \ar{dr}[swap]{(\partial_n\phi)^2} & & ND \ar{dl}{\phi^2} \\
& DD
\end{tikzcd}
\begin{tikzcd}[row sep=2.5em, column sep=2.5em]
& NNN \ar{dl}[swap]{\phi^2} \ar{rr}{(\partial_n\phi)^2} \ar{dd}[near start]{(\partial_n^2\phi+\cdots)^2}
& & NDN \ar{dd}{(\partial_n^2\phi+\cdots)^2} \ar{dl}[swap]{\phi^2} \\
DNN \ar[crossing over]{rr}[near start]{(\partial_n\phi)^2} \ar{dd}[swap]{(\partial_n^2\phi+\cdots)^2}
& & DDN \ar{dd}[near start]{(\partial_n^2\phi+\cdots)^2} \\
& NND \ar{rr}[near start]{(\partial_n\phi)^2} \ar{dl}[swap]{\phi^2}
& & NDD \ar{dl}{\phi^2} \\
DND \ar{rr}[swap]{(\partial_n\phi)^2} & & DDD \ar[crossing over, leftarrow]{uu}
\end{tikzcd}
\caption{
 \label{fig:flows}
 Some of the boundary RG flows induced in the $k=2$ system (left) and $k=3$ system (right).  There are additional flows, not shown, along the diagonals of the square and the diagonals of the faces of the hypercube, induced by $c\lsp \Phi^{(k,q)} \Phi^{(k,q')}$, $q \neq q'$, type deformations.
 }
\end{figure}

A nice way to visualize this flow is in terms of a $k$-dimensional hypercube.  We place the various choices
$X_0 X_1 \cdots X_{k-1}$
 of boundary condition
 at the corners of the cube and join them by edges corresponding to the relevant operators that induce the boundary RG flow.
 The cases $k=2$ and $k=3$ are shown in Figure \ref{fig:flows}.

 One could ask about a more general quadratic boundary deformation $c\lsp  \Phi^{(k,q)} \Phi^{(k,q')}$, for $q \neq q'$.  The first comment is that
 we must select $q$ and $q'$ such that neither the boundary conditions $\Phi^{(k,q)} = 0$ nor $\Phi^{(k,q')} = 0$ apply.
 If they do, $\Phi^{(k,q)} \Phi^{(k,q')}$ is an example of a redundant operator, i.e.\ one that vanishes by the (boundary) equations of motion.
 We still need to consider the remaining cases where neither $\Phi^{(k,q)}$ nor $\Phi^{(k,q')}$ are used as a boundary condition.
 Two of the boundary equations of motion will be modified by the additions $c\lsp \Phi^{(k,q)}$ and $c\lsp \Phi^{(k,q')}$, which in the limit $c \to \pm \infty$
 classically should set both of these fields to zero at a critical point\footnote{  In the context of the path integral and interactions, the story is more complicated.  For the $\Box \phi$ theory
  in four dimensions with a $\phi^4$ bulk interaction and $c\phi^2$ boundary term, $c>0$ is associated with Dirichlet boundary conditions while at the level of mean field theory, $c<0$ produces the extraordinary phase transition where
  $\langle \phi \rangle \sim 1/x_n$
  \cite{Lubensky:1975zz}.
  In our analysis, we see a difference between positive and negative $c$ in the Green's function.  In appendix \ref{app:RGflow}, we show that there is a pole in the Green's function for a negative value of $c \sim k_\parallel^{2(k-q)-1}$.
  }.  By adding such boundary deformations and tuning $c$ to be large, we can now additionally move along the diagonals of the faces of the hypercube.  Furthermore,
 we can move in either direction along the edges or diagonals, toward the
 IR or toward the UV, depending on the mass dimension of $c$.

\subsubsection*{Higher derivative fermion BCFT}\label{sec:fermion-primaries}

We will be less thorough in our analysis of the higher derivative fermion theories and limit our analysis to the theory
in $d$ dimensions which satisfies the Dirac-type equation $\slashed{\partial}^3 \psi = 0$.
In particular, we consider the action
\begin{align}
S = \frac{i}{2} \int d^d x \left( \bar{\psi} \lsp\slashed{\partial}^3 \psi + \overline{\slashed{\partial}^3 \psi}\lsp \psi \right) \, .
\end{align}
As in the higher derivative scalar theory, a remarkable simplification happens when we write the variation of the action
in terms of the fermion boundary primaries (\ref{fermionbryprimaries}).
The bulk part of the variation leads to the
Dirac-type equation of motion $\slashed{\partial}^3 \psi = 0$ while the boundary contribution can be written in the form
\begin{align}\label{eq:fermion-variation-boundary}
\delta S|_{\rm bry} = \frac{i}{2} \int_{x_n=0} d^{d-1} x \biggl( & + \overline{\Psi}^{(1,0)} \gamma_n \delta \Psi^{(1,2)}
-\delta\overline{\Psi}^{(1,0)} \gamma_n \Psi^{(1,2)}  \nonumber \\
& + \overline{\Psi} ^{(1,1)}\gamma_n \delta \Psi^{(1,1)}
- \delta \overline{\Psi} ^{(1,1)}\gamma_n  \Psi^{(1,1)}  \\
& + \overline{ \Psi}^{(1,2)} \gamma_n \delta \Psi^{(1,0)}
-  \delta \overline{\Psi}^{(1,2)} \gamma_n \Psi^{(1,0)} \nonumber
\biggr) \, ,
\end{align}
where $\overline{\Psi}^{(k,q)}=(\Psi^{(k,q)})^\dagger \gamma_0$.
This expression requires some further processing, because setting all the $\Psi^{(k,q)}$ to zero is too strong a boundary condition.
It is useful to decompose the boundary primaries into eigenvectors
with respect to the $\gamma_n$ matrix, which reduces them to representations of the boundary spin group.
We have $\Psi^{(k,q)} = \Psi_+^{(k,q)} + \Psi_-^{(k,q)}$ where $ \gamma_n \Psi_\pm^{(k,q)} = \pm \Psi_\pm^{(k,q)}$.
As we work in Lorentzian signature and $\overline \psi$ involves a $\gamma^0$ matrix, the nonvanishing inner products are
$\overline \Psi^{(k,q)}_\mp \Psi^{(k,q')}_\pm$.

If we study the contribution $ \overline{\Psi}_+^{(k,q)} \delta \Psi_-^{(k,2k-q)}
- \delta\overline{ \Psi}_+^{(k,q)}  \Psi_-^{(k,2k-q)}$ to $\delta S|_{\rm bry}$, we find we can add a boundary contribution to the
original action of the form
\begin{align}
S_{\rm bry} =  \pm \frac{i}{2}\int d^{d-1} x \, \overline \Psi^{(k,q)}_+ \Psi^{(k,2k-q)}_-\,,
\end{align}
which will leave
either $\overline{\Psi}_+^{(k,q)} \delta \Psi_-^{(k,2k-q)}$ or $- \delta\overline{ \Psi}  _+^{(k,q)} \Psi_-^{(k,2k-q)}$
on the boundary.
Depending on the choice of sign, then, we obtain either the free boundary condition $\overline \Psi^{(k,q)}_+ = 0 $ or the free boundary
condition $\Psi^{(k,2k-q)}_- = 0$.
In fact a reality constraint on the action means that if we add $\overline \Psi^{(k,q)}_+ \Psi^{(k,2k-q)}_-$ to the action, we also must add $\overline \Psi^{(k,2k-q)}_- \Psi^{(k,q)}_+$.  This reality constraint in turn means that if we set $\Psi^{(k,q)} = 0$ as a boundary condition, then we must also set $\overline \Psi^{(k,q)} = 0$, which is consistent with the definition of the Dirac adjoint.

  While we have only presented the analysis for the $\slashed{\partial}^3 \psi$ theory, a similar story
holds true for the original Dirac fermion $\slashed{\partial} \psi = 0$, and we believe these boundary conditions will
satisfy the variational principle for all $\slashed{\partial}^{2k+1} \psi = 0$ theories.

Similar to the $NDN\cdots$ notation for the higher derivative scalar theory, we can introduce a notation for the fermion boundary conditions, $+\!-\!-\!+ \cdots$, where the sign indicates which $\gamma_n$ chirality of the boundary primary we set to zero.
For the $k=0$ theory, we have $\pm$ boundary conditions which corresponds to setting
$\Psi^{(0,0)}_+ = 0$ or $\Psi^{(0,0)}_- = 0$.  For the $k=1$ theory, we have instead eight boundary conditions corresponding to the three pairs of boundary primaries
$(\Psi^{(1,0)}_+, \Psi^{(1,2)}_-)$, $(\Psi^{(1,2)}_+,\Psi^{(1,0)}_-)$, and $(\Psi^{(1,1)}_+, \Psi^{(1,1)}_-)$.
For example, $+\!+\!-$ would correspond to setting
\begin{align}
    \Psi^{(1,0)}_+ = 0 \ ,  \; \; \; \Psi^{(1,2)}_+ = 0 \ ,\; \; \; \Psi^{(1,1)}_- = 0 \ .
\end{align}

Again similar to the scalar story, there are a variety of marginal, relevant and irrelevant boundary perturbations we can add
to the theory to induce a flow between the boundary conditions.  Generically, given a set of boundary conditions, if
$\overline \Psi^{(k,q)}_\pm \Psi^{(k,q')}_{\mp}$ is nonzero, then we anticipate that adding
\begin{align}\label{eq:fermion-boundary-double-trace}
c \int d^{d-1} x \, \left(\overline \Psi^{(k,q)}_\pm \Psi^{(k,q')}_{\mp} + \overline \Psi^{(k,q')}_{\mp} \Psi^{(k,q)}_\pm \right)
\end{align}
to the boundary in the limit $c \to \infty$ will change the boundary conditions to $\Psi^{(k,q)}_\pm =0$, and $\Psi^{(k,q')}_{\mp} =0$.\footnote{ It would be interesting to think about how the story is further enriched by the possibility of a $\gamma_5$ matrix in even $d$.
 In particular, it is known that a  $\gamma_5$ matrix allows for a family of boundary conditions in the single derivative Dirac
 fermion case, $(1 + e^{i\beta \gamma_5} \gamma_n) \psi = 0$ where $\beta \in {\mathbb R}$.  We
 leave this story for the future.
}

In the case of the $\slashed{\partial}^3 \psi = 0$ theory,
the eight boundary conditions break up into four groups related by these types of boundary flows.
The two cases $+\!+\!+$ and $-\!-\!-$ are isolated.  There is no possible boundary mass term to write down because such a
boundary mass must involve a primary of each $\gamma_n$-chirality, $+$ and $-$.

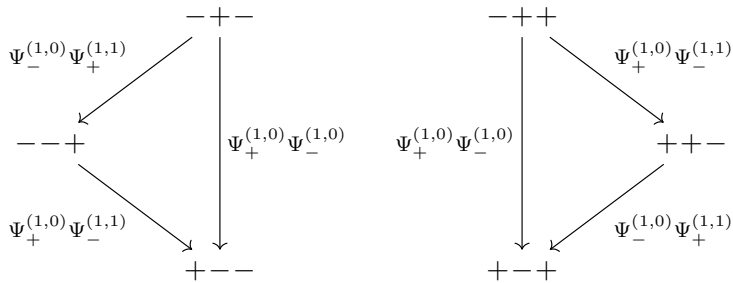
\begin{figure}
\centering
\begin{tikzcd}[row sep=3em, column sep=2.5em]
& -\!+\!- \ar{dl}[swap]{\Psi_-^{(1,0)}\Psi_+^{(1,1)}} \ar{dd}[]{\Psi_+^{(1,0)}\Psi_-^{(1,0)}} \\
-\!-\!+ \ar{dr}[swap]{\Psi_+^{(1,0)}\Psi_-^{(1,1)}} \\
& +\!-\!-
\end{tikzcd}
\hspace{6pt}
\begin{tikzcd}[row sep=3em, column sep=2.5em]
-\!+\!+ \ar{dr}[]{\Psi_+^{(1,0)}\Psi_-^{(1,1)}} \ar{dd}[swap]{\Psi_+^{(1,0)}\Psi_-^{(1,0)}} \\
& +\!+\!- \ar{dl}{\Psi_-^{(1,0)}\Psi_+^{(1,1)}} \\
+\!-\!+
\end{tikzcd}
\caption{
The RG flows triggered by relevant boundary deformations in the $\slashed{\partial}^3 \psi$ theory.
\label{fig:fermionflows}
}
\end{figure}

Then we have the group $+\!-\!+$, $+\!+\!-$, and $-\!+\!+$ which are related by deformations, as is the
group $-\!+\!-$, $-\!-\!+$, and $+\!-\!-$.
For example, if we examine the first group, we can flow from
$-\!+\!+$ to $+\!+\!-$ by adding $\overline \Psi^{(1,1)}_- \Psi^{(1,0)}_+ + \text{c.c.}$  Alternately, we can flow from $-\!+\!+$ to $+\!-\!+$ by adding $\overline \Psi^{(1,0)}_- \Psi^{(1,0)}_+ + \text{c.c.}$  Finally, we can flow from $+\!+\!-$ to $+\!-\!+$ by adding $\overline \Psi^{(1,0)}_- \Psi^{(1,1)}_+ + \text{c.c.}$
The flows are pictured in Figure \ref{fig:fermionflows}.

The results for the $k=0$ and $k=1$ theories obey a pattern that we conjecture
the $\slashed{\pd}^{2k+1}$ theory will follow. From the boundary contribution to the variation, we find a vanishing condition for pairs of boundary spinor primaries that we order in a similar manner as above, i.e.
\begin{align*}
(\Psi^{(k,0)}_+,\Psi^{(k,2k)}_-),\,(\Psi^{(k,2k)}_+,\Psi^{(k,0)}_-),\,(\Psi^{(k,1)}_+,\Psi^{(k,2k-1)}_-),\,\ldots,(\Psi^{(k,k)}_+,\Psi^{(k,k)}_-)\,.
\end{align*}
We can easily see that we have $2^{2k+1}$ conformal boundary conditions. Among all of the boundary conditions there will always be two isolated boundary conditions for any $k\geq 0$:
\begin{align*}\underbrace{+\ldots+}_{2k+1}\qquad \text{and}\qquad \underbrace{-\ldots-}_{2k+1}\,.\end{align*}
The rest of the boundary conditions can be grouped into sets that are connected by RG flows induced by deformations of the form of \eq{fermion-boundary-double-trace}.  However, note that \eq{fermion-boundary-double-trace} clearly does not change the relative number of $+$'s and $-$'s of the boundary condition along the flow. Thus, we find that, in addition to the two isolated boundary conditions, there are $2k$ directed graphs.  The vertices of a given graph have the same number of $+$'s and $-$'s and the edges correspond to relevant boundary deformations.

\subsection{A puzzle with \texorpdfstring{$T^{na}$}{Tna}}

Cardy's conformal boundary condition was that $T^{na}|_{\rm bry} = 0$ \cite{Cardy:1984bb}.
The condition is ambiguous because the stress tensor can be ``improved''.  However, in our case, there appears to be
an unambiguous choice of stress tensor, as discussed in refs. \cite{Osborn:2016bev,Stergiou:2022qqj}.  The claim of ref.~\cite{Stergiou:2022qqj} is that there should be a unique
stress tensor that is a primary operator with respect to the generators of the conformal group, both for the higher
derivative scalar and fermion theories.  Moreover, this stress tensor agrees with what one could derive starting
from the Weyl invariant curved space generalization of these higher derivative theories and the corresponding GJMS operator generalization of $\Box^k$ (and its equivalent for the Dirac operator).  This stress tensor is conserved, traceless, and annihilated by the special conformal generators $K^\mu$.

Given a stress tensor, the candidate generators of the conformal group can be written simply as integrals over codimension one
spatial slices.  For convenience, we choose the slice $t=0$:
\begin{align}
\label{PDKdef}
\begin{split}
P^a &= \int_{t=0} d^{d-1} x \, T^{a t}  + \int_{t,\, x_n=0} d^{d-2} x \, \tau^{a t}  \ , \\
D &= \int_{t=0} d^{d-1} x \, x_\mu  T^{\mu t} + \int_{t,\, x_n=0} d^{d-2}  x \, x_a \tau^{a t} \ , \\
K^a &= \int_{t=0} d^{d-1} x \, (2x^a x_\nu - x^2 \delta^a_\nu) T^{\nu t}
+ \int_{t,\, x_n=0} d^{d-2} x \, (2x^a x_b - x^2 \delta^a_b) \, \tau^{ bt} \ .
\end{split}
\end{align}
For reasons that will become clear, we allow for a purely boundary contribution to the stress tensor $\tau_{ab}$ that we assume
to be traceless and symmetric although not conserved.
We can then ask when are these generators independent of time?  Running through the usual manipulations for the conservation of a current, we find a residual term on the boundary $x_n=0$:
\begin{align}
\begin{split}
\frac{d P^a}{dt} &= \int_{t, \, x_n=0} d^{d-2}x \, \left[ T^{n a} + \partial_t \tau^{at}  \right] \ , \\
\frac{d D}{dt} &= \int_{t,\, x_n=0} d^{d-2} x \, \left[ x_a T^{a n} + \partial_t (x_a \tau^{at} ) \right] \ , \\
\frac{d K^a}{dt} &= \int_{t, \, x_n=0} d^{d-2} x \, \left[ (2 x^a x_b - x^2 \delta^a_b) T^{b n} + \partial_t (2 x^a x_b - x^2 \delta^a_b) \tau^{bt} \right]\ .
\end{split}
\end{align}
The simplest solution is to set $T^{an} = 0$ on the boundary and also to remove $\tau^{ab}$.  However, there is a possibility
of a more complicated solution.  Namely, suppose $T^{an} = -\partial_b \tau^{ab}$ on the boundary.
Then tracelessness and symmetry of $\tau^{ab}$ mean
that we can commute $\partial_b$ through $x_a$ in the case of the dilatation operator
and through $(2 x^a x_b - x^2 \delta^a_b)$ for special conformal transformations.  The total time derivative drops out of the integral,
leaving a total divergence in the remaining spatial directions, and the integral in each case will vanish.

Let us look at the $\Box^2 \phi = 0$ theory.  The unique stress tensor is\footnote{For the special case $d=2$, this improved stress tensor is ill-dedfined.  The unimproved stress tensor from Noether's theorem is finite, but cannot be made traceless \cite{Karananas:2015ioa,Nakayama:2016cyh}.
}
\begin{align}
T_{na} &= \frac{1}{2(d-1)} \biggl[
(d-4) \phi \Box_\parallel \partial_n \partial_a \phi + d (\partial_n \partial_a \phi) (\partial_n^2 \phi) - (d-2) (\partial_n \phi) \partial_n^2 \partial_a \phi
+(d-4) \phi \partial_n^3 \partial_a \phi \nonumber \\
& - (d+2) (\partial_n \Box_\parallel \phi) (\partial_a \phi)
- (d+2) (\partial_n^3 \phi) (\partial_a \phi) + 4 (\partial_n\partial_a \partial^b \phi) (\partial_b   \phi)
- \frac{4d}{d-2} (\partial_n \partial^b \phi) (\partial_b \partial_a \phi)\nonumber \\
&
+ \frac{d(d+2)}{d-2} (\partial_n  \partial_a \phi) (\Box_\parallel \phi)
-(d+2) (\partial_n\phi) (\Box_\parallel \partial_a \phi) \biggr] \ .
\end{align}
When written in terms of the boundary primaries, this becomes
\begin{equation}
\begin{split}
\label{eq:Tna}
T_{na} =
\frac{1}{2(d-1)} &\biggl(
-(d+2) \Phi^{(2,3)} \partial_a \Phi^{(2,0)} +(d-4) \Phi^{(2,0)} \partial_a \Phi^{(2,3)}   \\
&
-d \Phi^{(2,2)} \partial_a \Phi^{(2,1)} +(d-2) \Phi^{(2,1)} \partial_a \Phi^{(2,2)}
\biggr)
- \partial^b \tau_{ba} \ .
\end{split}
\end{equation}
There is some ambiguity in how to isolate $\tau_{ab}$, but if one insists that it is both symmetric and traceless, then it takes
the unique form
\begin{align}
\tau_{ab} &= \frac{1}{d-2} \biggl(
d \, \Phi^{(2,1)} \partial_a \partial_b \Phi^{(2,0)} + (d-4) \Phi^{(2,0)} \partial_a \partial_b \Phi^{(2,1)} \nonumber \\
& \hspace{1in} - d \, \partial_a \Phi^{(2,1)} \partial_b \Phi^{(2,0)} - d \, \partial_a \Phi^{(2,0)} \partial_b \Phi^{(2,1)} \biggr)  \\
& - \frac{\delta_{ab}}{(d-1)(d-2)} \biggl( d\,  \Phi^{(2,1)} \Box_\parallel \Phi^{(2,0)} + (d-4)  \Phi^{(2,0)} \Box_\parallel \Phi^{(2,1)}
- 2d \, \partial^c \Phi^{(2,1)} \partial_c \Phi^{(2,0)} \biggr) \ . \nonumber
\end{align}
Note that $\tau_{ab}$ is {\it not} divergence-free, $\partial^a \tau_{ab} \neq 0$.

While the $ND$, $DN$, and $DD$ boundary conditions are clearly compatible with the vanishing of $T^{na}$ on the boundary,
the $NN$ boundary condition is not.
The $\Phi^{(2,0)} \Phi^{(2,1)}$ dependent terms will not vanish in this case.
These terms can be combined into an object $\tau_{ab}$ that is symmetric, traceless, and has the right
scaling dimension to be a boundary stress tensor.  However, it is not conserved.

In order to better characterize
$\tau_{ab}$, we compute
the boundary
three-point function $\langle \tau_{ab} \Phi^{(2,0)}\Phi^{(2,1)} \rangle$.
Conformal invariance restricts the form of the correlation function \cite{Osborn:1993cr}:
\begin{align}
\left\langle \tau_{ab}(x_\parallel) \Phi^{(2,0)}(y_\parallel) \Phi^{(2,1)}(z_\parallel) \right\rangle = c_{\tau 01} \frac{ |y_\parallel-z_\parallel|^2 }
{
|x_\parallel-y_\parallel|^{d-2} |x_\parallel-z_\parallel|^{d}
}
\left( \hat V_a \hat V_b - \frac{1}{d-1} \delta_{ab} \right),
\end{align}
where
\begin{align}
V^a = \frac{y^a - x^a}{|y_\parallel-x_\parallel|^2} - \frac{z^a - x^a}{|z_\parallel-x_\parallel  |^2} \ ,
\end{align}
and the hat indicates unit normalization, $\hat V^a = V^a /|V|$.   For this theory, we find,
using Wick's theorem and the $NN$ Green's function derived later in section \ref{sec:twopoint}, that
\begin{align}
c_{\tau 01} = 2d(4-d) 2^d \kappa^2 \ ,
\end{align}
where $\kappa$ fixes the normalization of the
$\langle \phi \phi \rangle$ two-point
function.\footnote{Our calculation assumed $d \neq 2$ or 4.
Note these cases are special in part
because the
Green's functions will have logarithmic behavior.
}

Another interesting quantity is the two point function
\begin{align}
\langle \tau_{ab}(x_\parallel) \tau_{cd}(0) \rangle  = \frac{c_{\tau \tau}}{|x_\parallel|^{2d-2}} \left( \frac{1}{2} (I^{ad}(x_\parallel) I^{bc}(x_\parallel) + I^{ac}(x_\parallel) I^{bd}(x_\parallel)) -
\frac{1}{d} \delta^{ab} \delta^{cd} \right) \ .
\end{align}
We are using the so-called inversion tensor $I^{ab}(x_\parallel) = \delta^{ab} - 2 \frac{x^a x^b}{x_\parallel^2}$.
In our case, we find
\begin{align}
c_{\tau \tau} = \frac{2^{d+4} \kappa^2 d(d-1)(d-4)}{d-2} \ .
\end{align}
This two-point function takes the same form no matter if $\tau^{ab}$ is conserved or not.  Thus we find that
$\partial_a \langle \tau^{ab}(x_\parallel) \tau^{cd}(0) \rangle = 0$, which has the curious consequence that
$\langle T^{na}(x) T^{nb}(0) \rangle$ evaluated on the boundary will also vanish.

Normally, for a conserved stress tensor, there would be a Ward identity that relates $\langle \tau_{ab} \Phi^{(2,0)} \Phi^{(2,1)} \rangle$
to $\langle \Phi^{(2,0)} \Phi^{(2,1)}\rangle$.  In particular, the coefficient $c_{\tau 01}$ we computed above would be
proportional to the normalization of  $\langle \Phi^{(2,0)}(y_\parallel) \Phi^{(2,1)}(z_\parallel) \rangle$.
Thus if ${\langle \Phi^{(2,0)}\Phi^{(2,1)} \rangle}$ vanishes, as indeed happens in this case because the two operators
are primaries of different dimension,
then so should the three point function.
However, this Ward identity argument fails
because $\tau_{ab}$ is not conserved and does not generate translations on its own.

We conclude this section by considering the case of more general boundary conditions, namely the ones corresponding to setting the boundary operators
\beq\label{eq:NN_primaries_sec}
\tilde \Phi^{(2,2)} \equiv (\partial_n^2 +\nu \lsp\Box_{\parallel})\phi  , \qquad \tilde \Phi^{(2,3)} \equiv (\partial_n^3 +\mu \lsp\partial_n\Box_{\parallel})\phi  \,
\eeq
to zero.
Note that these are not boundary primaries unless $\nu = -1$ and $\mu = 3$. We can repeat the analysis above by requiring that the boundary stress tensor is symmetric. Dropping the tildes for convenience, this gives
\beq
\begin{split}
\tau_{ab} = &\ \frac{1}{2(d-2)} \Big[ a_1 (\partial_a \partial_b \Phi^{(2,0)})\Phi^{(2,1)} + a_2 \Phi^{(2,0)}\partial_a \partial_b \Phi^{(2,1)}
+ a_3 \partial_{(a}\Phi^{(2,0)} \partial_{b)}\Phi^{(2,1)}+ \\
& +\frac{\delta_{ab}}{d-1} \Big(a_4 \Phi^{(2,0)}\Box_\parallel \Phi^{(2,1)} +  a_5 \Box_\parallel \Phi^{(2,0)} \Phi^{(2,1)} +a_6  \partial_c \Phi^{(2,0)} \partial^c \Phi^{(2,1)} \Big) \Big]\,,
\end{split}
\eeq
where
\begin{align}
& a_1 = \frac{(d-2) (2 \nu +\gamma-\nu  d+d+2)}{d-1}\,, \qquad a_2=\frac{-4 \nu -2 \gamma +d (4 \nu +\gamma -\nu  d+d-8)+4}{d-1}\,, \nonumber\\
& a_3 = \frac{2 \gamma  (d-2)-2 d (-\nu  (d-2)+d+2)}{d-1} \, , \qquad a_4 = 2 (-2 \nu +\gamma +2)-d (-2 \nu +\gamma -2)\, , \nonumber\\
& a_5 = -\gamma  (d-2) \, , \qquad a_6 = 4 \nu +4 \gamma +2 d (-\nu -\gamma +3)+4 \, ,
\end{align}
together with
\beq
\label{eq:beta_cond}
\mu+\nu = 2 \, .
\eeq
We find that the condition \eq{beta_cond} is necessary for $T_{na}$ to be written in the form of \eq{Tna}. In particular, if \eq{beta_cond} is not satisfied then translational invariance along the boundary is lost. Without further requirement we also see that the boundary stress tensor is not uniquely fixed but depends on an undetermined coefficient $\gamma$. Nicely, even though the trace of $\tau_{ab}$ is not vanishing for generic $\nu$, it is a total divergence. Indeed, we may write
\beq
\tau^a_a = -\partial^a j_a\ ,
\eeq
where
\beq
\begin{split}
& j_a = b_1  \Phi^{(2,0)}\partial_a \Phi^{(2,1)} + b_2 \partial_a  \Phi^{(2,0)}\Phi^{(2,1)}\,, \\
& b_1 = \frac{d (-\nu +\gamma -3)-2 \gamma }{2 (d-1)}\,, \qquad b_2 =- \frac{d(-\nu  -\gamma  +1)+2 \nu +2 \gamma +2}{2(d-1)}\, .
\end{split}
\eeq
This means that the boundary conditions with $\nu \ne -1$ are scale invariant but not conformal.
More precisely, we can add a boundary integral over $j^t$ to the definition of $D$ in eq.~(\ref{PDKdef}) to guarantee that $dD/dt = 0$.
The minimal condition to have a conformal boundary is that $j_a$ can be written as a total divergence, in which case a similar alteration can be made to the definition of $K^a$ to ensure that $dK^a /dt=0$. However imposing $j^a = \partial^a j$ can be accomplished if and only if $b_1 = b_2$, implying $\nu=-1$ in agreement with the discussion above.
Finally, if the latter condition is satisfied, then $\tau_{ab}$ can be improved to be traceless. This last condition requires $b_1 = b_2 = 0$, which uniquely fixes $\tau_{ab}$ giving $\gamma = 2d/(d-2)$.

\section{Characterizing higher derivative BCFTs}
\label{sec:apps}

In this section, we employ some of the lessons learned from our analysis of boundary conformal primaries in higher derivative free theories to characterize the boundary theory.  To that end, we study two particularly important quantities: the boundary free energy on $\mathds{HS}^d$ and the two-point function of the displacement operator in flat space.
As a prelude to the computation of the displacement two-point function, this section also contains a derivation of the
$\langle \phi \phi \rangle$ and $\langle \psi \overline \psi\rangle$ correlators in our higher derivative theories.

\subsection{Hemisphere free energy}
\label{sec:F}

In this section, our main focus will be the study of the boundary free energy, $F_\pd$, of higher derivative CFTs placed on the hemisphere $\mathds{HS}^{d}$.  In odd $d$, $F_\pd$ is directly related to the boundary A-type Weyl anomaly, and so provides crucial data for characterizing the boundary theory.  Moreover, it has been shown in unitary CFTs in both $d=3$ and $d=5$ that the boundary `$a$-theorem' holds \cite{Jensen:2015swa,Casini:2018nym,Wang:2021mdq}.

In the following subsections, we will first compute $F_\pd$ for the generalized Dirichlet ($DND\cdots$) and Neumann ($NDN\cdots$) boundary conditions, and then we will compute the variation in $F_\pd$ generated by relevant deformations triggered by boundary primaries.
The technique we use to compute the change in $F_\partial$ due to the boundary deformations was first
employed for double trace deformations in AdS/CFT
\cite{Gubser:2002vv, Allais:2010qq}, and because
of conformal symmetry the result is very similar as well
\cite{Herzog:2019bom}.
Finally, we will repeat the same analysis for higher-derivative free fermion theories. Our discussion will employ standard results for the spectrum of GJMS-type operators on $\mathds{S}^d$ \cite{branson1995sharp}.
Our results overlap to some extent with work of Dowker \cite{Dowker:2010qy,Dowker:2013mba,Dowker:2014rva,Dowker:2017hpm}.  We will make the points of overlap clearer below.

Our emphasis in the text is on the logarithmic divergence
in $F_\partial$ in odd $d$.  In Appendix \ref{app:constantterm},
we extend the analysis to the regularization independent
constant term that appears in even dimensional theories.
 This constant term is also believed to be monotonic under RG flow for unitary theories
\cite{Gaiotto:2014gha, Kobayashi:2018lil}.

As we will see, an interesting upshot is to show explicitly that for bulk non-unitary CFTs the boundary $a$-theorem does not necessarily hold,
both for the coefficient of the logarithmic term in odd dimensional cases and the constant term in even dimensional ones. The violation presumably occurs because these theories are non-unitary.

\subsubsection*{Scalars on $\mathds{HS}^d$}

We begin by computing $F_\pd$ for higher derivative scalars.  Our background geometry is taken to be the round metric on $\mathds{HS}^d$ with line element
\begin{align}\label{eq:sd-metric}
ds^2_d = R^2 ( d\theta^2_d + \sin^2\theta_d\lsp ds_{d-1}^2) \,,
\end{align}
where for notational simplicity we will call $\theta_d=\theta$ and take $\theta\in[0,\frac{\pi}{2}]$ such that the boundary is located at $\theta=\frac{\pi}{2}$.  When extending to the round metric on $\mathds{S}^d$, $\theta\in[0,\pi]$.
The quantity $R$ fixes the radius of the sphere.  Often we will set it to one.

Because
$\mathds{HS}^d$ is curved,
we use the action  \eq{scalar-box-action}
to start our spectral analysis.  Since $F_\pd$ must be independent of marginal boundary deformations, the boundary terms of the form eq.~\eqref{bryctrterm} will play no role here.  Recall that the general form of the spherical scalar GJMS operator on the unit $d$-sphere is given by \cite{branson1995sharp}
\begin{align}\label{eq:scalar-GJMS-operator}
\triangle_{2k} = \prod_{j=\frac{d}{2}}^{\frac{d}{2}+k-1}(-\Box_d + j(d-1-j))\,,
\end{align}
where the Laplace operator $\Box_d$ on $\mathds{S}^d$ with the round metric as in \eq{sd-metric} with $R=1$ takes the form
\begin{align}
\Box_d = \pd_\theta^2 +(d-1)\cot\theta\,\pd_\theta +\frac{1}{\sin^2\theta}\Box_{d-1}-\frac{d(d-2)}{4}\,.
\end{align}
Notice that the form \eq{scalar-GJMS-operator} implies that the scalar spherical harmonic eigenfunctions of $\Box_d$ are also eigenfunctions of $\triangle_{2k}$.

Performing the spectral analysis of $\triangle_{2k}$ on $\mathds{HS}^d$ requires a bit of care as the choice of boundary conditions applied at $\theta =\pi/2$ can change the allowed eigenvalues, their degeneracies, or even their form. Indeed, obtaining the spectrum of the GJMS operator for generic boundary condition is complicated.

A great simplification arises if we focus only on generalized Neumann ($NDN\cdots$) or generalized Dirichlet ($DND\cdots$) boundary conditions. In these cases, some of the eigenfunctions are removed from the spectrum. The remaining ones keep
the same form that they had on the sphere, and the eigenvalues are the same.\footnote{ In general on a curved space, we anticipate that our $\Phi^{(k,q)}$ boundary primaries should get additional contributions
 from curvature terms, as in e.g.\ \cite{case2018boundary}.  In the particular case of a hemisphere, simplifications should arise because the extrinsic curvature vanishes.  There can nevertheless be
  intrinsic curvature terms which we have not worked out in general.
 When we say $N$ and $D$ in this section, we implicitly assume our $\Phi^{(k,q)}$ can and have been improved
 by these curvature terms.
 }
 To see why, first recall that the solution of the eigenvalue problem for the ordinary Laplacian $\Box_d$ on $\mathds{S}^d$ is the generalized spherical harmonics $Y_{\vec \ell\,}(\vec{\theta}\, )$ labelled by the angular quantum numbers  $\vec{\ell} = (\ell_1,\ldots,\ell_{d-1},\,\ell_d)$ with $|\ell_1|\leq\ell_2\leq\ldots\leq\ell_d$ and  $\vec\theta = (\theta_1,\ldots,\theta_{d-1},\,\theta)$.  For notational convenience, we will relabel $\ell_d=\ell$ and $\ell_{d-1}=m$. In $d$-dimensions the eigenvalues $\Box_d Y_{\vec{\ell}\,}(\vec{\theta}\,) = - \ell(\ell+d-1)Y_{\vec{\ell\,}}(\vec{\theta}\,)$ for given $\ell$ have degeneracy
\begin{align}\label{eq:scalar-spherical-harmonic-degeneracy}
{\rm deg}(\ell) =\begin{pmatrix}d+\ell-1\\ d-1\end{pmatrix}+ \begin{pmatrix}d+\ell-2\\d-1\end{pmatrix} = \frac{(d+2\ell-1)\Gamma\left( d+\ell-1 \right)}{\Gamma\left(d \right)\Gamma\left(1+\ell \right)} \ .
\end{align}

In order to see how the degeneracies are lifted when imposing boundary conditions, it is useful introduce the `parity' operators $P_\theta: \theta\,\to \pi-\theta$.  The set of spherical harmonics can then be refined according to parity eigenvalue $P_\theta Y_{\vec{\ell}\,}(\vec{\theta}\,) = (-1)^{\ell-m}Y_{\vec{\ell}\,}(\vec{\theta}\,)$, which suggests a decomposition into parity even modes with  $\ell-m\in2\mathbb{Z}$ and parity odd modes with $\ell-m\in2\mathbb{Z}+1$.  Clearly for the spectrum of $\Box_d$, the parity odd modes will be those that satisfy Dirichlet boundary conditions due to vanishing at $\theta=\pi/2$ by construction, and the parity even modes obey Neumann boundary conditions.  Hence, the degeneracy of a mode with given $\ell$ is lifted according to the boundary conditions applied; for Neumann and Dirichlet boundary conditions, respectively,
\begin{align}\label{eq:scalar-hemispherical-harmonic-degeneracy}
{\rm deg}^{(N)}(\ell) = \begin{pmatrix}d+\ell-1\\ d-1\end{pmatrix},\qquad {\rm deg}^{(D)}(\ell)=\begin{pmatrix}d+\ell-2\\d-1\end{pmatrix}.
\end{align}

Returning to the spectral problem for $\triangle_{2k}$ on $\mathds{HS}^d$, it is easy to see that spherical harmonics corresponding to Neumann or Dirichlet conditions for $k=1$ also satisfy the generalized Neumann or Dirichlet conditions for generic $k$. Thus, we can simply use the degeneracy reported in \eq{scalar-hemispherical-harmonic-degeneracy}, while from \eq{scalar-GJMS-operator} the eigenvalues can be read off easily and with a bit of algebra put into the form
\begin{align}
\lambda_\ell = \prod_{j=0}^{k-1}\left(-\left(\ell+\frac{d-1}{2}\right)^2+\left(j+\frac{1}{2}\right)^2\right).
\end{align}

  We are now in a position to compute the boundary contribution to the hemisphere free energy for higher derivative scalars.  We can formally write the hemisphere free energy as
\begin{align}
F^{(X)}_{\mathds{HS}^d} = -\log Z^{({X})}= \frac{1}{2}\sum_{\ell=0}^\infty {\rm deg}^{(X)}(\ell)\log\la_\ell + c\,,
\end{align}
where $Z^{(X)}$ is the partition function with boundary conditions ${(X) = (N)}$ or $(D)$.\footnote{For simplicity in the notation, in this section we will call the generalized Neumann and Dirichlet boundary condition as $N$ and $D$, respectively.}
There are a variety of regularization schemes for isolating the anomalous part of $F_\pd$.  Here we take advantage of the fact that ${\rm deg}^{(D)}(\ell+1) = {\rm deg}^{(N)}(\ell)$ to write the difference
$F^{(N)}_{\mathds{HS}^d}-F^{(D)}_{\mathds{HS}^d} \equiv \Delta F^{({ND})}_\pd$ in the form
\begin{align}
\label{DeltaFND}
\Delta F^{({ND})}_\pd = -\frac{1}{2} \sum_{\ell=0}^\infty  \begin{pmatrix}d+\ell-1\\ d-1\end{pmatrix} \log \frac{\ell + \frac{d}{2} +k}{\ell + \frac{d}{2} - k} \, .
\end{align}

The anomalous part of $\Delta F^{({ND})}_\pd$ for odd $d$ is the log-divergent part. Its coefficient can be isolated by computing a large $\ell$ expansion of the summand.  For example, in the case $d=3$, the summand can be expanded as
\begin{align}
\frac{1}{4}(\ell+1)(\ell+2)  \log \frac{\ell + \frac{3}{2} +k}{\ell + \frac{3}{2} - k}  =  \frac{k \ell}{2} + \frac{3k}{4} + \frac{4k^3 - 3k}{24 \ell} + O(\ell^{-2}) \, .
\end{align}
The $O(\ell^{-1})$ term in this expansion allows us to read off the coefficient of the
log\footnote{The number is typically reported as the coefficient of the logarithm of a short-distance cut-off (see for example \cite{Chalabi:2021jud}).  Here we naturally find the coefficient of the logarithm of a large angular momentum cut-off, and so we include an extra minus sign.
}
\begin{align}\label{eq:bryF_d3}
\left. \Delta F^{({ND})}_\pd \right|_{\log} = \frac{k (4k^2 - 3)}{24} \, .
\end{align}
For $k=1$, the two theories are related by a relevant boundary mass deformation, and
the sign is consistent with the $a$-theorem in 3d BCFTs \cite{Jensen:2015swa}. Ref.~\cite{Jensen:2015swa} also considered
this particular example and computed the value $\frac{1}{24}$, in agreement with eq.~\eqref{eq:bryF_d3}.
The more general result for arbitrary $k$ can be found in table (12) of \cite{Dowker:2013mba}, where it is related to the determinant
of a Dirac type GJMS operator.

It will be useful to introduce a notation that tracks the value of $k$ and $d$ for this anomaly coefficient:
\begin{equation}
f^{ND}(d,k) \equiv  \left. \Delta F^{({ND})}_\pd \right|_{\log} \,,
\end{equation}
Repeating the same procedure for odd $d$ up to $d=11$, we find the results for the boundary anomaly reported in \tbl{scalar-ND-boundary-anomaly}. The results in the $k=1$ and $d=5$ case also match the literature \cite{FarajiAstaneh:2021foi,Wang:2021mdq}.
The result for $f^{ND}(d,k)$ for odd $d$
can be expressed as an integral over a polynomial:

\begin{align}\label{eq:fND_int}
f^{ND}(d,k) &= \frac{1}{(d-1)!} \int_0^k \frac{\Gamma\left(x+\frac{d}{2}\right)}{\Gamma\left(x - \frac{d}{2}+1 \right)} dx
= \frac{1}{(d-1)!} \int_0^k \prod_{j=1}^{d-1}
\left( x- \frac{d}{2}+j \right) dx \ .
\end{align}
The integrand has zeros at half integer
values of $k$ which in turn means $f^{ND}(d,k)$ has local maxima and minima at these values.\footnote{%
A closely related result appears as (13) of \cite{Dowker:2013mba}, in the context of relating the determinant of
Dirac GJMS operators to the determinant of ordinary GJMS operators on the sphere.
For $d=2k$ and half-integer $k$, the anomalous part of $F_{\mathds{HS}^d}^{(X)}$ with $(X) = (N)$ or $(D)$ was previously computed in ref.~\cite{Kislev:2022emm}. There, the authors find a similar integral expression for $F_{\mathds{HS}^d}^{(X)}$, or equivalently $f^{ND}(d,d/2)$, which agrees with ours for odd $d$.
}

 \begin{table}[t]
 \begin{center}
\begin{tabular}{|l|l|l|l|l|l|l|l|l|l|l|}
\hline
     & $k=1$                                           & $k=2$                                        & $k=3$                                        & $k=4$                                           & $k=5$                                                        & $k=6$                                             & $k=7$                                              \\ \hline
$d=3$  & $\frac{1}{24}$                                  & $\frac{13}{12}$                                & $\frac{33}{8}$                              & $\frac{61}{6}$                                 &$ \frac{485}{24}$ & $\frac{141}{4}     $            & $\frac{1351}{24}$                         \\ \hline
$d=5$  &$ -\frac{17}{5760}  $           & $\frac{103}{2880}   $          & $ \frac{741}{640}$          & $\frac{9223}{1440} $           & $\frac{25135}{1152} $                         &$ \frac{18381}{320}  $           &$ \frac{739081}{5760}$                       \\ \hline
$d=7$  & $\frac{367}{967680}  $         &$ -\frac{1061}{483840}$         &$ \frac{1111}{35840}   $    &$ \frac{295627}{241920}       $    &$ \frac{1706575}{193536}  $                     & $\frac{685123}{17920}  $      & $\frac{17292079}{138240}      $        \\ \hline
$d=9$  & $-\frac{27859}{464486400}  $   & $\frac{8603}{33177600} $      &$ -\frac{9257}{5734400} $ & $\frac{3194621}{116121600} $     &$ \frac{3393775}{2654208}    $               & $\frac{32535943}{2867200} $   & $\frac{3958017581}{66355200} $        \\ \hline
\end{tabular}
\end{center}
\caption{$\Delta F^{(ND )}\big|_{\log}$ for scalars with $\triangle_{2k}$ GJMS kinetic operator on $\mathds{HS}^d$ for various odd $d$ and $k$.}
\label{tab:scalar-ND-boundary-anomaly}
\end{table}

 Before moving on to consider the full space of conformal boundary conditions, it is useful to give a second interpretation to the quantity $\Delta F_\partial^{(ND)}$ that we just computed.  Instead of comparing the free energy of the generalized Neumann case with the free energy of the generalized Dirichlet case, we could have instead compared the free energy of the Dirichlet or Neumann case on the hemisphere with the total free energy on the sphere \cite{Gaiotto:2014gha}.  The idea is to
 regularize the
 free energy on the hemisphere by substracting half of the free energy on the sphere.  So for example, we could consider
 $F^{(X)}_ {\mathds{HS}^d} - \frac{1}{2} F_{\mathds{S}^d}$.  In fact, though, through the diagonalization of $\Box^{2k}$ that we
 just carried out, it is clear that  $F_{\mathds{S}^d} = F^{(N)}_ {\mathds{HS}^d} + F^{(D)}_ {\mathds{HS}^d}$.  Thus we have that this
 quantity regularized in comparison to the sphere is not independent of the quantity we just computed:
 \begin{align}
 F^{(N)}_ {\mathds{HS}^d} - \frac{1}{2} F_{\mathds{S}^d} = \frac{1}{2} \Delta F^{(ND)}_\partial =  -F^{(D)}_ {\mathds{HS}^d} + \frac{1}{2} F_{\mathds{S}^d}  \ .
 \end{align}

The space of conformal boundary conditions is much richer than simply fully Neumann and Dirichlet for $k>1$, and we would like to explore the change in boundary free energy between fixed points connected by flows triggered by any of the relevant boundary primary operators found in the previous section.  We will restrict ourselves in the following to considering only quadratic deformations of the form
\beq\label{eq:scalar-k-q-double-trace}
 S_c = c \int d^{d-1} x_\parallel \, \Phi^{(k,q)}\Phi^{(k,q)}\,,
 \eeq
  where $\Phi^{(k,q)}$ is a boundary primary like those found in the previous section.\footnote{ We could consider a more general deformation by
 $\Phi^{(k,q)}\Phi^{(k,q')}$ where $q\neq q'$.
 A generalization of the Hubbard-Stratonovich transformation  we are about
 to use can demonstrate such a deformation
 is equivalent to a deformation by
 $\Phi^{(k,q)}\Phi^{(k,q)}+\Phi^{(k,q')}\Phi^{(k,q')}$.  We do not give the details here as we will present a fermionic version of this argument in the next subsection.
 }
 Employing a Hubbard-Stratonovich transformation, we introduce an auxiliary field $\sigma$ and rewrite the deformed theory as
   \beq
 \begin{split}
 Z[c] & =  \frac{1}{\mathcal{Z}_\sigma}  \int D \sigma \,D \phi  \, \exp \left\{ - S_0  - \frac{1}{16 \, c} \int d^{d-1} x_\parallel \,\sigma^2  - \frac{i}{2} \int d^{d-1} x_\parallel \, \sigma\Phi^{(k,q)}\right\}  \\
 &  = \frac{Z[0]}{\mathcal{Z}_\sigma}  \int D \sigma \, e^{- \frac{1}{16 \, c} \int d^{d-1} x_\parallel \, \sigma^2}  \, \left< e^{   - \frac{i}{2} \int d^{d-1} x_\parallel \, \sigma\Phi^{(k,q)} } \right>_0,
 \end{split}
 \eeq
 where $\left<\ldots\right>_0$ denotes the correlation function being computed in the undeformed path integral, $Z[0]$, and we defined
\begin{align}
\CZ_\sigma\equiv \int D\sigma\, e^{-\frac{1}{16\, c}\int d^{d-1}x_\parallel \sigma^2}\,,
\end{align}
with the $\sigma$-integration parallel to the real axis. Since the operator $\Phi^{(k,q)}$ is linear in $\phi$, the expectation value can be computed using Wick's theorem
 \beq
 \begin{split}
 &\left< e^{   - \frac{i}{2} \int d^{d-1} x_\parallel \, \sigma \Phi^{(k,q)} } \right>_0  =\\
 &=\sum_{j=0}^{\infty} (-1)^j\frac{(2j-1)!!}{(2j)!\,4^j}  \left[ \int d^{d-1}x_\parallel \, d^{d-1} y_\parallel  \, \sigma(x_\parallel)  \left< \Phi^{(k,q)} (x_\parallel) \Phi^{(k,q)}(y_\parallel) \right>_0 \sigma (y_\parallel)  \right]^j \\
 & =  \exp \left\{ -\frac{1}{8} \int d^{d-1}x_\parallel \, d^{d-1} y_\parallel  \, \sigma(x_\parallel)  \left<\Phi^{(k,q)} (x_\parallel)\Phi^{(k,q)}(y_\parallel) \right>_0 \sigma (y_\parallel)  \right\} .
 \end{split}
 \eeq
Integrating out $\sigma$, we arrive at
 \beq
 \frac{Z[c]}{Z[0]} = \frac{1}{\sqrt{\text{det}(2 c \, \mathcal{G}^{(k,q)}+1)}}\,,
 \eeq
 where we defined
 \beq
 \mathcal{G}^{(k,q)}(x_\parallel,\,y_\parallel)\equiv \left<\Phi^{(k,q)}(x_\parallel) \Phi^{(k,q)}(y_\parallel) \right>_0 \,.
 \eeq
Thus the variation in the hemisphere free energy is given by
 \beq
 \label{eq:DeltaF-trace}
 \Delta F_\pd=  \frac{1}{2} \Tr \log\left(  2 c  \, {\cal{G}}^{(k,q)}(x_\parallel,\,y_\parallel)+ 1 \right),
 \eeq
 where the UV boundary condition is taken to be Neumann with respect to the $\Phi^{(k,q)}$ primary and
 the IR fixed point will be Dirichlet. For relevant deformations, such that the coupling $c$ has positive mass dimension $2(k-q)-1>0$, the IR limit corresponds to sending $c$ to infinity. In this limit, we can neglect the contribution of the identity in \eq{DeltaF-trace} and interpret the scale set by $c$ as the UV cutoff of the IR theory.

  In the IR limit, and this point will be crucial for us, $\Delta F_\pd$
 depends on $k$ and $q$ only through the conformal dimension of $\Phi^{(k,q)}$ which is
  the quantity $\Delta = \frac{d-2k+2q}{2}$.
Indeed, by conformal invariance, the two-point function ${\cal{G}}^{(k,q)}$ can in general be written
 \beq
 {\cal{G}}^{(k,q)}(x_{\parallel},y_{\parallel}) = \frac{1}{R^{2 \Delta}} \frac{1}{s^{2\Delta}(x_{\parallel},y_{\parallel})}\,,
 \eeq
where $s(x_{\parallel},y_{\parallel})$ is the invariant distance on the boundary $\mathds{S}^{d-1}$ of unit radius.
 Similar to the $f^{ND}(d,k)$ notation we introduced above, let us introduce now also
 \begin{equation}
\left.  \Delta F_\pd\right|_{\log} \equiv f(d,k-q) \ .
 \end{equation}
 Thus the free energy difference from a $\Phi^{(k+1,q+1)}$ deformation is the same as that caused by a
  $\Phi^{(k,q)}$ deformation.

 Now if we start with generalized Neumann boundary conditions $NDN\cdots$ and add all the possible relevant deformations,
 we should end up in the same place that we would if we started with generalized Dirichlet boundary conditions $DND\cdots$
 and again added all the possible relevant deformations, namely the $DDD\cdots$ case.  This physical expectation means that
 \begin{align}
 \label{pseudocharacteristic}
 \begin{split}
 f^{ND}(d,k) &=
 {\sum_{j=0}^{\lceil \frac{k}{2} \rceil -1} f(d,k -2j)  - \sum_{j=0}^{\lfloor \frac{k}{2} \rfloor-1} f(d,k - 2j-1)}
= \sum_{q=1}^{k} (-1)^{k-q} f(d,q) \ ,
 \end{split}
 \end{align}
 where the first and second sums are the contributions from all flows starting with generalized Neumann and generalized Dirichlet boundary conditions, respectively. (A related expression appears as (16) of \cite{Dowker:2017hpm}.)
 We learn from this short computation that we do not need to evaluate
 $\Tr \log {\cal{G}}^{(k,q)}(x_\parallel,\,y_\parallel)$.  The hard work in computing the $f^{ND}(d,k)$ is enough to extract
 the $f(d,k-q)$.  We just need to add the neighboring  $f^{ND}$:
 \begin{equation}
  f(d,k) = f^{ND}(d,k) +  f^{ND}(d,k-1)  \ .
 \end{equation}
 It is possible to compute $\Tr \log {\cal{G}}^{(k,q)}(x_\parallel,\,y_\parallel)$ directly using the methods of ref.~\cite{Gubser:2002vv}.
 We have checked that the answer is the same, but prefer the derivation here as it avoids many technical details.

 The expression (\ref{pseudocharacteristic}) is curious,
 especially given its relation to the intrinsic Euler density
 on the boundary when $d$ is odd through the trace anomaly.
 The alternating sum is reminiscent of the Euler characteristic, which can be expressed as an alternating
 sum over the Betti numbers.  Of course the boundary topology
 of the hemisphere is trivial, but the result is suggestive
 that some kind of topology can be associated to this space of RG flows for which the quantity
 $f^{ND}(d,k)$ is an invariant.

From the $f^{ND}(d,k)$ in eq.~\eqref{eq:fND_int}, we find for $\kappa_q =k-q$ the following closed form expression as an integral over a polynomial:
\begin{align}
f(d,\kappa_q) &= \frac{1}{(d-1)!} \int_\frac{1}{2}^{\kappa_q} \frac{(2x-1) \Gamma \left( x+ \frac{d}{2}-1\right)}{\Gamma\left(x - \frac{d}{2}+1\right)} dx \ .
\end{align}
Closely related expressions appear in
\cite{Gubser:2002vv} and \cite{Herzog:2019bom}.
That we are able to compute the change in free energy purely from the dimension of a boundary two-point function means
that the change is independent to some extent of the theory under consideration.  Ref.\ \cite{Gubser:2002vv} considers a
massive scalar in anti-de Sitter space subject to the Klein-Gordon equation, while ref.~\cite{Herzog:2019bom} considers a Weyl
equivalent set-up on a hemisphere.  Even though neither consider a higher derivative theory, the change in the hemisphere free energy is computed in a similar way.

As a concrete example, the results of the deformation of the boundary theory by the boundary scalar mass term, i.e.\ the $q=0$ case, for general $k$ are contained in \tbl{scalar-RX-boundary-anomaly}. Since non-zero $q$ enters as a shift $k\to k-q$, the effect on \tbl{scalar-RX-boundary-anomaly} by adding a $q>0$ deformation is to shift the $k^{\rm th}$ column to the $(k+q)^{\rm th}$ column.

\begin{table}[t]
\begin{center}
\begin{tabular}{|l|l|l|l|l|l|l|l|l|l|l|}
\hline
     & $k=1$                                           & $k=2$                                        & $k=3$                                        & $k=4$                                           & $k=5$                                                        & $k=6$                                             & $k=7$                                              \\ \hline
$d=3$  & $\frac{1}{24}$                                  & $\frac{9}{8}$                                & $\frac{125}{24}$                              & $\frac{343}{24}$                                 &$ \frac{243}{8}$ & $\frac{1331}{24}     $            & $\frac{2197}{24}$                         \\ \hline
$d=5$  &$ -\frac{17}{5760}  $           & $\frac{21}{640}   $          & $ \frac{1375}{1152}$          & $\frac{43561}{5760} $           & $\frac{18063}{640} $                         &$ \frac{456533}{5760}  $           &$ \frac{1069939}{5760}$                       \\ \hline
$d=7$  & $\frac{367}{967680}  $         &$ -\frac{13}{7168}$         &$ \frac{5575}{193536}   $    &$ \frac{34643}{27648}       $    &$ \frac{359829}{35840}  $                     & $\frac{45529517}{967680}  $      & $\frac{31608239}{193536}      $        \\ \hline
$d=9$  & $-\frac{27859}{464486400}  $   & $\frac{1143}{5734400} $      &$ -\frac{25175}{18579456} $ & $\frac{1718381}{66355200} $     &$ \frac{7489989}{5734400}    $               & $\frac{5864733391}{464486400} $   & $\frac{32976945833}{464486400}  $        \\ \hline
\end{tabular}
\end{center}
\caption{$\Delta F_\pd\big|_{\log}$ for scalars with $\triangle_{2k}$ GJMS kinetic operator on $\mathds{HS}^d$ for various odd $d$ and $k$ with boundary scalar mass deformation ($q=0$).}\label{tab:scalar-RX-boundary-anomaly}
\end{table}

Lastly in this subsection, we note that the results in tables  \ref{tab:scalar-ND-boundary-anomaly} and \ref{tab:scalar-RX-boundary-anomaly} imply that the boundary $a$-theorem \cite{Jensen:2015swa,Casini:2018nym,Wang:2021mdq} can be violated for general $k$.  Explicitly the integrated boundary A-type Weyl anomaly is captured by $F_\pd|_{\log}$. Consider the scalar theory in $d=5$ where $F_\pd|_{\log}= -4a$. Starting with conformal boundary conditions in the UV and following the flow triggered by \eq{scalar-k-q-double-trace} to some IR conformal boundary condition, the boundary $a$-theorem $a_{\text{UV}}-a_{\text{IR}}\geq 0$ implies $\Delta F_\pd|_{\log}\leq 0$, which when $q=0$ is only satisfied for $k=1$. This demonstrates that the {\textit{boundary}} $a$-theorem does not necessarily hold when the {\textit{bulk}} theory is non-unitary.

\subsubsection*{Fermions on $\mathds{HS}^d$}

We follow the same logic that we used for the higher derivative scalar theory, applied now to a higher derivative theory
of free Dirac fermions on $\mathds{HS}^d$.
We first calculate the free energy associated with the boundary conditions that allow eigenspinors of the Dirac operator
to be eigenspinors of the higher derivative theory.  Then we calculate the free energy for more general boundary conditions
by computing how the free energy changes under a quadratic boundary deformation.

Similar to what happens for the scalar on the sphere, eigenspinors
of the Dirac operator are also eigenspinors
of the spinor GJMS
operator ${\triangle\!\!\!\!\slash\,}_{2k+1}$.  Our first task is to identify what boundary conditions these eigenspinors can satisfy on the hemisphere.  There will in general be eigenspinors on the hemisphere that satisfy more general boundary conditions and that do not descend from eigenspinors on the sphere, but we will have to compute their free energy from the
quadratic deformations we consider later.

Our first claim is that, given a collection of eigenspinors with the same eigenvalue with respect to  ${\triangle\!\!\!\!\slash\,}_{1}$,
 the degenerate set divide up into two equal parts, one of which satisfies
$\Pi_+ \psi_+ = 0$ and the other satisfies $\Pi_- \psi_- = 0$.  As these eigenspinors are dealt with at length in the literature, we leave a detailed justification of this claim to the references (see e.g.\ \cite{Herzog:2019bom}).

Our next claim is that taking a normal derivative will flip the boundary chirality of the eigenspinor.  If $\psi_+$ is annihilated by $\Pi_+$ on the equator, then $\partial_n \psi_+$ will be annihilated by $\Pi_-$ and so on.
This claim follows by applying the Dirac operator on $\psi_\pm$ at the equator and noting that $\Pi_\pm \gamma_a = \gamma_a \Pi_\mp$.
In general, we expect $\Pi_\pm (\partial_n)^{2j} \psi_\pm = 0$ and $\Pi_\mp (\partial_n)^{2j+1} \psi_\pm = 0$.
In the notation developed in section \ref{sec:primaries}, we expect a $\psi_+$ eigenspinor will satisfy ${+\!+\!-\!-\!+\!+\!-\!-\cdots}$ boundary conditions while $\psi_-$ will satisfy the opposite ${-\!-\!+\!+\!-\!-\!+\!+\cdots}$ boundary conditions.

Similar to the notation we introduced for the scalars, let $F_{\mathds{HS}^d}^{(\pm)}$ describe the hemisphere free energy
for these boundary conditions  ${+\!+\!-\!-\!+\!+\!-\!-\cdots}$ and ${-\!-\!+\!+\!-\!-\!+\!+\cdots}$.
From the construction above, we can draw two conclusions about these quantities. Firstly, since the eigenvalues and degeneracies for both boundary conditions are the same, $F_{\mathds{HS}^d}^{(+)} = F_{\mathds{HS}^d}^{(-)}$. Secondly, together these two sets of eigenspinors diagonalize ${\triangle\!\!\!\!\slash\,}_{2k+1}$ on the sphere, which implies $F_{\mathds{HS}^d}^{(+)} + F_{\mathds{HS}^d}^{(-)} = F_{\mathds{S}^d}$.
Thus, the regularized hemisphere free energy associated with the $(+)$ and $(-)$ boundary conditions must therefore vanish:
\begin{align}
\Delta F_\partial^{(+-)} = F_{\mathds{HS}^d}^{(+)} - F_{\mathds{HS}^d}^{(-)} = 2 \left(  F_{\mathds{HS}^d}^{(+)}  - \frac{1}{2}  F_{\mathds{S}^d} \right) = 0\,.
\end{align}

For any $k$, we can again use a Hubbard-Stratonovich transformation to study the flow triggered by a boundary deformation quadratic in the boundary spinor primaries found in \sn{fermion-primaries}.  That is, we add to the action in \eq{fermion-GJMS-action} the deformation
\begin{align}\label{eq:fermion-double-trace-action}
S_{\text{def}} = c \int d^{d-1}x_\parallel \left(i \overline{\Psi}^{(k,q)}_\pm\Psi^{(k,q^\prime)}_\mp +\text{c.c.}\right).
\end{align}
 Introducing auxiliary boundary spinors $\bar{\eta}_\pm$, $\eta_\mp$, we can write the deformed theory as
\begin{align}
Z[h] = \frac{Z[0]}{\CZ_\eta} \int D\eta D\psi \exp\left[\int dx\,  (i \bar\eta_\pm \eta_\mp + i h' \bar\eta_\pm\Psi^{(k,q^\prime)}_\mp + i h \overline{\Psi}^{(k,q)}_\pm\eta_\mp+\text{c.c.})\right]~,
\end{align}
where $Z[0]$ is the path integral in the undeformed theory, we define
\begin{align}
\CZ_\eta = \int D\eta_\mp D\bar\eta_\pm\, e^{-i \int dx\,   \bar\eta_\pm \eta_\mp } \ ,
\end{align}
and $c = h h'$. The couplings $h$ and $h'$ have mass dimension $\frac{d-1}{2} - \Delta_q$ and $ \frac{d-1}{2} - \Delta_{q'}$, where $\Delta_q$ and $\Delta_{q'}$ are the conformal dimensions of $\Psi^{(k,q)}_\pm$ and $\Psi^{(k,q')}_\pm$, respectively.  To relieve the notation, as all integrals here are over the parallel directions,
we will suppress the ${}_\parallel$ subscript and also write $\int d^{d-1} x_\parallel$ and $\int dx$.
Integrating over $\psi$ and $\bar\psi$ introduces an expectation value of the exponential function:
\begin{align}
Z[h] = \frac{Z[0]}{\CZ_\eta} \int D\eta  \Big\langle \exp\Big[\int dx (i \bar\eta_\pm \eta_\mp + i h' \bar\eta_\pm {\Psi}^{(k,q^\prime)}_\mp+ i h \overline{\Psi}^{(k,q)}_\pm\eta_\mp+\text{c.c.})\Big]\Big\rangle_0~.
\end{align}
We can bring the expectation value into the exponent by expanding out the original exponential and using
Wick's Theorem:
\begin{align}
Z[h] &= \frac{Z[0]}{\CZ_\eta} \int D\eta \exp\Big[\int dx\, dy\, \Big(i \bar\eta_\pm(x)\delta(x-y)\eta_\mp(y)+i \bar\eta_\mp(x)\delta(x-y)\eta_\pm(y) \\
&\hspace{1cm}- h'^2 \bar\eta_\pm(x)\langle {\Psi}^{(k,q^\prime)}_\mp(x) \overline{\Psi}^{(k,q^\prime)}_\mp(y)\rangle\eta_\pm(y)- h^2 \bar\eta_\mp(x)\langle {\Psi}_\pm^{(k,q)}(x) \overline{\Psi}_\pm^{(k,q)}(y)\rangle\eta_\mp(y)\Big)\Big]\,. \nonumber
\end{align}
This integral has a Gaussian form with an exponent
\begin{align}
\left(
\begin{array}{cc}
\bar \eta_\pm & \bar \eta_\mp
\end{array}
\right)
\left(
\begin{array}{cc}
-h'^2 \mathfrak{G}^{(k,q^\prime)}_\mp & i \\
i & -h^2 \mathfrak{G}^{(k,q)}_\pm
\end{array}
\right)
\left(
\begin{array}{c}
 \eta_\pm \\ \eta_\mp
\end{array}
\right)
\end{align}
where the boundary Green's function
\begin{align}
\mathfrak{G}^{(k,q)}_\pm(x,y) \equiv \left<  \overline{\Psi}^{(k,q)}_\pm(x) {\Psi}^{(k,q)}_\pm(y)\right>.
\end{align}
The change in the boundary sphere free energy is given by the determinant of this $2 \times 2$ matrix:
\begin{align}
\Delta F_\pd=-\Tr\log\left(1 + c^2 \mathfrak{G}^{(k,q^\prime)}_\mp(x, y) \mathfrak{G}^{(k,q)}_\pm(x, y)\right) \ ,
\end{align}
which we will now analyze in the large $c$ limit.

 The spectrum of the fermionic operator $\mathfrak{G}_\mp^{(k,q)}(x_\parallel,y_\parallel)$ can be found by expanding on the basis of spinor spherical harmonics.  The eigenvalues of $\mathfrak{G}_\mp^{(k,q)}$ on the unit $\mathds{S}^{d-1}$ are (up to $\ell$ independent factors) \cite{Allais:2010qq}
\begin{align}
\lambda_{{(k,q)}}(\ell)\propto
\frac{\Gamma(\ell+\Delta_q +\frac{1}{2})}{\Gamma(\ell+d-\Delta_q-\frac{1}{2})}\,.
\end{align}
The eigenvalue degeneracies are
\begin{align}
{\rm deg}_{\mf{G}}(\ell)  = 2^{\lfloor\frac{d-1}{2}\rfloor} \frac{\G(\ell+d-1)}{\G(\ell+1)\G(d-1)}\,.
\end{align}
In the large $c$ limit, the anomalous part of the boundary free energy is then read off from the log-divergent part of
\begin{align}
\Delta F_\pd = - \sum_{\ell=0}^\infty {\rm deg}_{\mf{G}}(\ell)\left(\log\left(h^2\,\lambda_{{(k,q)}}(\ell)\right)+\log\left(h'^2\,\lambda_{{(k,q^\prime)}}(\ell)\right)\right).
\end{align}

Let us consider just the $(k,q)$ term for simplicity; adding back in the $(k,q^\prime)$ contribution is trivial. Using generalized zeta function regularization, we first note that  $h^2$ has dimension $d-1-2\Delta_q$, and we rewrite the log
\begin{align}
    -\sum_{\ell=0}^\infty {\rm deg}_{\mf{G}}(\ell)\log(\mu^{d-1-2\Delta_q}\lambda_{(k,q)}(\ell)) = \frac{d}{ds}\sum_{\ell=0}^\infty {\rm deg}_{\mf{G}}(\ell) (\mu^{d-1-2\Delta_q}\lambda_{(k,q)}(\ell))^{-s}\Big|_{s=0}\,,
\end{align}
where $\mu$ is a mass scale.
The part of the sum contributing to the anomaly is then the coefficient of the $\log \mu$ term
\begin{align}
    -\sum_{\ell=0}^\infty {\rm deg}_{\mf{G}}(\ell)\log(\mu^{d-1-2\Delta_q}\lambda_{(k,q)}(\ell))\Big|_{\rm log} = (d-2\Delta_q-1)\zeta_{\lambda_{(k,q)}}(0)\log \mu\,,
\end{align}
where we define the generalized zeta function
\begin{align}
    \zeta_{\lambda_{(k,q)}}(s) \equiv \sum_{\ell =0}^\infty {\rm deg}_{\mf{G}}(\ell) \left(\frac{\Gamma(\ell+\Delta_q+\frac{1}{2})}{\Gamma(\ell+d-\Delta_q-\frac{1}{2})}\right)^{-s}\,,
\end{align}
whose value at $s=0$ can be thought of as the regularized number of eigenvalues.
Combining everything together, the anomalous part of the change in the boundary free energy is
\begin{align}\begin{split}\label{eq:fermion-kqkqp-boundary-anomaly}
\Delta F_{\pd}|_{{\log}} =~& 2\sum_{\ell=0}^\infty {\rm deg}_{\mf{G}}(\ell)(k-q)\left(\frac{\Gamma(\ell+\frac{d}{2}-k+q)}{\Gamma(\ell+\frac{d}{2}+k-q)}\right)^{-s}\Big|_{s=0}\\
&+2\sum_{\ell=0}^\infty {\rm deg}_{\mf{G}}(\ell)(k-q^\prime)\left(\frac{\Gamma(\ell+\frac{d}{2}-k+q^\prime)}{\Gamma(\ell+\frac{d}{2}+k-q^\prime)}\right)^{-s}\Big|_{s=0}\,,
\end{split}\end{align}
where we set $\Delta_q = (d-2k-1)/2 +q$.

We still need to regularize the sums to compute the boundary anomaly.  We fix the scheme where we first separate out the $\ell=0$ modes, expand in large $\ell$, perform the sum, and then take the $s\to 0$ limit. The terms that contribute to the anomaly are those that diverge in the large $\ell$ limit (see appendix C of ref.~\cite{Brust:2016gjy}).  Noting that the $k$-dependent part of the sum scales as
\begin{align}
\frac{\Gamma(\ell+\frac{d}{2}-k+q)}{\Gamma(\ell+\frac{d}{2}+k-q)}\sim \ell^{-2k+2q}(1+\CO(\ell^{-1}))\,,
\end{align}
 we can easily read off the anomalous part of $\Delta F_\pd$.

 \begin{table}[t]
\begin{center}
\begin{tabular}{|c|c|c|c|c|c|c|c|c|c|}
\hline
& $k=\frac{1}{2} $  & $k=1$   & $k=\frac{3}{2}$   & $k=2$ & $k=\frac{5}{2}$ & $k=3$ & $k=\frac{7}{2}$ & $k=4$ \\ \hline
$d=3$   &$-\frac{1}{3}$ & $\frac{1}{3}$ & $3$ & $\frac{26}{3}$ & $\frac{55}{3}$ &$ 33$ & $\frac{161}{3}$ & $\frac{244}{3}$ \\ \hline
$d=5$  &$\frac{11}{90}$  &$- \frac{17}{360}$ & $-\frac{3}{10}$ & $\frac{103}{180}$ & $\frac{95}{18}$ &$ \frac{741}{40}$ &  $\frac{4277}{90} $ &$\frac{9223}{90} $ \\ \hline
$d=7$ &$-\frac{191}{3780}$ & $\frac{367}{30240}  $ &$ \frac{13}{140}$ &$ -\frac{1061}{15120}$ &$- \frac{275}{756} $ & $ \frac{1111}{1120} $ &$ \frac{5257}{540}  $ &$ \frac{295627}{7560} $ \\ \hline
$d=9$ &$\frac{2497}{113400}$  & $-\frac{27859}{7257600}  $   & $-\frac{7}{200} $ & $\frac{8603}{518400} $ & $\frac{425}{4536} $ &$ -\frac{9257}{89600} $ & $-\frac{8183}{16200} $   &$\frac{3194621}{1814400} $ \\ \hline
\end{tabular}
\end{center}
\caption{\label{tab:fermion-k0k0-boundary-anomaly}  $\Delta F_{\pd}|_{{\log}}$ for the boundary RG flow in the $\slashed{\triangle}_{2k+1}$ theory induced by the $\Psi^{(1,0)}_\pm\Psi^{(1,0)}_\mp$ type deformation.}
\end{table}

In order to illustrate the method, let us consider $k=1$, $d=3$ with $q=q^\prime=0$, where
\begin{align}
\Delta F_\pd|_{\log} = 2^{\lfloor\frac{d+3}{2}\rfloor}\left(1 +\sum_{\ell=1}^\infty(\ell+1)\ell^{2s}\left(1+\frac{2}{\ell}+\frac{3}{4\ell^2}\right)^{s}\right)\Big|_{s\to0}\,.
\end{align}
For convenience in taking the large $\ell$ limit, we have factored out $\ell^{2(k-q)s}$ and $\ell^{2(k-q^\prime)s}$ from the $(\ldots)^{s}$ terms.  Expanding in large $\ell$ to $O(\ell^{-d})$
\begin{align}
    \left(1+\frac{2}{\ell}+\frac{3}{4\ell^2}\right)^{s} = 1+\frac{2s}{\ell}+\frac{s(8s-5)}{4\ell^2}+\frac{s(s-1)(8s-7)}{6\ell^3}+\ldots.
\end{align}
Inserting this expansion back into the sum over $\ell$ and taking the $s\to 0$ limit, we find
\begin{align}
    \Delta F_\pd|_{\log} = \frac{1}{3}~.
\end{align}

Comparing with our results for the scalar, we find the following general expression for the free energy difference in odd $d$, where we denote $\k_{q} = k-q$:
\begin{align}
\Delta F_\partial|_{\rm log} =
2^{-\frac{d+1}{2}} \left(
f^{ND}(d,\kappa_q) + f^{ND}(d, \kappa_{q'}) \right) \ .
\end{align}
 As a concrete examples, the values for $\Delta F_{\pd}|_{{\log}}$ for $q=q^\prime=0$ and $q=0,\,q^\prime=1$, have been compiled in tables~\ref{tab:fermion-k0k0-boundary-anomaly} and \ref{tab:fermion-k0k1-boundary-anomaly}, respectively, up to $d=9$ and $k=4$. Note that, for $k=1$ the values of $\Delta F_\pd|_{\log}$ for  $q=q^\prime =0$  are exactly double the values for $q=0,\,q^\prime=1$.  This should be expected because the diagrams of the flows in \fig{fermionflows} should commute.  Further, since we have the explicit value for the boundary sphere free energy for the $\pm\pm\mp$ boundary condition, i.e.\ $\Delta F^{(+-)}_\pd =0$, the values in table~\ref{tab:fermion-k0k1-boundary-anomaly} (up to sign) give the boundary sphere free energy for $\pm\mp\pm$ and $\pm\mp\mp$ boundary conditions, and not just the difference. Additionally, as is obvious from \eq{fermion-kqkqp-boundary-anomaly} at $k=0$, $\Delta F_{\pd}|_{\log}=0$ follows from the $q=q^\prime=0$ boundary deformation being marginal. As was the case for the higher derivative scalar theory, we also see that in $d=5$ the boundary $a$-theorem can be violated in $k>0$ fermionic theories along flows with generic $q$ and $q^\prime$.

\begin{table}[t]
\begin{center}
\begin{tabular}{|c|c|c|c|c|c|c|c|c|}
\hline
 & $k=1$ & $k=\frac{3}{2}$ & $k=2$ & $k=\frac{5}{2}$ & $k=3$ & $k=\frac{7}{2}$ & $k=4$ \\ \hline
$d=3$ & $\frac{1}{6}$ & $\frac{4}{3}$ & $\frac{9}{2}$ &$\frac{32}{3}$ & $\frac{125}{6}$ &$ 36   $ & $\frac{343}{6}$ \\ \hline
$d=5$    &$ -\frac{17}{720}  $ & $-\frac{4}{45}   $ & $\frac{21}{80}   $ & $\frac{112}{45} $ &$ \frac{1375}{144}$   &  $\frac{132}{5} $ &$\frac{43561}{720} $ \\ \hline
$d=7$ & $\frac{367}{60480}  $ &$ \frac{4}{189}$ &$- \frac{13}{448}$    &$ -\frac{128}{945} $ & $ \frac{5575}{12096} $ &$ \frac{164}{35} $ &$ \frac{34643}{1728} $       \\ \hline
$d=9$ & $-\frac{27859}{14515200} $ & $-\frac{92}{14175} $ & $\frac{1143}{179200} $ & $\frac{416}{14175} $ &$ -\frac{25175}{580608} $ & $-\frac{36}{175} $   &$\frac{1718381}{2073600} $ \\ \hline
\end{tabular}
\end{center}
\caption{\label{tab:fermion-k0k1-boundary-anomaly}  $\Delta F_{\pd}|_{{\log}}$ for the boundary RG flow in the $\slashed{\triangle}_{2k+1}$ theory induced by the $\Psi^{(1,0)}_\pm\Psi^{(1,1)}_\mp$ type deformation.}
\end{table}

\subsection{Two-point functions}
\label{sec:twopoint}

We would like to compute the possible $\langle \phi(x) \phi(x') \rangle$ and $\langle \psi(x) \overline \psi(x') \rangle$ two-point functions
in the $\Box^k \phi$ and $\slashed{\partial}^{2k+1} \psi$ theories that are compatible with boundary conformal symmetry.
We will start with the scalar case.

Boundary conformal symmetry
means the two point function should have the form
\begin{align}
\langle \phi(x) \phi(x') \rangle = \frac{f(\xi)}{ (4x_n x'_n)^{\frac{d}{2}-k}}\,, \; \; \mbox{where} \; \;
\xi = \frac{(x-x')^2}{4x_n x'_n}
\end{align}
is the conformally invariant cross-ratio. The Green's function has the property that
\begin{align}
\label{Greenseq}
 (-\Box)^k \langle \phi(x) \phi(x') \rangle  = \delta^{(d)} (x-x') \ .
\end{align}
In terms of $f(\xi)$ and $\xi$, this partial differential equation
boils down to a $2k$-order ordinary differential equation.
For example, in the
$k=1$ and $k=2$ cases, the differential equations are
\begin{align}
k=1 &: \; \xi(1+\xi) f''(\xi) + \frac{d}{2} (1+2 \xi) f'(\xi) + \frac{d(d-2)}{4} f(\xi) = 0 \ , \\
k=2 &: \;
 \xi^2 (1+\xi)^2 f^{(4)}(\xi) + (d+2) \xi (1+\xi)(1+2\xi) f^{(3)}(\xi) + \frac{d(d+2)}{4} (1+6 \xi(1+\xi)) f''(\xi)  \nonumber \\
&\quad  + \frac{d(d^2-4)}{4} (1+2\xi) f'(\xi) + \frac{d(d-4)(d^2-4)}{16} f(\xi) = 0 \ .
\end{align}
This differential equation in general has the $2k$ solutions
\begin{align}\label{eq:2_pt_f}
f(\xi) = \sum_{j=0}^{k-1} \left( \frac{c_j}{\xi^{\frac{d}{2}-k +j}} + \frac{b_j}{(1+\xi)^{\frac{d}{2}-k+j}} \right) \ .
\end{align}
Note the $d=2k$, $2k-2$, $2k-4$, \ldots cases are special: the solutions become degenerate and one needs to introduce logarithms.

To check that $\Box^k \langle \phi(x) \phi(x') \rangle=0$ away from the coincident limit, it is useful to perform a Fourier
transform.  The Fourier transformed version of the two point function has the structure
\begin{align}
\int d^{d-1} x_\parallel  \,\langle \phi(x) \phi(x') \rangle e^{-i p_\parallel \cdot x_\parallel}  = \sum_{j=0}^{k-1}
 (x_n x'_n)^j \left( c'_j e^{p |x_n - x'_n|}  +b'_j e^{-p (x_n + x'_n)} \right) ,
\end{align}
where the coefficients $b'_j$ and $c'_j$ are proportional to $b_j$ and $c_j$, respectively, and $p=|p_\parallel|$.
Indeed, the Fourier transformed Laplacian $(-p^2 + \partial_n^2)$
acting once on each term in this sum will reduce the $x_n$ degree by one.
Thus any polynomial in $x_n$ of degree less than $k$ will be annihilated by $\Box^k$.

A moment's consideration reveals that the $\delta^{(d)}(x-x')$ in the definition of the Green's function can come only from
 the $c_0$ term in eq.~\eqref{eq:2_pt_f}.  The remaining
 $c_j$ will lead to different singular behavior in the coincident limit, and so we can safely set them to zero.
 The $(1+\xi)$ type terms on the other hand have singularities only in the nonphysical region,
 where $x_n + x_n' <0$, and so
 we keep them.

While $c_0$ is fixed by the normalization of the Dirac delta in (\ref{Greenseq}), the remaining $b_j$, $j=0, \ldots, k-1$,
 must be fixed by boundary conditions at $x_n = 0$.  We have then in general a $k$-dimensional family of boundary conditions.
The variational principle will pick out two special values for each $b_j$.
Concretely, if we impose a boundary condition by setting certain boundary conformal primary to zero, then an appropriate combination of derivatives of the two-point function must vanish in the boundary limit. This fixes all the $b_j$ uniquely.
For example, in the $k=2$ case, we find
\begin{subequations}\label{eq:scalar_prop}
\begin{align}
\label{DNprop}DN: \; f(\xi) &= \kappa \left( \frac{1}{\xi^{\frac{d-4}{2}}} - \frac{1}{(1+\xi)^{\frac{d-4}{2}}}\right)  \ , \\
\label{NDprop}ND: \; f(\xi) &= \kappa \left(\frac{1}{\xi^{\frac{d-4}{2}}} + \frac{1}{(1+\xi)^{\frac{d-4}{2}}}  \right) \ , \\
\label{DDprop}DD: \; f(\xi) &=  \kappa \left(\frac{1}{\xi^{\frac{d-4}{2}}} - \frac{d-4}{2} \frac{1}{(1+\xi)^{\frac{d-2}{2}}} - \frac{1}{(1+\xi)^{\frac{d-4}{2}}}  \right) \ , \\
\label{NNprop}NN: \; f(\xi) &=  \kappa \left(\frac{1}{\xi^{\frac{d-4}{2}}} + \frac{d-4}{2} \frac{1}{(1+\xi)^{\frac{d-2}{2}}} + \frac{1}{(1+\xi)^{\frac{d-4}{2}}}  \right) \ ,
\end{align}
\end{subequations}
where the normalization $\kappa = \Gamma \left(\frac{d}{2}-k\right)/2^{2k}\pi^{\frac{d}{2}} \Gamma(k)$ is chosen to guarantee eq.~(\ref{Greenseq}).

The fermion two-point function is more intricate to construct. The constraint of boundary conformal invariance means that
\cite{Herzog:2022jlx}
\begin{align}
\label{eq:genpsipsi}
\langle \psi(x) \overline \psi(x') \rangle
= \frac{\gamma_\mu (x-x')^\mu f(\xi) + \gamma_n \gamma_\mu (x' - \tilde x)^\mu g(\xi)}{(4 x_n x_n')^{\Delta +\frac{1}{2}}}\ ,
\end{align}
where $\Delta = \frac{d-1}{2}-k$ and $\tilde x = (-x_n, x_\parallel)$, in contrast with $x = (x_n, x_\parallel)$.
Sparing the reader the details, we find that
\begin{subequations}
\begin{align}
f(\xi) &= \sum_{j=0}^{k-1} \left( \frac{a_j}{\xi^{\frac{d}{2}-k+j}} + \frac{b_j}{(1+\xi)^{\frac{d}{2}-k+j}} \right) + \frac{a_k}{\xi^\frac{d}{2}} \ , \\
g(\xi) &= \sum_{j=0}^{k-1} \left( \frac{c_j}{\xi^{\frac{d}{2}-k+j}} + \frac{d_j}{(1+\xi)^{\frac{d}{2}-k+j}} \right) + \frac{d_k}{(1+\xi)^\frac{d}{2}} \ .
\end{align}
\end{subequations}
The boundary condition in the coincident limit means we need to set $a_j$ and $c_j$ all to zero, except for $a_0$.
The remaining $b_j$ and $d_j$ are then set from the boundary conditions at $x_n = 0$.

In the case of the $k=1$ theory, there are eight possible Green's functions to match with the eight conformal boundary conditions.  The general Green's function in this case will take the form \eq{genpsipsi} with
\begin{subequations} \label{eightfoldGreens}
\begin{align}
f(\xi) &= \kappa \left( \frac{1}{\xi^{\frac{d}{2}-1}} + \frac{b_0}{(1+\xi)^{\frac{d}{2}-1}} \right)  , \\
g(\xi) &= \kappa \left( \frac{d_0}{(1+\xi)^{\frac{d}{2}-1}} + \frac{d_1}{(1+\xi)^{\frac{d}{2}}} \right) .
\end{align}
\end{subequations}
The values of the constants $b_0$, $d_0$, and $d_1$ are given in table \ref{table:eightfoldway}.
As discussed earlier,
the $\pm\!\pm\!\pm$ theories are isolated, while the others are connected by RG flows as shown in Figure~\ref{fig:fermionflows}.

\begin{table}
    \centering
    \begin{tabular}{c|ccc}
        &  $b_0$ & $d_0$ & $d_1$ \\
      \hline
      $\pm\!\pm\!\pm$ & $-4+\frac{4}{d}$ & $\mp \left(3-\frac{4}{d}\right)$ & $\mp \frac{(d-2)^2}{d}$ \\
      $+\!-\!\pm$ & $-1$ &0 & $ \pm \frac{d-2}{2}$ \\
      $\pm\pm\mp$ & 0 &  $\pm 1$ &  0 \\
      $-\!+\!\pm$ & 1 & $\mp \frac{4(d-1)}{3d-2}$ & $\mp \frac{(d-2)^2}{2(3d-2)}$
    \end{tabular}
    \caption{The coefficients for specifying the fermion Green's function eq.\ (\ref{eightfoldGreens}) in the $\slashed{\partial}^3 \psi$ theory for the eight different possible conformal boundary conditions.}
    \label{table:eightfoldway}
\end{table}

\subsection{Displacement operator two-point function}
\label{sec:DD}

In this subsection, we consider one of the basic observables that can be used to characterize the boundary theory in a BCFT: the two-point function of the displacement operator.  The displacement operator is defined as $\CD\equiv T_{nn}\big|_{\pd\CM}$. Thus, utilizing the form of the bulk energy-momentum tensor for $d$-dimensional higher derivative free CFTs, we compute $\left<\CD\CD\right>$ for
the higher derivative scalars and fermions.  We focus on the cases $k=2$ and $k=3$ for the scalars and $k=1$ for the fermions.
(The scalar $k=1$ and fermion $k=0$ cases are well known \cite{McAvity:1993ue}.)
The coefficient of $\left<\CD\CD\right>$ is related to a boundary Weyl anomaly in the special cases of $d=3$ \cite{Herzog:2017xha, Herzog:2017kkj} and $d=5$ \cite{Chalabi:2021jud}.

\subsubsection*{$\langle \CD\CD\rangle$ for the $\Box^k\phi$ theory}

The two-point function of the displacement operator is fixed by conformal invariance and reads
\beq
\left< \mathcal{D}(x_\parallel) \mathcal{D}(0) \right> = \frac{c_{\mathcal{DD}}}{|x_\parallel|^{2d}}\,.
\eeq
For the $\Box^2$ theory, $T_{nn}$ may be written as \cite{Osborn:2016bev,Stergiou:2022qqj}
\begin{align}
\label{Tnnd5k2}
T_{n n} = ~& \frac{1}{d-1}\Big( \partial_\rho \phi \partial^\rho \Box \phi+ 2 \partial^2_n \partial_\rho \phi \partial^\rho \phi- (d+2) \partial_n \Box \phi \partial_n \phi + \frac{d-4}{2} \partial_n^2 \Box \phi \phi \\\nonumber
&
+\frac{1}{d-2}\Big(\frac{d(d+2)}{2} \Box \phi \partial^2_n \phi - \frac{d+2}{2} \Box \phi \Box \phi   + 2\partial_\rho \partial_\sigma \phi  \partial^\rho \partial^\sigma \phi - 2d\partial_n \partial_\rho \phi \partial_n \partial^\rho \phi \Big)\Big)\,.
\end{align}
The two-point function of the displacement operator can be easily computed by using Wick's theorem and the form of the propagators in eq.\ \eqref{eq:scalar_prop}. We find the same result for the $DN$ and $ND$ boundary conditions, namely
\beq\label{k2cDD}
c_{\mathcal{DD}} = -16 (d-4)^2 (d-2) (d+4) \kappa ^2,
\eeq
while for $DD$ and $NN$ we get
\beq
\label{cDD-DDNN}
c_{\mathcal{DD}} = 32 (d-4)^2 (d-2)^2 \kappa ^2\, .
\eeq
Note that for the $DN$ and $ND$ cases $c_{\mathcal{DD}}$ is negative, while for the $DD$ and $NN$ it is positive.
Unitarity would guarantee that $c_{\mathcal{DD}}$ is always positive.

Still using Wick contractions and the explicit form of $T_{nn}$ for the $\Box^3$ theory
\cite{Osborn:2016bev,Stergiou:2022qqj}, we calculate
\beq\label{k3cDD}
c_{\mathcal{DD}}=384 (d-6)^2 (d-4) (d-2) (d+4) (d+6) \kappa ^2
\eeq
for $DND$ and $NDN$ boundary conditions. For the $\Box^k\phi$ theory with boundary conditions $DND\cdots$ and $NDN\cdots$, the pattern of \eqref{k2cDD} and \eqref{k3cDD} suggests
\beq
c_{\mathcal{DD}}=2^{k^2+k-1} k \lsp(d-2 k)^2 \big(1-\tfrac12d\big)_{k-1} \big(\tfrac12d+2\big)_{k-1}\kappa^2\,,
\eeq
where $(a)_x$ is the Pochhammer symbol.
We were led to this result through comparison
with the coefficient of the stress tensor two-point function, which has been computed for
general $d$
\cite{Osborn:2016bev}.
Note this result naively seems to vanish in $d = 2(k-n)$ dimensions for $n$ a non-negative integer.  In fact, with our choice of normalization for $\kappa$ --
below eq.\ (\ref{eq:scalar_prop}) --  
$c_{\mathcal DD}$ is finite in $d = 2k$ and diverges when $n$ is positive. 
The corresponding divergence in $\kappa$ in these cases correlates with the 
appearance of logarithms in the propagator.  These theories are all non-unitary cousins of the scalar field in two dimensions.

We would like to give a couple of additional results in the $d=5$ and $k=2$
case.
In the appendix, we derive the $\langle \phi \phi \rangle$ two-point function for  general scale invariant boundary conditions \eq{gen_prop_k2d5}.  A special case of that
boundary condition is consistent with having a symmetric $\Box^2$
operator.
Even though the boundary generically breaks conformal invariance in these cases, our philosophy is to continue to use
the bulk stress tensor (\ref{Tnnd5k2}) because locally, at sufficient distance from the boundary, the conformal invariance should be restored.
We then use Wick's theorem to compute $c_{\mathcal{DD}}$ from
these more general expressions for $\langle \phi \phi \rangle$.
From the propagator \eq{gen_prop_k2d5}, we find
\beq
c_{\mathcal{DD}} = -\frac{3 \left(-8 a^2+9 a b+16 a c-6 b^2-12 b c+c^2+9\right)}{32 \pi ^4 }\,.
\eeq
Specializing to the boundary conditions in the introduction,
\beq
\partial_n^2 \phi +\nu \partial_{\parallel}^2 \phi= 0\,, \qquad
\partial_n^3 \phi +(2-\nu)
\partial_{\parallel}^2 \partial_n \phi  = 0 \,,
\eeq
the displacement two-point function becomes
\beq
 c_{\mathcal{DD}} = \frac{-3+3 \nu  (88+\nu(86+3(-16+\nu)\nu))}{8 \pi ^4 (\nu -1)^2 (3 +\nu)^2}\,.
 \eeq
 For $\nu = -1$, we match the $NN$ case for $d=5$ (\ref{cDD-DDNN}), as expected.
 Also, for $\nu \to \infty$, we match the $DD$ case as well.
 We do not have a physical expectation for why this expression diverges at $\nu = 1$ and $\nu = -3$
 although one should note that the underlying
 $\langle \phi \phi \rangle$ correlator diverges for these values.

\subsubsection*{$\langle \CD\CD\rangle$ for the $\slashed{\pd}^3\psi$ theory}

In this section, we consider the $k=1$ free fermion theory with generic boundary conditions leaving $b_0$, $d_0$, and $d_1$ unspecified, and compute $c_{\CD\CD}$.  The symmetric, conserved, off-shell stress tensor for this theory on flat space can be found in ref.~\cite{Stergiou:2022qqj}, which we won't reproduce in complete detail here.  In order to compute the displacement operator two-point function, we only need the form of $T_{nn}$, which after a bit of massaging becomes
\begin{align}\label{eq:Tnn-fermion}
\hspace{-1cm}T_{nn}&= \frac{(d+1)(\Box_\parallel\bar\psi\G_n\pd_n\psi-\pd_n\bar\psi\G_n\Box_\parallel\psi)}{2(d-1)}+\frac{(\bar\psi\G_n\pd_n\Box_\parallel\psi -\pd_n\Box_\parallel\bar\psi\G_n\psi)}{d-1}\nonumber\\
&\quad+\frac{d(\pd_a\bar\psi\G^a\Box_\parallel\psi -\Box_\parallel\bar\psi\G^a\pd_a\psi)}{(d-1)(d-2)}+\frac{(d-3)(\Box_\parallel\pd_a\bar\psi\G^a\psi-\bar\psi\G^a\pd_a\Box_\parallel\psi)}{2(d-1)}\nonumber\\
&\quad-\frac{(d^2-3d+4)(\pd_n\bar\psi\G_n\pd_n^2\psi-\pd_n^2\bar\psi\G_n\pd_n\psi)}{(d-1)(d-2)}+\frac{(d+3)(\pd^a\bar\psi\G_n\pd_n\pd_a\psi-\pd_n\pd^a\bar\psi\G_n\pd_a\psi)}{2(d-1)}\nonumber\\
&\quad-\frac{(d^2-3d+6)(\pd_n\bar\psi\G^a\pd_n\pd_a\psi -\pd_n\pd_a\bar\psi\G^a\pd_n\psi)}{2(d-1)(d-2)}+\frac{2(\pd_a\pd_b\bar\psi\G^b\pd^a\psi -\pd_a\bar\psi\G^b\pd_b\pd^a\psi)}{(d-1)(d-2)}\nonumber\\
&\quad+\frac{1}{2}(\pd_n^2\bar\psi\G^a\pd_a\psi +\pd_n^2\pd_a\bar\psi\G^a\psi-\bar\psi\G^a\pd_n^2\pd_a\psi -\pd_a\bar\psi\G^a\pd_n^2\psi)~.
\end{align}

The details of the computation of the two-point function are straightforward and not particularly interesting.
We use Wick's theorem with the propagator \eq{genpsipsi} in
combination with a point splitting procedure.  Leaving the boundary conditions unspecified, the final result for the coefficient in $\left<\CD\CD\right>$ is
\begin{align}
\hspace{-0.5cm}c_{\CD\CD} = -\frac{2^{\lfloor\frac{d-7}{2}\rfloor}\Gamma^2\left(\frac{d}{2}\right)}{(d-2)^2\pi^d}& \Big(d^3 \left(6 b_0^2 + d_0^2 + 1 \right) -  d^2 \left (18 b_0^2 + d_0^2 - 4 d_0 d_1 + 1 \right) \\\nonumber&\hspace{-0.5cm}+ 4 d \left(3 b_0^2 - 5 d_0^2 - 11d_0 d_1 - 9 d_1^2 - 5 \right) +
 36 \left(d_0^2 + 2 d_0 d_1 + d_1^2 + 1 \right) \Big).
\end{align}

The results for the conformal boundary conditions in table \ref{table:eightfoldway} are displayed in \tbl{cdd-fermion}.  Note that, as with the scalars, $c_{\CD\CD}$ does not have definite sign for all conformal boundary conditions, which again signals an absence of unitarity in the two-point function.

\begin{table}
\centering
\begin{tabular}{|c|c|}
\hline
$\pm\pm\pm$&$-(d-1) \left(41d-70\right)/ (d-2)$\\\hline
$+-\pm$& $d-9  $\\\hline
$\pm\pm\mp$ & $ -\left(d^2+d-18\right) /(d-2)$\\ \hline
$-+\pm$&$-\left(39 d^4-97 d^3-58 d^2+180 d-72\right)/(d-2) (3d-2)^2$\\\hline
\end{tabular}
\caption{For the corresponding conformal boundary condition in the left column, we report the values of $\pi^d c_{\CD\CD}/\big(2^{\lfloor(d-5)/2\rfloor}\Gamma^2(d/2)\big)$ for the $\slashed{\pd}^3\psi$ theory in the right column. }\label{tab:cdd-fermion}
\end{table}

\section{A curious duality}
\label{sec:duality}

In this section we discuss a duality involving a pair of identical higher derivative theories.  For simplicity, we will focus on the scalar case.
A closely related duality for (two derivative) massive scalars in anti-de Sitter space was pointed out first to our knowledge in a work by
Witten \cite{Witten:2001ua}.
The connection to BCFT was later emphasized by one of us in ref.~\cite{Herzog:2021spv} and also in ref.~\cite{DiPietro:2020fya}.
The duality we discuss below generalizes to the
$\slashed{\partial}^{2k+1} \psi$ theories as well.

Let us consider a pair of $\Box^k \phi$ theories in the presence of a boundary at $x_n=0$.
We will assume that for one of the scalars, the boundary condition for a particular boundary primary is $\Phi^{(k,q)} = 0$.
Call that scalar $\phi_D$.  For the other scalar, we then insist on the conjugate boundary condition $\Phi^{(k,2k-q)} = 0$.
We call that scalar $\phi_N$.
From our analysis in section~\ref{sec:primaries}, we know that there will be a piece in the variation of the boundary action that looks like
\begin{align}
\delta S_q = \int d^{d-1} x_\parallel \, \left[\delta \Phi^{(k,q)}_N \Phi_N^{(k,2k-q)} - \Phi^{(k,q)}_D \delta \Phi_D^{(k,2k-q)} \right] \ .
\end{align}
Allowing the variations to be arbitrary, we deduce the corresponding boundary equations of motion
$\Phi^{(k,2k-q)}_N=0$ and $\Phi^{(k,q)}_D=0$.

Given this starting point, we can consider the exactly marginal boundary deformation
\begin{align}\label{eq:duality_def1}
 S_{\rm def} = g \int d^{d-1} x_\parallel  \, \Phi_N^{(k,q)} \Phi_D^{(k,2k-q)} \ .
\end{align}
(Note that if we try to construct a marginal deformation using $\Phi_D^{(k,q)}$ or $\Phi_N^{(k,2k-q)}$, it will vanish
by the boundary equations of motion and hence is redundant.)
 The variation now takes the more complicated form
 \begin{align}
 \delta S_q + \delta S_{\rm def} = \int d^{d-1} x_\parallel \left[ \delta \Phi_N^{(k,q)} (  \Phi^{(k,2k-q)}_N  + g  \Phi_D^{(k,2k-q)} )
 + \delta \Phi_D^{(k, 2k-q)} ( -\Phi^{(k,q)}_D + g \Phi_N^{(k,q)})  \right] .
 \end{align}
In the limit $g \to \infty$, this deformation will act to switch the boundary conditions on this pair of boundary primaries:
$\Phi^{(k,2k-q)}_N=0 \rightarrow \Phi^{(k,q)}_N = 0$ and $\Phi^{(k,q)}_D =0 \rightarrow \Phi^{(k,2k-q)}_D = 0$.  Neumann
becomes Dirichlet and Dirichlet becomes Neumann.

To make precise the duality we have in mind, let us start from the $g \to \infty$ limit, where this original boundary condition is flipped,
and ask how to get back to the original theory.  For this conjugate theory, from our earlier analysis, we have the following
boundary contributions to the variation of the action:
\begin{align}
\delta \tilde S_q = \int d^{d-1} x_\parallel \, \left[- \Phi^{(k,q)}_N  \delta \Phi_N^{(k,2k-q)} + \delta \Phi^{(k,q)}_D  \Phi_D^{(k,2k-q)} \right] .
\end{align}
To this action, we add instead
\begin{align}\label{eq:duality_def2}
\tilde S_{\rm def} = \tilde g \int d^{d-1} x_\parallel  \, \Phi_D^{(k,q)} \Phi_N^{(k,2k-q)} \, ,
\end{align}
and the variation now is
\begin{align}
\delta \tilde{S}_q + \delta \tilde{S}_{\rm def} = \int d^{d-1} x_\parallel \left[ \delta \Phi_N^{(2k-q,q)} ( - \Phi^{(k,q)}_N  + \tilde g  \Phi_D^{(k,q)} )
 + \delta \Phi_D^{(k, q)} ( \Phi^{(k,2k-q)}_D + \tilde g \Phi_N^{(k,2k-q)})  \right] .
\end{align}
In other words, we find exactly the same boundary equations of motion as before, provided we identify $g = \tilde g^{-1}$.
This is the duality we have in mind.  We have a pair of theories with conjugate boundary conditions on one of the boundary primaries.
The conjugate nature allows for an exactly marginal deformation which can then be used to swap this pair of boundary conditions.
We claim the two theories can be identified under
this quadratic marginal deformation, provided one sets $g =  \tilde g^{-1}$.

These pairs of conjugate scalar fields can arise very naturally in certain circumstances.
Consider a  $\Box^k \phi$ theory with no boundary and introduce a trivial interface at $x_n=0$.
The interface divides the fields into $\tilde \phi_R$ living on the right side, with $x_n>0$, and $\tilde \phi_L$ fields living on the left side, with $x_n<0$.
To ensure the interface is trivial, we must have continuity of the field and its derivatives at the interface
\[
\tilde \phi_L |_{x_n=0} = \tilde \phi_R|_{x_n=0} \ , \; \; \;
 \partial_{n} \tilde\phi_L |_{x_n=0} =  \partial_{n} \tilde \phi_R|_{x_n=0}  \ , \; \; \;
\partial^2_{n} \tilde \phi_L |_{x_n=0} =  \partial^2_{n} \tilde \phi_R|_{x_n=0}  \ , \; \mbox{etc.}
\]
Having split the field in this way, we are free invoke the folding trick and place both fields on the same side of the trivial interface, $x_n>0$, converting an interface theory into a boundary theory, $\phi_L(-x_n) = \tilde \phi_L(x_n)$ and $\phi_R(x_n) = \tilde \phi_R(x_n)$.
When we fold, the $x_n$ direction for the fields on the left is inverted, which will flip the sign of the $\partial_{n}$ derivatives, and the
matching conditions become
\[
 \phi_R |_{x_n=0} =  \phi_L|_{x_n=0} \ , \; \; \;
 \partial_{n} \phi_R |_{x_n=0} =  -\partial_{n} \phi_L|_{x_n=0}  \ , \; \; \;
\partial^2_{n}  \phi_R |_{x_n=0} =  \partial^2_{n} \phi_L|_{x_n=0}  \ , \; \mbox{etc.}
\]
This complicated looking set of relations can be simplified in a surprising way by considering the following linear combinations of the
fields:
\begin{align}
\phi_D = \frac{1}{\sqrt{2}} \left( \phi_R - \phi_L \right) \ , \; \; \;
\phi_N = \frac{1}{\sqrt{2}} \left( \phi_R + \phi_L \right) .
\end{align}
The square root factor is to keep the same normalization of the two point function in the coincident limit.  We have used the subscript
$D$ to emphasize that $\phi_D|_{x_n=0} =0$ satisfies a Dirichlet condition and $N$ to emphasize that $\partial_{n} \phi_N|_{x_n=0} = 0$
satisfies a Neumann condition.   More precisely, in earlier notation,
$\phi_D$ will have a $DND\cdots$ type boundary condition while $\phi_N$ will have a $NDN\cdots$ boundary condition. We are thus left with a pair of scalars with conjugate boundary conditions for which one can construct marginal boundary deformations of the form of \eq{duality_def1}.

\section{Outlook}
\label{sec:discussion}
This work provides a careful look at a class of free BCFTs.
We were able to characterize the conformal boundary conditions for the $\Box^k \phi$ and $\slashed{\partial}^{2k+1} \psi$ theories and the flows between them.  We also computed the hemisphere free energy and the displacement operator two-point function.
We conclude by mentioning some future directions: higher derivative theories of gravity, interactions, and the
peculiar nature of these theories in small, integer numbers of dimensions.

\subsection*{Quantum gravity}

The analysis of
theories that are more than quadratic in time derivatives is an important challenge in the context of quantum gravity.
String theories, while leading to Einstein gravity at low energy scales, naturally produce higher derivative corrections to Einstein's equations that become important at high scales.  Alternate quantum gravity scenarios, such as conformal gravity or
brane worlds, are faced with similar higher derivative issues.

Our biharmonic theory $\Box^2 \phi$ features
prominently as a toy model in discussions of conformal gravity and brane worlds (see \cite{Bender:2008gh,Smilga:2017arl}
and references therein).
Decomposing this biharmonic theory into momentum modes parallel to the boundary and treating the normal direction
to the boundary as time,
the Lagrangian becomes closely related to
that of the Pais-Uhlenbeck oscillator \cite{Pais:1950za},
\begin{align}
L = \ddot z^2 - (\omega_1^2 + \omega_2^2) \dot z^2 + \omega_1^2 \omega_2^2  z^2 \ ,
\label{PUoscillator}
\end{align}
 in the case where the frequencies are degenerate, $\omega_1 = \omega_2$.
 The non-degenerate case is already
a challenging example to give a quantum mechanical interpretation, as it naively has an unbounded Hamiltonian,
with eigenenergies $E_{nm} = \left(n+\frac{1}{2} \right) \omega_1 - \left(m + \frac{1}{2} \right) \omega_2$ for
$m$ and $n$ positive integers.  The $m$ dependent sign can be flipped but at the price of
rendering the corresponding eigenfunctions non-normalizable.
The degenerate case requires special treatment, as the canonical transformation employed to rewrite the Hamiltonian
as a pair of harmonic oscillators becomes singular when $\omega_1 = \omega_2$.

The authors of
refs.\ \cite{Bender:2008gh,Smilga:2017arl} do not appear to agree on how to make sense of
(\ref{PUoscillator}).  Ref.\ \cite{Bender:2008gh} invokes the machinery of PT symmetric quantum mechanics,
changing the norm of the Hilbert space and the reality properties of $z$ to both flip the sign of $m$ in $E_{nm}$
while simultaneously keeping the eigenfunctions normalizable.  Ref.\ \cite{Smilga:2017arl}, critical of what they
perceive as an overly radical reframing of the original physics problem in \cite{Bender:2008gh},
instead stresses that for a free theory,
there is nothing wrong with negative energies per se.  The issues will come later with interactions.  At least classically,
ref.\ \cite{Smilga:2017arl} provides some evidence that within a certain regime of parameter space these higher derivative theories can have stable dynamics.  Given tunneling, whether this stable behavior persists quantum mechanically is an open question.

A complementary line of work is ref.\ \cite{Brust:2016gjy} where the authors study $\Box^k \phi$ theories more generally
(but without boundaries).
Their motivation is again quantum gravity, but from a holographic perspective.
As quantum gravity in anti-de Sitter space can be defined through a unitary CFT living on its boundary,
there is hope that de Sitter space can be defined similarly.  An important difference however is that
any such CFT is likely to be non-unitary.
These $\Box^k \phi$ are some of the simplest non-unitary CFTs and so they are perhaps a good place to
start an investigation into de Sitter holography.\footnote{%
 See \cite{Boyle:2021jaz} for another recent
 use of the $\Box^2 \phi = 0$ equation in 
 the context of cosmology.
}

Against  this backdrop,
one of our motivations for a boundary approach to these higher derivative theories is a perhaps naive hope
that the instabilities can be cured through boundary conditions.
For example ref.\ \cite{Maldacena:2011mk} argues that the ghosts of conformal gravity can be eliminated and familiar black hole solutions obtained in de Sitter space if appropriate Neumann boundary conditions are chosen.
In our simpler case, the boundary primaries
$\Phi^{(2,2)}$ and $\Phi^{(2,3)}$ for the biharmonic theory can be cast as canonically conjugate momenta to the
fields $\phi$ and $\partial_n \phi$.  Indeed, as we saw in the introduction, there is some flexibility in how we define
$\Phi^{(2,2)}$ and $\Phi^{(2,3)}$, parametrized by the Poisson ratio $\nu$.  Only in the case $\nu = -1$ do we recover
full conformal symmetry while elasticity theory bounds the Poisson ratio $-1 \leq \nu \leq \frac{1}{2}$.
This flexibility in defining the conjugate momenta is absent in the discussions of  \cite{Bender:2008gh,Smilga:2017arl}, and we wonder if restoring this Lorentzian version of
Poisson's ratio will lead to a better physical interpretation of the biharmonic theory.

To see how this flexibility emerges, let us review the Hamiltonian formalism for the $\Box^2 \phi$ theory and its corresponding
Ostrogradski instability.  We start with the Lagrangian density
\begin{align}
{\mathcal L}_0 = \frac{1}{2} \left( \Box \phi \right)^2 = \frac{1}{2} \left( \ddot \phi^2 - 2 (\partial_\parallel^2 \phi) \ddot \phi
+ (\partial_\parallel^2 \phi)^2 \right) \ ,
\end{align}
where we have defined $\dot \phi \equiv \partial_t \phi$.
Using a Lagrange multipler, we introduce a new field $\chi$ which on-shell will be identified with $\dot \phi$:
\begin{align}
{\mathcal L} = \frac{1}{2} \left( \dot \chi^2 - 2 \dot \chi (\partial_\parallel^2 \phi) + (\partial_\parallel^2 \phi)^2 \right)
+ \lambda (\chi - \dot \phi) + \alpha \partial_t (\chi \partial_\parallel^2 \phi) \ ,
\end{align}
where we have left ourselves the freedom of adding a total derivative with a free parameter $\alpha$.
We can identify two canonical momenta
\begin{align}
\begin{split}
P_\chi &\equiv  \frac{\partial {\mathcal L}}{\partial \dot \chi} = \dot \chi +(\alpha-1)\partial_\parallel^2 \phi \ , \\
P_\phi &\equiv \frac{\partial{\mathcal L}}{\partial \dot \phi} = - \lambda + \alpha \partial_\parallel^2 \chi \ .
\end{split}
\end{align}
An Euler Lagrange equation
$\ddot \chi - \partial_\parallel^2 \dot \phi = \lambda$
along with the constraint $\chi = \dot \phi$ can be used to re-express the canonical momenta:
\begin{align}
\begin{split}
 P_\chi &= \ddot \phi + (\alpha-1) \partial_\parallel^2 \phi \ , \\
 P_\phi &=  - \dddot \phi + (\alpha + 1) \partial_\parallel^2 \dot \phi \ .
 \end{split}
\end{align}
Up to a minus sign from a factor of $g_{tt}$,
these are precisely the boundary conditions we identified in the introduction from insisting that $\Box^2$ be a symmetric operator.
The choice which leads to conformal boundary conditions is $\alpha=2$, and the Poisson ratio can be identified as
$\nu = 1-\alpha$.  The choice $\alpha = 0$ made by refs.\  \cite{Bender:2008gh,Smilga:2017arl} is curiously outside the physical range for the Poisson ratio although the significance of this fact, if any, in this Lorentzian setting remains to be determined.

The Hamiltonian density takes the form
\begin{align}
{\mathcal H} &= P_\phi \dot \phi + P_\chi \dot \chi - {\mathcal L} \nonumber \\
&= \frac{1}{2} P_\chi^2 + (1-\alpha) P_\chi \partial_\parallel^2 \phi +P_\phi \chi  + \frac{1}{2} \alpha(\alpha-2)
(\partial_\parallel^2 \phi)^2 - \alpha \chi \partial_\parallel^2 \chi \ .
\end{align}
The Ostrogradski instability is the fact that the Hamiltonian is linear in $P_\phi$ and thus unbounded below.
Note that choosing $\alpha=2$ or zero means the Hamiltonian is at most quadratic in spatial derivatives.

\subsection*{Interactions}

We should also make some comments about interactions.  With an unbounded spectrum, the worry is that interactions
between the modes can easily lead to runaway behavior while conserving energy.
At least classically, however, these systems can be stable.
For example, the degenerate Pais-Uhlenbeck oscillator
\begin{align}
L = \ddot z^2 - 2\omega^2 \dot z^2 + \omega^4  z^2 - \frac{1}{2} \alpha z^4
\end{align}
 with a quartic interaction
was studied in ref.\ \cite{Smilga:2004cy} using numerics.  For $\alpha>0$, initial conditions in the neighborhood of
 $z(0) = \dot z(0) = \ddot z(0) = \dddot z(0) = 0$ lead to stable time evolution.

 In the Euclidean context, interactions in the presence of boundaries will likely lead to a further proliferation in the types of allowed
 conformal boundary conditions.
 We know for the $\Box \phi$ theory that a quartic interaction $\phi^4$ can produce
 so-called extraordinary boundary conditions \cite{Lubensky:1975zz}.
 The classically marginal $\phi^4$ interaction in four dimensions
 supports a scaling solution $\phi \sim x_n^{-1}$.
 In the language of boundary critical phenomena, as one decreases the temperature,
 the surface orders before the
 bulk, and then at the bulk critical point, there is already a
 one-point function for the scalar field $\langle \phi \rangle \sim x_n^{-1}$ supported by the interaction.
 In de Sitter and anti-de Sitter space, the Weyl factor cancels against the $x_n^{-\Delta}$ behavior of the one point function;
 the extraordinary transition gets mapped to a state
 with a constant one-point function in curved space-time.

 It will be interesting to investigate these interactions in more detail.  A finger counting exercise
reveals that classically marginal interactions support $\phi \sim x_n^{-\Delta}$ behavior in a variety of other
cases as well.  For example, for the $\Box^2 \phi$ theory
a $\phi^{10}$ interaction in $d=5$, $\phi^6$ in $d=6$, $\phi^4$ in $d=8$, and $\phi^3$ in $d=12$
all support this type of near boundary behavior.  More general types of interactions
that involve derivatives of the scalar field as well, in particular so called shift-symmetric theories \cite{Safari:2021ocb}
invariant under $\phi \to \phi + \text{constant}$, may lead to novel surface behavior.
The generalization of Liouville theory in two dimensions to higher dimensions may also be an interesting way to
add interactions to these systems with boundaries
\cite{Levy:2018bdc}.

\subsection*{Low dimensions}

A final topic which requires a few closing remarks is the peculiar behavior that emerges in these theories in low numbers of dimensions.  While we emphasized in the introduction the connection between the biharmonic theory and
deformations of a thin plate, the $d=2$ dimensional case does not have a traceless stress tensor.  We did not actually
show in this case how conformal symmetry can be realized.  Ref.\ \cite{Brust:2016gjy} discusses some of the exotic features of this theory in the absence of a boundary -- that the single trace Hilbert space is finite dimensional and that the theory has comparatively few primary operators and nonzero correlation functions.   The underlying issue is that the two point function of the primary operator $\langle \phi(x) \phi(0) \rangle \sim x^2$ is a monomial.  As a result, there are only a few nonzero two-point functions in the associated conformal multiplet; acting thrice with momentum kills the correlation function.
The emphasis in our work was on higher dimensional theories where the correlation functions die off rather than grow with distance and these truncation issues do not occur.
 It would be interesting to revisit these low dimensional theories, the interplay between scale and conformal invariance, and
how a boundary does or does not affect the correlation functions and operator spectrum.

\section*{Acknowledgments}
The authors thank Bobby Acharya, Dio Anninos, 
Samuel Bartlett-Tisdall, Hans Werner Diehl, Stuart Dowker, Sarben Sarkar, and Volodia Schaub for discussions.
 A.C.\ and K.R.\ would like to thank King's College London for hospitality, where a part of this work was completed. A.C.\ is funded by DFF-FNU through grant number 1026-00103B and by the Royal Society award RGF/EA/180098.
  C.P.H.\ is funded by a Wolfson Fellowship from the Royal Society
  and by the U.K.\ Science \& Technology Facilities Council Grant ST/P000258/1.
    K.R.\ is funded by the Rhodes Trust via a Rhodes Scholarship.
    B.R.\ is funded in part by the KU Leuven C1 grant ZKD1118 C16/16/005 and by            the INFN.
    J.S.\ is funded by the Knut and Alice Wallenberg Foundation under grant KAW 2021.0170, VR grant 2018-04438 and Olle Engkvists Stiftelse grant n. 2180108.
    A.S.\ is funded by the Royal Society under grant URF{\textbackslash}R1{\textbackslash}211417.

\appendix

\section{Details on higher-derivative scalars with boundaries}\label{app:scalar-appendix}

In this appendix we present a momentum space derivation of the two-point function for
the $\Box^k\phi$ theories that complements the real space derivation presented in the text.  By going to momentum space, we are able to derive more general two-point functions that break conformal invariance but preserve dilatations.  We also provide a detailed discussion of the two-point function along the boundary RG flows of section~\ref{sec:scalar-primaries}.

\subsection{Propagator}\label{app:scalar-propagator-appendix}

The bulk propagator in the absence of a boundary can be easily found by using the momentum-space representation, namely
\begin{equation}
\left< \phi(x) \phi(y) \right>\equiv G(x,y) = \int \frac{d^d p}{(2\pi)^d} \frac{e^{i p \cdot (x-y)}}{p^{2k}}  = \frac{\Gamma \left( \frac{d-2k}{2} \right)}{2^{2k} \pi^{d/2} \Gamma(k)} \frac{1}{|x-y|^{d-2k}}\,.
\end{equation}
In the presence of a boundary, we can find the propagator by solving
\beq
\label{eq:green_equation}
 \left( - \partial^2 \right)^k G(x_{\parallel}, z_1,z_2) = \d^{(d-1)} (x_{\parallel}) \d (z_1- z_2)
\eeq
for the Green's function, where in this appendix we set $x_n \equiv z$ for convenience.
Due to translational invariance along the parallel directions to the boundary we can write
\beq
G(x_\parallel, z_1,z_2) = \int \frac{d^{d-1}p_{\parallel}}{(2\pi)^{d-1}} f(p_{\parallel}, z_1,z_2) e^{i p_{\parallel} \cdot x_\parallel} ,
\eeq
and by plugging this form into the differential equation we obtain
\beq
\left( p_{\parallel}^2 - \partial_{z_1}^2 \right)^k f(p_{\parallel}, z_1,z_2)  = \d (z_1 - z_2) \ .
\eeq
First of all, we study the homogeneous equation, whose general solution reads
\beq
f(p_{\parallel}, z_1,z_2) = \sum_{l=0}^{k-1} \left[ A_l(z_2) z_1^l e^{- p_{\parallel} z_1} + B_l(z_2) z_1^l e^{+ p_{\parallel} z_1} \right] \ ,
\eeq
where by a slight abuse of notation we write $p_\parallel = |p_\parallel|$.

Next we have to consider the two regions $z_1 > z_2$ and $z_1< z_2$ separately. We call the coefficients of the former region $A_{>,l}$ and $B_{>,l}$ while the one of the latter $A_{<,l}$ and $B_{<,l}$. Let us first focus on the region $z_1 > z_2$. In this case regularity at infinity forces $B_{>,l}  = 0$. In contrast, for the region $z_1 < z_2$ there are no conditions so far. Thus,
\beq
f(p_\parallel, z_1,z_2)  =
\begin{cases}
 \sum_{l=0}^{k-1}  A_{>,l}(z_2) z_1^l e^{- p_\parallel z_1}  \,, \qquad &z_1 > z_2\,,  \\
 \sum_{l=0}^{k-1} \left[ A_{<,l}(z_2) z_1^l e^{- p_\parallel z_1} + B_{<,l}(z_2) z_1^l e^{+ p_\parallel z_1} \right], \qquad &z_1 < z_2 \,.
\end{cases}
\eeq
To solve the inhomogeneous equation we will need to require continuity up to the $(2k-2)$-th derivative on $z_1$, while there will be a discontinuity in the $(2k-1)$-th one. Thus, we impose
\beq
\left.  \partial^i_{z_1} f \right|_{z_1 = z_2}^+ = \left.  \partial^i_{z_1}  f \right|_{z_1 = z_2}^- ,  \qquad i = 0,\dots, 2 k -2\,,
\eeq
and
\beq
\left. \partial_{z_1}^{2k-1} f \right|_{z_1 = z_2}^+ - \left. \partial_{z_1}^{2k-1} f \right|_{z_1 = z_2}^- = (-1)^k \,.
\eeq

To be concrete, we first specialize to $k=2$. In that case, the solution to the above algebraic system reads
\beq
\begin{split}
& A_{<,0}(z_2)= A_{>,0}(z_2)+\frac{  p_\parallel z_2-1}{4 p_\parallel^3} e^{p_\parallel z_2} \,, \qquad A_{<,1}(z_2) = A_{>,1}(z_2) - \frac{e^{p_\parallel z_2}}{4 p_\parallel^2} \,,\\
& B_{<,0}(z_2)= \frac{  p_\parallel z_2 + 1}{4 p_\parallel^3} e^{- p_\parallel z_2} \,, \qquad B_{<,1}(z_2) =  - \frac{e^{- p_\parallel z_2}}{4 p_\parallel^2}\,,
\end{split}
\eeq
which implies
\beq
\begin{split}
& f(p_\parallel, z_1,z_2)  = \\
& \begin{cases}
A_{<,0}(z_2) e^{- p_\parallel z_1} -  \frac{  p_\parallel z_2-1}{4 p_\parallel^3} e^{p_\parallel(z_2 - z_1)}  + A_{<,1}(z_2) e^{- p_\parallel z_1} z_1 + \frac{e^{p_\parallel(z_2-z_1)}}{4 p_\parallel^2} z_1\,, \qquad &z_1 > z_2\,,  \\
A_{<,0}(z_2) e^{- p_\parallel z_1} + A_{<,1}(z_2) e^{- p_\parallel z_1} z_1 + \frac{  p_\parallel z_2 + 1}{4 p_\parallel^3} e^{-p_\parallel (z_2-z_1)}   - \frac{e^{- p_\parallel(z_2-z_1)}}{4 p_\parallel^2} z_1\,, \qquad &z_1 < z_2 \,.
\end{cases}
\end{split}
\eeq
At this point we impose symmetry under the exchange $z_1 \leftrightarrow z_2$. This leads to the equation
\beq
A_{<,0}(z_2)e^{- p_\parallel z_1} + A_{<,1}(z_2)e^{- p_\parallel z_1}  z_1 = A_{<,0}(z_1)e^{- p_\parallel z_2} + A_{<,1}(z_1)e^{- p_\parallel z_2}  z_2 \,,
\eeq
with general solution
\beq
A_{<,0}(z_2) = \frac{a  + c\, p_\parallel z_2 }{p_\parallel ^3} e^{- p_\parallel z_2}\,, \qquad A_{<,1}(z_2) = \frac{b\, p_\parallel z_2  + c  }{p_\parallel ^2}  e^{- p_\parallel z_2} \,.
\eeq
Thus, the final result is
\beq
\label{eq:general_f}
f(p_\parallel, z_1,z_2)  =  \frac{ p_\parallel | z_1 - z_2| +1}{4 p_\parallel^3}  e^{-p_\parallel |z_1 -z_2| } +  \frac{  a + c \, p_\parallel (z_1+ z_2 ) + b \, p_\parallel^2 z_1 z_2}{4 p_\parallel^3} e^{-p_\parallel (z_1+z_2)}  \,.
\eeq

Repeating the same analysis for $k=3$, we instead obtain
\beq
\label{eq:general_fk3}
\begin{split}
& f(p_\parallel, z_1,z_2)  =   e^{-p_\parallel | z_1- z_2| }\frac{ p_\parallel | z_1- z_2|  (p_\parallel |
   z_1- z_2| +3)+3}{16 p_\parallel^5} \\
 & + e^{-p_\parallel (z_1+z_2)}  \frac{ a+b\, p_\parallel (z_1+ z_2)+c \,
   p_\parallel^2 \left(z_1^2+ z_2^2\right)+e \, p_\parallel^2 z_1 z_2+g \,
   p_\parallel^3 z_1 z_2 (z_1+z_2)+h \, p_\parallel^4 z_1^2
   z_2^2}{p_\parallel^5}   .
\end{split}
\eeq
Generalizing to arbitrary $k$, we find
\beq
\label{eq:general_fkn}
\begin{split}
 f(p_\parallel, z_1,z_2)  = \frac{ | z_1- z_2|
   ^{k-\frac{1}{2}} K_{\frac{1}{2}-k}(p_\parallel | z_1- z_2|
   )}{2^{k - \frac{1}{2}}\sqrt{\pi } \, \Gamma (k) p_\parallel^{k -\frac{1}{2}}}  + \frac{e^{-p_\parallel (z_1+z_2)}}{p_\parallel^{2 k -1}}
 \left[ \sum_{i,j=0}^{k -1 } a_{ij} p_\parallel^{i+j} z_1^i z_2^j \right]  ,
\end{split}
\eeq
where $a_{ij}$ is a $k \times k$-symmetric matrix. We stress that up to now the only requirement is that the bulk equation of motion \eq{green_equation} is satisfied. In particular, we did not require either conformal or scale invariant boundary conditions.

If we impose scale invariance, then the coefficients $a_{ij}$ are pure numbers independent of the momentum $p_\parallel$.
As an example we consider the case $k=2$ and $d=5$, where going back to position space gives the propagator
\beq
\label{eq:gen_prop_k2d5}
G(x,y) = \frac{1}{16 \pi^2} \frac{1}{|x - y|} + \frac{a |x - \tilde y |^3 +b \,  x_n y_n
   \left(|x-\tilde y |+x_n+y_n \right)+c \,
   (x_n + y_n) |x - \tilde y|^2}{16 \pi
   ^2 |x - \tilde y |^3
   \left(|x - \tilde y | + x_n + y_n \right)} \,.
\eeq
While this Green's function is scale invariant, it does not in general respect conformal symmetry, as we will show in the next subsection.  The conformal boundary conditions in the main part of the text correspond to the following assignments for $a$, $b$, and $c$:
\begin{align}
\begin{array}{c|rrr}
& a & b & c \\
\hline
        DD &   -1 & -2 &  -1 \\
        DN & -1 &  0 &  -1 \\
        ND &  1 &  0 &  1
        \end{array}
\end{align}
For the scale but not necessarily conformally invariant generalized $NN$ boundary condition discussed around \eq{NN_primaries_sec}, we find in terms of $\nu$
\begin{align}
    NN: \; a = -\frac{5+2 \nu + \nu^2}{(\nu-1)(\nu+3)} \ , \; \; \;
    b = - \frac{2(\nu-1)}{\nu+3} \ , \; \; \;
    c = - \frac{\nu-1}{\nu+3} \ ,
\end{align}
where the conformal choice is $\nu= -1$.
Evidently this Green's function is not well defined for $\nu = -3$ and $\nu = 1$.

\subsection{Constraint due to conformal symmetry}\label{app:conformal-constraints-appendix}

In this section we discuss the constraints of conformal symmetry on the boundary conditions from the point of view of the propagator. A convenient way to proceed is by employing the Ward identity associated with the special conformal transformation, which reads
\beq
\left[ \sum_{j=1}^n \left( 2 \Delta_j x_j^\mu + 2 x_j^\mu x_j^\nu \frac{\partial}{\partial x_j^\nu} - x^2_j \frac{\partial}{\partial x_{j \mu}}\right) \right] \left< \CO_1(x_1) \CO_2(x_2)\dots \CO_n(x_n) \right> = 0\,.
\eeq
In the presence of a boundary, only the generators along the parallel directions are preserved. In momentum space, the Ward identity for the two-point function $\CO_1 = \CO_2 =\CO$ becomes
\beq
\begin{split}
\left[ 2(\Delta -d+1) \frac{\partial}{\partial p_a} - 2 p^b \frac{\partial}{\partial p_b} \frac{\partial}{\partial p_a} + p^a \frac{\partial}{\partial p^b}\frac{\partial}{\partial p_b} + \right. &\\
\left.  + 2 z_1 \frac{\partial}{\partial p_a} \frac{\partial}{\partial z_1} - z_1^2 p^a + z_2^2 p^a  \right]&f(p, z_1,z_2)  = 0\,,
\end{split}
\eeq
where for simplicity we drop the subscripts and write $p^a \equiv p^a_{\parallel}$, and $a,b=0,\dots,d-2$ run over the preserved special conformal generators, which are along the directions parallel to the boundary. The discussion here is an adaptation of \cite{Bzowski:2013sza} to the case with a boundary.
The above equation implies
\beq
\sum_{i,j = 0}^{k-1} a_{ij} \left[ A_{ij} z_1^i z_2^j + B_i\, |p| z_1^{i+1} z_2^{j} - B_j\, |p|z_1^i z_2^{j+1} \right] =0\,,
\eeq
where we defined

\beq
A_{ij} \equiv (i-j) (1+i+j-2k), \qquad B_i \equiv 2(k -i-1)\,.
\eeq
Since $z_1, z_2$ can assume any positive value, we need to set to zero all the possible combinations of $z_1^i z_2^j$. This gives the following conditions
\begin{subequations}
\begin{align}
(a_{10}-a_{00})(k -1) &= 0\,,  \\
a_{i0}A_{i0}+a_{i-1,0}B_{i-1} &= 0\,,&  i &\in (2,k -1)\,, \\
a_{ij}A_{ij} + a_{i-1,j}B_{i-1}-a_{i,j-1}B_{j-1} &= 0\,,&  i,j &\in(1,k -1)\,.
\end{align}
\end{subequations}
We note that the first two equations impose $k-1$ conditions, while the last one corresponding to a $(k-1)\times (k-1)$ antisymmetric matrix gives $(k-1)(k-2)/2$ conditions. Thus, we got $(k-1)k/2$ conditions in total. Finally, since the $a_{ij}$ matrix contains $k(k+1)/2$ free coefficients, we see that conformal invariance leaves us with precisely $k$ undetermined constants.

Now we discuss some specific cases:
\begin{itemize}
\item For $k=1$, there are no conditions. Indeed, the first one disappears due to the combination $k-1$, while the other ones are absent. Thus, we are left with one free coefficient.
\item For $k=2$ there is only the first condition, which gives $a_{10}= a_{00}$ (or $a=c$ in the notation of eq.~\eqref{eq:general_f}), in agreement with what we found in the section~\ref{sec:twopoint}.
\item For $k = 3$ we get $a_{10}-a_{00} = 0$, $a_{10}-3 a_{20} = 0$, and $a_{12}+2 a_{02}-a_{11}=0$. In terms of the coefficient of the propagator \eq{general_fk3} those are $a=b$, $b=3c$, and $2c+g -e = 0$.
\end{itemize}

As an example, we write down the propagator for $k=2$ and generic $d$, which we used in the main text to compute the two-point function of the displacement operator. It reads
\beq
\begin{split}
G(x_{\parallel}, z_1,z_2) &= \frac{\Gamma \left( \frac{d-4}{2} \right)}{16 \pi^{d/2}} \left[ \frac{1}{|x-y|^{d-4}} +\frac{a}{|x-\tilde y|^{d-4}} + (d-4) b \frac{z_1 \, z_2}{|x-\tilde y|^{d-2}}\right]  \\
& = \frac{\Gamma \left( \frac{d-4}{2} \right)}{16 \pi^{d/2}} \frac{1}{\left(4 z_1 z_2\right)^{\frac{d-4}{2}} } \left[ \frac{1}{\xi^{\frac{d-4}{2}}} +\frac{a}{\left(1+ \xi\right)^{\frac{d-4}{2}}} + \frac{(d-4)}{4}  \frac{b}{\left( 1+\xi\right)^{\frac{d-2}{2}}}\right],
\end{split}
\eeq
where we defined
\beq
\xi \equiv \frac{(x-y)^2}{4 z_1 \, z_2} \,. \nonumber
\eeq
$ND$ and $DN$ boundary conditions correspond to $a=+1$ and $a=-1$, respectively, and $b=0$.  $NN$ and $DD$ correspond to $a =\pm 1, \, b = \pm 2$.

\subsection{RG flows and two-point functions}
\label{app:RGflow}

Here, we consider a BCFT containing the boundary primary of level $(k,q)$. When $0\le q <2k-1$, the quadratic boundary deformation
\beq
S_c^{(q)} = c \int d^{d-1} x_{\parallel} \, \Phi^{(k,q)}\Phi^{(k,q)}
\eeq
is relevant, and we expect it to trigger an RG flow from the Neumann-type boundary condition $\Phi^{(k,2k-q-1)}=0$ to the Dirichlet-type boundary condition $\Phi^{(k,q)}=0$. Since the deformation is quadratic, we can solve the theory exactly, obtaining the propagator and other interesting correlators analytically for any value of the coupling $c$.

For simplicity we normalize the boundary primary fields such that
\beq
\left<  \Phi^{(k,q)} \left(x_{\parallel}\right)\Phi^{(k,q')} (0)\right>_0 = \frac{\delta_{q,q'}}{\left|x_{\parallel}\right|^{d-2(k-q)}}
\eeq
and
\beq
\left<  \phi(x) \Phi^{(k,q)} (0)\right>_0 = \frac{C_{k,q}}{z^{-q}\left(x^2_{\parallel} + z^2\right)^{\frac{d-2(k-q)}{2}}}\,,
\eeq
where $\langle\,\cdot\,\rangle_0$ indicates that the correlation functions are computed in the undeformed theory, i.e.\ for $c=0$. To obtain the propagator of the deformed theory, we expand the exponential in the path-integral corresponding to the deformation and successively apply Wick's theorem as in ref.~\cite{Bianchi:2021snj}. We obtain
\begin{equation}
\label{eq:prop_expansion}
\begin{split}
\left<\phi(x)\phi(0)\right>_c=\left<\phi(x)\phi(0)\right>_0+ \sum_{n=1}^\infty (-c)^n  \int \prod_{i=1}^n d^{d-1} x_{\parallel i} \frac{\left<\phi(x)\Phi^{(k,q)}(x_{ \parallel 1} )\right>_0  \left<\Phi^{(k,q)}(x_{\parallel n})\phi(0)\right>_0}{\prod_{j=1}^{n-1}|x_{\parallel j} -x_{\parallel j+1} |^{d-2(k-q)}}\,.
\end{split}
\end{equation}
At this point we find it convenient to go to Fourier space, namely
\beq
\frac{1}{\left|x_{\parallel}\right|^{d-2(k-q)}} = A \int \frac{d^{d-1} p_{\parallel}}{(2\pi)^{d-1}} \, \frac{e^{i p_{\parallel} \cdot x_{\parallel}}}{\left| p_{\parallel}\right|^{2(k-q)-1}}\,, \qquad A = \frac{4^{k-q} \pi^{\frac{d-1}{2}}\Gamma \left( k-q-\frac{1}{2}\right) }{2\,\Gamma\left( \frac{d}{2}-k+q\right)}\,,
\eeq
and
\beq
\frac{1}{\left( x^2_{\parallel} + z^2 \right)^{\frac{d-2(k-q)}{2}}} =  \int \frac{d^{d-1} p_{\parallel}}{(2\pi)^{d-1}} \, \frac{e^{i p_{\parallel} \cdot x_{\parallel}}}{\left| p_{\parallel}\right|^{2(k-q)-1}} g\left(\left|p_{\parallel}\right| z\right),
\eeq
with
\beq
g\left(\left|p_{\parallel}\right| z\right) = \frac{2\pi^{\frac{d-1}{2}}}{\Gamma\left( \frac{d}{2} -k+q \right)} \left(2 |p_{\parallel}|z \right)^{k-q-\frac{1}{2}} K_{k-q-\frac{1}{2}} \left( |p_{\parallel}|z \right).
\eeq
In Fourier space the integrals and the sum in \eq{prop_expansion} can be performed giving
\begin{equation}
\label{eq:final}
\begin{split}
&\left<\phi(x_1)\phi(0,z_2)\right>_c=\left<\phi(x_1)\phi(0,z_2)\right>_0 \\
&- C_{k,q}^2  \int  \frac{d^{d-1} p_{\parallel}}{(2\pi)^{d-1}} \, e^{i p_{\parallel}\cdot x_{||}} \frac{c \, z_1^q z_2^q }{p_{\parallel}^{2(k-q)-1}+A \, c }\,\frac{1}{\left|
p_{\parallel}\right|^{2(k-q)-1}}g\left(\left|p_{\parallel}\right| z_1\right)g\left(\left|p_{\parallel}\right| z_2\right).
\end{split}
\end{equation}
This is the general result valid for any $k$, $q$, and $c$.

As an example, let us consider the case $k=2$ with $NN$ boundary condition and the deformation corresponding to $(k,q)=(2,0)$. In this case we find $C^2_{2,0}=8 \pi^{d/2}/\Gamma(d/2-2)$ and $A=4 \pi ^{d/2}/\Gamma \left(\frac{d}{2}-2\right)$. Thus, we find
\begin{equation}
\begin{split}
\left<\phi(x_1)\phi(0,z_2)\right>_c=\ &\left<\phi(x_1)\phi(0,z_2)\right>_0 \\
&-  \frac{8\pi^{\frac{d}{2}}}{\,\Gamma \left(\frac{d}{2}-2\right) } \int  \frac{d^{d-1} p_{\parallel}}{(2\pi)^{d-1}} \, e^{i p_{\parallel}\cdot x_{\parallel}} \frac{c \, e^{- \left|p_{\parallel}\right|\left(z_1+z_2\right)} }{\left|p_{\parallel}\right|^{3}+A \, c }\, \frac{\left(1+ \left|p_{\parallel}\right| z_1\right) (1 + \left|p_{\parallel}\right| z_2)}{\left|p_{\parallel} \right|^3} .
\end{split}
\end{equation}
The interesting limit is $c \rightarrow +\infty$, where the theory is expected to become conformal. Indeed, it is easy to see that we obtain
\begin{equation}
\label{eq:final_q0}
\begin{split}
\left<\phi(x_1)\phi(0,z_2)\right>_c=\frac{\Gamma \left( \frac{d-4}{2} \right)}{16 \pi^{d/2}}  \frac{1}{|x-y|^{d-4}} -  \int  \frac{d^{d-1} p_{\parallel}}{(2\pi)^{d-1}} \, e^{i p_{\parallel}\cdot x_{\parallel}}  \frac{1+ \left| p_{\parallel} \right| (z_1+z_2)}{2\,\left|p_{\parallel}\right|^3}e^{- \left|p_{\parallel}\right|(z_1+z_2)} ,
\end{split}
\end{equation}
which is exactly the propagator in the $DN$ case, as expected.

Another interesting correlator is the two-point function of the boundary primary labeled by $(k,q)$. By repeating the above procedure one finds
\begin{equation}
\label{eq:final_primary}
\begin{split}
\left<\Phi^{(k,q)}(x_{\parallel})\Phi^{(k,q)}(0)\right>_c= A \, \int \frac{d^{d-1} p_{\parallel}}{(2\pi)^{d-1}} \frac{e^{i p_{\parallel}\cdot x_{\parallel}}}{\left|p_{\parallel}\right|^{2(k-q)-1}+A \, c } \,.
\end{split}
\end{equation}
The claim is that at short distances we recover the unperturbed two-point function corresponding to the level $(k,q)$, while at large separations we expect to find the power-law behavior of the conjugate field with level $(k,2k-q-1)$.
If we consider the endpoint of the flow, where we have the nonzero operator
$\Phi^{(k,2k-q-1)}$, we can walk back up the flow a little ways by introducing the irrelevant deformation $c^{-1} \Phi^{(k,2k-q-1)} \Phi^{(k,2k-q-1)}$ to the Lagrangian.  As we saw in section \ref{sec:primaries}, this deformation leads to the boundary equation of motion $\Phi^{(k,q)} + c^{-1} \Phi^{(k,2k-q-1)} = 0$.  This boundary equation of motion lets us read off the perturbative result in the large $c$ limit
\begin{align}
    \langle \Phi^{(k,q)}(x_\parallel) \Phi^{(k,q)}(0) \rangle \sim
    c^{-2} \langle \Phi^{(k,2k-q-1)}(x_\parallel) \Phi^{(k,2k-q-1)}(0) \rangle \ ,
\end{align}
which we will now demonstrate explicitly using (\ref{eq:final_primary}).

Firstly, we take advantage of the spherical symmetry of the boundary and rewrite the above correlator in terms of the Hankel transform
\begin{equation}
\label{eq:Hankel}
\begin{split}
\left<\Phi^{(k,q)}(x_{\parallel})\Phi^{(k,q)}(0)\right>_c \propto \frac{1}{|x_{\parallel}|^{\frac{d-3}{2}}} \int_0^{+\infty} \frac{d |p_{\parallel}|}{(2\pi)^{d-1}} \frac{|p_{\parallel}|^{\frac{d-1}{2}}}{\left|p_{\parallel}\right|^{2(k-q)-1}+A \, c } J_{\frac{d-3}{2}} \left(|p_{\parallel}||x_{\parallel}| \right)  \,.
\end{split}
\end{equation}
Expanding the rational part of the integrand for small $|p_\parallel|$,
\[
 \frac{|p_{\parallel}|^{\frac{d-1}{2}}}{\left|p_{\parallel}\right|^{2(k-q)-1}+A \, c } =
 \frac{1}{Ac} \left( |p_\parallel|^{\frac{d-1}{2}} - \frac{|p_\parallel|^{\frac{d-3}{2} + 2(k-q)}}{Ac} + \ldots \right) \ ,
\]
we can employ the following standard integral
\begin{equation}
\label{eq:bessel_int}
\int_0^\infty d t \, t^\mu  J_\nu (tx)= \frac{2^\mu \Gamma\left(\frac{1+\mu+\nu}{2} \right)}{x^{1+\mu} \Gamma\left(\frac{1-\mu+\nu}{2} \right)} \ .
\end{equation}
Integrating $|p_\parallel|^{\frac{d-1}{2}}$ against the Bessel function gives a contact term that we ignore, but the second term in the expansion gives the result of interest:
\begin{equation}
\label{eq:final_Hankel}
\begin{split}
\left<\Phi^{(k,q)}(x_{\parallel})\Phi^{(k,q)}(0)\right>_c \sim \frac{1}{c^2} \frac{1}{|x_{\parallel}|^{d+2(k-q-1)}}  \,,
\end{split}
\end{equation}
where we are ignoring overall constant factors and subleading contributions at large separations. Note that if the power of $p_\parallel$ in the denominator of the integrand in \eq{Hankel} were an even positive number, then the integral \eq{bessel_int} would vanish due to a pole in the $\Gamma$-function in the denominator. This is consistent with the asymptotic behaviour of the usual scalar propagator $(p^2 + m^2)^{-1}$, which, being exponentially decreasing in real space, cannot be obtained by a perturbative expansion.

\section{The constant term in \texorpdfstring{$\Delta F_\partial$}{DeltaF\_partial}}
\label{app:constantterm}

In the text, we emphasized the coefficient of the logarithmic divergence in $\Delta F_\partial$ that occurs in odd bulk dimensions $d$.
This coefficient has been important because of its connection to boundary contributions to the anomaly in the trace of the stress tensor.
In even dimensions, there is a constant term which is also believed to be regularization independent and should obey a
similar monotonicity condition, at least for unitary theories \cite{Gaiotto:2014gha}.   Isolating this term is more involved
than the procedures used in the text to compute the coefficient of the logarithmic divergence.  Here, we use zeta function regularization to compute $\Delta F^{(ND)}_\partial$ for the $\Box^k \phi$ theories.

A general integral formula for this constant term is available as eq.\ (24) of \cite{Dowker:2010qy}. 
We content ourselves here to reproduce a few specific cases (see table \ref{table:zeta}).
Our method fails to produce a finite result when $k \geq \frac{d}{2}$.
This bound is interesting because it corresponds to theories where there is an obstruction to writing a stress tensor that
is also a conformal primary \cite{Stergiou:2022qqj}.

\begin{table}
\begin{center}
\[
\begin{array}{c|ccc}
& k=1 & k=2 & k=3
\\
\hline
d=4 & \frac{\zeta(3)}{8 \pi^2} \\
d=6 & -\frac{\pi^2 \zeta(3) + 3 \zeta(5)) }{96 \pi^4} & \frac{5 \pi^2 \zeta(3) - 3 \zeta(5)}{48 \pi^4}  \\
d=8 & \frac{8 \pi^4 \zeta(3) + 30 \pi^2 \zeta(5) + 45 \zeta(7)}{5760 \pi^6} & -\frac{22 \pi^4 \zeta(3) +60\pi^2 \zeta(5) - 45 \zeta(7) }{2880 \pi^6} & -\frac{56 \pi^4 \zeta(3) + 70\pi^2  \zeta(5) - 15 \zeta(7)}{640 \pi^6}
\end{array}
\]
\end{center}
\caption{The constant term in $\Delta F^{(ND)}_\partial$ for the $\Box^k\phi$ theories for the first few values of $k$ and $d$.
The $d=4$, $k=1$ result appears in \cite{Klebanov:2011gs,Gaiotto:2014gha}. 
The $d=6$, $k=1$ result is (14) of \cite{Dowker:2014rva}. 
\label{table:zeta}}
\end{table}

We will illustrate how to compute $\Delta F^{(ND)}_\partial$ for the $\Box \phi$ theory in $d=4$, which recovers a result in  \cite{Klebanov:2011gs,Gaiotto:2014gha}.  
The remaining values in table \ref{table:zeta} follow from analogous if more involved calculations.
The starting point is the sum (\ref{DeltaFND}) in the special case $k=1$ and $d=4$:
\[
\Delta F^{(ND)}_\partial = -\frac{1}{12} \sum_{\ell=0}^\infty (\ell+1)(\ell+2)(\ell+3) \log \frac{\ell+3}{\ell+1} \ .
\]
We regularize the logarithm using zeta functions:
\[
\Delta F^{(ND)}_\partial =\frac{1}{12} \frac{d}{ds} \sum_{\ell=0}^\infty  (\ell+1)(\ell+2)(\ell+3)
\left[ \frac{1}{(\ell+3)^s} - \frac{1}{(\ell+1)^s} \right]_{s=0} \ .
\]
Evaluating the sum then gives
\[
\Delta F^{(ND)}_\partial = \frac{\zeta(3)}{8 \pi^2} \ .
\]
The positive sign here is consistent with the monotonicity of this quantity proposed in ref.~\cite{Gaiotto:2014gha}.
That the sign alternates as we move from column to column is in contradiction with a monotonic
behavior under RG flow.  Presumably this failure of monotonicity is correlated with the loss of unitarity for the $k>1$ theories.

 Following a similar strategy to what is in the text, we could then in principle go further and determine the change $\Delta F_\partial$ induced by a relevant boundary deformation and also
similar data for the fermionic $\slashed{\partial}^{2k+1}\psi$ theories.  While we include this appendix
 to illustrate the control we have over these theories, as it is not clear to us yet what lessons to
extract from this data, we will leave these further tasks as an exercise for the reader.

\bibliographystyle{JHEP}
\bibliography{HD-BCFT}

\providecommand{\href}[2]{#2}\begingroup\raggedright\begin{thebibliography}{10}

\bibitem{germain1821recherches}
S.~Germain, \emph{Recherches sur la th{\'e}orie des surfaces {\'e}lastiques}.
\newblock V. Courcier, 1821.

\bibitem{lagrange1828note}
J.~Lagrange, \emph{Note communiqu{\'e}e aux commissaires pour le prix de la
  surface {\'e}lastique {D{\'e}cembre} 1811}, {\emph{Ann. Chimie Physique}
  {\bfseries 39} (1828) }.

\bibitem{poisson1829memoire}
S.~Poisson, \emph{Memoire sur l’{\'e}quilibre et le movement des corps
  {\'e}lastique. {L’Acad{\'e}mie Royale des Sciences}}.
\newblock 1829.

\bibitem{timoshenko1951theory}
S.~Timoshenko and J.~N. Goodier, \emph{Theory of Elasticity}.
\newblock McGraw-Hill, 1951.

\bibitem{slaughter2012linearized}
W.~S. Slaughter, \emph{The linearized theory of elasticity}.
\newblock Springer Science \& Business Media, 2012.

\bibitem{landau1986theory}
L.~D. Landau, E.~M. Lifshitz, A.~M. Kosevich and L.~P. Pitaevskii, \emph{Theory
  of elasticity: volume 7}, vol.~7.
\newblock Elsevier, 1986.

\bibitem{rayleigh1893xxxviii}
L.~Rayleigh, \emph{{XXXVIII}. {On} the flow of viscous liquids, especially in
  two dimensions}, \href{https://doi.org/10.1080/14786449308620489}{\emph{The
  London, Edinburgh, and Dublin Philosophical Magazine and Journal of Science}
  {\bfseries 36} (1893) 354--372}.

\bibitem{burkhardt2010fluctuations}
T.~W. Burkhardt, Y.~Yang and G.~Gompper, \emph{Fluctuations of a long,
  semiflexible polymer in a narrow channel},
  \href{https://doi.org/10.1103/physreve.82.041801}{\emph{Physical Review E}
  {\bfseries 82} (2010) 041801},
  [\href{https://arxiv.org/abs/arXiv:1008.1594}{{\ttfamily arXiv:1008.1594}}].

\bibitem{Diehl:2003kv}
H.~W. Diehl, S.~Rutkevich and A.~Gerwinski, \emph{{Surface critical behavior at
  m axial Lifshitz points: Continuum models, boundary conditions and two loop
  renormalization group results}},
  \href{https://doi.org/10.1088/0305-4470/36/46/c01}{\emph{J. Phys. A}
  {\bfseries 36} (2003) L243--L248},
  [\href{https://arxiv.org/abs/cond-mat/0303148}{{\ttfamily
  cond-mat/0303148}}].

\bibitem{Diehl:2004bjl}
H.~W. Diehl, \emph{{Bulk and boundary critical behavior at {Lifshitz} points}},
  \href{https://doi.org/10.1007/BF02704584}{\emph{Pramana} {\bfseries 64}
  (2005) 803--816}, [\href{https://arxiv.org/abs/cond-mat/0407352}{{\ttfamily
  cond-mat/0407352}}].

\bibitem{diehl2006boundary}
H.~Diehl, M.~Shpot and P.~Prudnikov, \emph{Boundary critical behaviour at
  m-axial {Lifshitz} points of semi-infinite systems with a surface plane
  perpendicular to a modulation axis},
  \href{https://doi.org/10.1088/0305-4470/39/25/s09}{\emph{J. Phys. A}
  {\bfseries 39} (2006) 7927},
  [\href{https://arxiv.org/abs/cond-mat/0512681}{{\ttfamily
  cond-mat/0512681}}].

\bibitem{love1888xvi}
A.~E.~H. Love, \emph{{XVI}. the small free vibrations and deformation of a thin
  elastic shell},
  \href{https://doi.org/10.1098/rsta.1888.0016}{\emph{Philosophical
  Transactions of the Royal Society of London.(A.)} (1888) 491--546}.

\bibitem{kirchoff1876vorlesungen}
G.~Kirchoff, \emph{Vorlesungen {\"u}ber mathemathische {Physik}}.
\newblock Teubner, 1876.

\bibitem{kirchoff1850balance}
G.~Kirchoff, \emph{About the balance and the movement of an elastic disc},
  {\emph{Journal of Pure and Applied Mathematics (Crelle’s J)} {\bfseries 40}
  (1850) 51--88}.

\bibitem{greaves2013poisson}
G.~N. Greaves, \emph{Poisson's ratio over two centuries: challenging
  hypotheses}, \href{https://doi.org/10.1098/rsnr.2012.0021}{\emph{Notes and
  Records of the Royal Society} {\bfseries 67} (2013) 37--58}.

\bibitem{case2018boundary}
J.~S. Case, \emph{Boundary operators associated with the {Paneitz} operator},
  {\emph{Indiana University Mathematics Journal} (2018) 293--327},
  [\href{https://arxiv.org/abs/arXiv:1509.08342}{{\ttfamily
  arXiv:1509.08342}}].

\bibitem{CaseLuo}
J.~S. Case and W.~Luo, \emph{{Boundary Operators Associated With the
  Sixth-Order GJMS Operator}},
  \href{https://doi.org/10.1093/imrn/rnz121}{\emph{International Mathematics
  Research Notices} {\bfseries 2021} (07, 2019) 10600--10653},
  [\href{https://arxiv.org/abs/arXiv:1810.08027}{{\ttfamily
  arXiv:1810.08027}}].

\bibitem{Fradkin:1982xc}
E.~S. Fradkin and A.~A. Tseytlin, \emph{{Asymptotic Freedom in Extended
  Conformal Supergravities}},
  \href{https://doi.org/10.1016/0370-2693(82)91018-8}{\emph{Phys. Lett. B}
  {\bfseries 110} (1982) 117--122}.

\bibitem{Fradkin:1981jc}
E.~S. Fradkin and A.~A. Tseytlin, \emph{{One Loop Beta Function in Conformal
  Supergravities}},
  \href{https://doi.org/10.1016/0550-3213(82)90481-3}{\emph{Nucl. Phys. B}
  {\bfseries 203} (1982) 157--178}.

\bibitem{graham1992conformally}
C.~R. Graham, R.~Jenne, L.~J. Mason and G.~A. Sparling, \emph{Conformally
  invariant powers of the {Laplacian}, {I}: Existence},
  \href{https://doi.org/10.1112/jlms/s2-46.3.557}{\emph{Journal of the London
  Mathematical Society} {\bfseries 2} (1992) 557--565}.

\bibitem{fischmann2013conformal}
M.~Fischmann, \emph{On conformal powers of the {Dirac} operator on spin
  manifolds},  \href{https://arxiv.org/abs/arXiv:1311.4182}{{\ttfamily
  arXiv:1311.4182}}.

\bibitem{holland2001conformally}
J.~Holland and G.~Sparling, \emph{Conformally invariant powers of the ambient
  {Dirac} operator},  \href{https://arxiv.org/abs/math/0112033}{{\ttfamily
  math/0112033}}.

\bibitem{deBerredo-Peixoto:2001uob}
G.~de~Berredo-Peixoto and I.~L. Shapiro, \emph{{On the High derivative
  fermionic operator and trace anomaly}},
  \href{https://doi.org/10.1016/S0370-2693(01)00801-2}{\emph{Phys. Lett. B}
  {\bfseries 514} (2001) 377--384},
  [\href{https://arxiv.org/abs/hep-th/0101158}{{\ttfamily hep-th/0101158}}].

\bibitem{Stergiou:2022qqj}
A.~Stergiou, G.~P. Vacca and O.~Zanusso, \emph{{Weyl covariance and the energy
  momentum tensors of higher-derivative free conformal field theories}},
  \href{https://doi.org/10.1007/JHEP06(2022)104}{\emph{JHEP} {\bfseries 06}
  (2022) 104}, [\href{https://arxiv.org/abs/2202.04701}{{\ttfamily
  2202.04701}}].

\bibitem{Osborn:2016bev}
H.~Osborn and A.~Stergiou, \emph{{C$_{T}$ for non-unitary CFTs in higher
  dimensions}}, \href{https://doi.org/10.1007/JHEP06(2016)079}{\emph{JHEP}
  {\bfseries 06} (2016) 079},
  [\href{https://arxiv.org/abs/1603.07307}{{\ttfamily 1603.07307}}].

\bibitem{OsbornLectures}
H.~Osborn, \emph{{Lectures on Conformal Field Theories in more than two
  dimensions}},
  \href{https://arxiv.org/abs/http://www.damtp.cam.ac.uk/user/ho/CFTNotes.pdf}{{\ttfamily
  http://www.damtp.cam.ac.uk/user/ho/CFTNotes.pdf}}.

\bibitem{Ferrari:1995gc}
F.~Ferrari, \emph{{Biharmonic conformal field theories}},
  \href{https://doi.org/10.1016/0370-2693(96)00677-6}{\emph{Phys. Lett. B}
  {\bfseries 382} (1996) 349--355},
  [\href{https://arxiv.org/abs/hep-th/9507142}{{\ttfamily hep-th/9507142}}].

\bibitem{Karananas:2015ioa}
G.~K. Karananas and A.~Monin, \emph{{Weyl vs. Conformal}},
  \href{https://doi.org/10.1016/j.physletb.2016.04.001}{\emph{Phys. Lett. B}
  {\bfseries 757} (2016) 257--260},
  [\href{https://arxiv.org/abs/1510.08042}{{\ttfamily 1510.08042}}].

\bibitem{Nakayama:2016cyh}
Y.~Nakayama, \emph{{Interacting scale invariant but nonconformal field
  theories}}, \href{https://doi.org/10.1103/PhysRevD.95.065016}{\emph{Phys.
  Rev. D} {\bfseries 95} (2017) 065016},
  [\href{https://arxiv.org/abs/1611.10040}{{\ttfamily 1611.10040}}].

\bibitem{Riva:2005gd}
V.~Riva and J.~L. Cardy, \emph{{Scale and conformal invariance in field theory:
  A Physical counterexample}},
  \href{https://doi.org/10.1016/j.physletb.2005.07.010}{\emph{Phys. Lett. B}
  {\bfseries 622} (2005) 339--342},
  [\href{https://arxiv.org/abs/hep-th/0504197}{{\ttfamily hep-th/0504197}}].

\bibitem{branson2001conformally}
T.~Branson and A.~Gover, \emph{Conformally invariant non-local operators},
  \href{https://doi.org/10.2140/pjm.2001.201.19}{\emph{Pacific Journal of
  Mathematics} {\bfseries 201} (2001) 19--60}.

\bibitem{gover2018conformal}
A.~Gover and L.~Peterson, \emph{Conformal boundary operators, {T}-curvatures,
  and conformal fractional {Laplacians} of odd order},
  \href{https://arxiv.org/abs/arXiv:1802.08366}{{\ttfamily arXiv:1802.08366}}.

\bibitem{Cardy:1984bb}
J.~L. Cardy, \emph{{Conformal Invariance and Surface Critical Behavior}},
  \href{https://doi.org/10.1016/0550-3213(84)90241-4}{\emph{Nucl. Phys. B}
  {\bfseries 240} (1984) 514--532}.

\bibitem{Herzog:2017kkj}
C.~Herzog, K.-W. Huang and K.~Jensen, \emph{{Displacement Operators and
  Constraints on Boundary Central Charges}},
  \href{https://doi.org/10.1103/PhysRevLett.120.021601}{\emph{Phys. Rev. Lett.}
  {\bfseries 120} (2018) 021601},
  [\href{https://arxiv.org/abs/1709.07431}{{\ttfamily 1709.07431}}].

\bibitem{Bianchi:2015liz}
L.~Bianchi, M.~Meineri, R.~C. Myers and M.~Smolkin, \emph{{R\'enyi entropy and
  conformal defects}},
  \href{https://doi.org/10.1007/JHEP07(2016)076}{\emph{JHEP} {\bfseries 07}
  (2016) 076}, [\href{https://arxiv.org/abs/1511.06713}{{\ttfamily
  1511.06713}}].

\bibitem{Herzog:2017xha}
C.~P. Herzog and K.-W. Huang, \emph{{Boundary Conformal Field Theory and a
  Boundary Central Charge}},
  \href{https://doi.org/10.1007/JHEP10(2017)189}{\emph{JHEP} {\bfseries 10}
  (2017) 189}, [\href{https://arxiv.org/abs/1707.06224}{{\ttfamily
  1707.06224}}].

\bibitem{Chalabi:2021jud}
A.~Chalabi, C.~P. Herzog, A.~O'Bannon, B.~Robinson and J.~Sisti, \emph{{Weyl
  anomalies of four dimensional conformal boundaries and defects}},
  \href{https://doi.org/10.1007/JHEP02(2022)166}{\emph{JHEP} {\bfseries 02}
  (2022) 166}, [\href{https://arxiv.org/abs/2111.14713}{{\ttfamily
  2111.14713}}].

\bibitem{FarajiAstaneh:2021foi}
A.~Faraji~Astaneh and S.~N. Solodukhin, \emph{{Boundary conformal invariants
  and the conformal anomaly in five dimensions}},
  \href{https://doi.org/10.1016/j.physletb.2021.136282}{\emph{Phys. Lett. B}
  {\bfseries 816} (2021) 136282},
  [\href{https://arxiv.org/abs/2102.07661}{{\ttfamily 2102.07661}}].

\bibitem{branson1995sharp}
T.~P. Branson, \emph{Sharp inequalities, the functional determinant, and the
  complementary series},
  \href{https://doi.org/10.1090/S0002-9947-1995-1316845-2}{\emph{Transactions
  of the American Mathematical Society} {\bfseries 347} (1995) 3671--3742}.

\bibitem{Gaiotto:2014gha}
D.~Gaiotto, \emph{{Boundary F-maximization}},
  \href{https://arxiv.org/abs/1403.8052}{{\ttfamily 1403.8052}}.

\bibitem{Jensen:2015swa}
K.~Jensen and A.~O'Bannon, \emph{{Constraint on Defect and Boundary
  Renormalization Group Flows}},
  \href{https://doi.org/10.1103/PhysRevLett.116.091601}{\emph{Phys. Rev. Lett.}
  {\bfseries 116} (2016) 091601},
  [\href{https://arxiv.org/abs/1509.02160}{{\ttfamily 1509.02160}}].

\bibitem{Wang:2021mdq}
Y.~Wang, \emph{{Defect a-theorem and a-maximization}},
  \href{https://doi.org/10.1007/JHEP02(2022)061}{\emph{JHEP} {\bfseries 02}
  (2022) 061}, [\href{https://arxiv.org/abs/2101.12648}{{\ttfamily
  2101.12648}}].

\bibitem{Kobayashi:2018lil}
N.~Kobayashi, T.~Nishioka, Y.~Sato and K.~Watanabe, \emph{{Towards a
  $C$-theorem in defect CFT}},
  \href{https://doi.org/10.1007/JHEP01(2019)039}{\emph{JHEP} {\bfseries 01}
  (2019) 039}, [\href{https://arxiv.org/abs/1810.06995}{{\ttfamily
  1810.06995}}].

\bibitem{Witten:2001ua}
E.~Witten, \emph{{Multitrace operators, boundary conditions, and AdS / CFT
  correspondence}},  \href{https://arxiv.org/abs/hep-th/0112258}{{\ttfamily
  hep-th/0112258}}.

\bibitem{ostrogradsky1850memoire}
M.~Ostrogradsky, \emph{{M\'emoires sur les \'equations diff\'erentielles,
  relatives au probl\`eme des isop\'erim\`etres}}, {\emph{Mem. Acad. St.
  Petersbourg} {\bfseries 6} (1850) 385--517}.

\bibitem{Mack:1975je}
G.~Mack, \emph{{All unitary ray representations of the conformal group SU(2,2)
  with positive energy}},
  \href{https://doi.org/10.1007/BF01613145}{\emph{Commun. Math. Phys.}
  {\bfseries 55} (1977) 1}.

\bibitem{Lubensky:1975zz}
T.~C. Lubensky and M.~H. Rubin, \emph{{Critical phenomena in semi-infinite
  systems. 2. Mean-field theory}},
  \href{https://doi.org/10.1103/PhysRevB.12.3885}{\emph{Phys. Rev. B}
  {\bfseries 12} (1975) 3885--3901}.

\bibitem{Osborn:1993cr}
H.~Osborn and A.~C. Petkou, \emph{{Implications of conformal invariance in
  field theories for general dimensions}},
  \href{https://doi.org/10.1006/aphy.1994.1045}{\emph{Annals Phys.} {\bfseries
  231} (1994) 311--362},
  [\href{https://arxiv.org/abs/hep-th/9307010}{{\ttfamily hep-th/9307010}}].

\bibitem{Casini:2018nym}
H.~Casini, I.~Salazar~Landea and G.~Torroba, \emph{{Irreversibility in quantum
  field theories with boundaries}},
  \href{https://doi.org/10.1007/JHEP04(2019)166}{\emph{JHEP} {\bfseries 04}
  (2019) 166}, [\href{https://arxiv.org/abs/1812.08183}{{\ttfamily
  1812.08183}}].

\bibitem{Gubser:2002vv}
S.~S. Gubser and I.~R. Klebanov, \emph{{A Universal result on central charges
  in the presence of double trace deformations}},
  \href{https://doi.org/10.1016/S0550-3213(03)00056-7}{\emph{Nucl. Phys. B}
  {\bfseries 656} (2003) 23--36},
  [\href{https://arxiv.org/abs/hep-th/0212138}{{\ttfamily hep-th/0212138}}].

\bibitem{Allais:2010qq}
A.~Allais, \emph{{Double-trace deformations, holography and the c-conjecture}},
  \href{https://doi.org/10.1007/JHEP11(2010)040}{\emph{JHEP} {\bfseries 11}
  (2010) 040}, [\href{https://arxiv.org/abs/1007.2047}{{\ttfamily 1007.2047}}].

\bibitem{Herzog:2019bom}
C.~P. Herzog and I.~Shamir, \emph{{On Marginal Operators in Boundary Conformal
  Field Theory}}, \href{https://doi.org/10.1007/JHEP10(2019)088}{\emph{JHEP}
  {\bfseries 10} (2019) 088},
  [\href{https://arxiv.org/abs/1906.11281}{{\ttfamily 1906.11281}}].

\bibitem{Dowker:2010qy}
J.~S. Dowker, \emph{{Determinants and conformal anomalies of GJMS operators on
  spheres}}, \href{https://doi.org/10.1088/1751-8113/44/11/115402}{\emph{J.
  Phys. A} {\bfseries 44} (2011) 115402},
  [\href{https://arxiv.org/abs/1010.0566}{{\ttfamily 1010.0566}}].

\bibitem{Dowker:2013mba}
J.~S. Dowker, \emph{{Spherical Dirac GJMS operator determinants}},
  \href{https://doi.org/10.1088/1751-8113/48/2/025401}{\emph{J. Phys. A}
  {\bfseries 48} (2015) 025401},
  [\href{https://arxiv.org/abs/1310.5563}{{\ttfamily 1310.5563}}].

\bibitem{Dowker:2014rva}
J.~S. Dowker, \emph{{The boundary F-theorem for free fields}},
  \href{https://arxiv.org/abs/1407.5909}{{\ttfamily 1407.5909}}.

\bibitem{Dowker:2017hpm}
J.~S. Dowker, \emph{{a-F interpolation with boundary}},
  \href{https://arxiv.org/abs/1709.08569}{{\ttfamily 1709.08569}}.

\bibitem{Kislev:2022emm}
A.~C. Kislev, T.~Levy and Y.~Oz, \emph{{Odd dimensional nonlocal Liouville
  conformal field theories}},
  \href{https://doi.org/10.1007/JHEP07(2022)150}{\emph{JHEP} {\bfseries 07}
  (2022) 150}, [\href{https://arxiv.org/abs/2206.10884}{{\ttfamily
  2206.10884}}].

\bibitem{Brust:2016gjy}
C.~Brust and K.~Hinterbichler, \emph{{Free \ensuremath{\square}$^{k}$ scalar
  conformal field theory}},
  \href{https://doi.org/10.1007/JHEP02(2017)066}{\emph{JHEP} {\bfseries 02}
  (2017) 066}, [\href{https://arxiv.org/abs/1607.07439}{{\ttfamily
  1607.07439}}].

\bibitem{Herzog:2022jlx}
C.~P. Herzog and V.~Schaub, \emph{{Fermions in Boundary Conformal Field Theory
  : Crossing Symmetry and $\epsilon$-Expansion}},
  \href{https://arxiv.org/abs/2209.05511}{{\ttfamily 2209.05511}}.

\bibitem{McAvity:1993ue}
D.~M. McAvity and H.~Osborn, \emph{{Energy momentum tensor in conformal field
  theories near a boundary}},
  \href{https://doi.org/10.1016/0550-3213(93)90005-A}{\emph{Nucl. Phys. B}
  {\bfseries 406} (1993) 655--680},
  [\href{https://arxiv.org/abs/hep-th/9302068}{{\ttfamily hep-th/9302068}}].

\bibitem{Herzog:2021spv}
C.~P. Herzog and V.~Schaub, \emph{{A sum rule for boundary contributions to the
  trace anomaly}}, \href{https://doi.org/10.1007/JHEP01(2022)121}{\emph{JHEP}
  {\bfseries 01} (2022) 121},
  [\href{https://arxiv.org/abs/2107.11604}{{\ttfamily 2107.11604}}].

\bibitem{DiPietro:2020fya}
L.~Di~Pietro, E.~Lauria and P.~Niro, \emph{{3d large $N$ vector models at the
  boundary}},
  \href{https://doi.org/10.21468/SciPostPhys.11.3.050}{\emph{SciPost Phys.}
  {\bfseries 11} (2021) 050},
  [\href{https://arxiv.org/abs/2012.07733}{{\ttfamily 2012.07733}}].

\bibitem{Bender:2008gh}
C.~M. Bender and P.~D. Mannheim, \emph{{Exactly solvable PT-symmetric
  Hamiltonian having no Hermitian counterpart}},
  \href{https://doi.org/10.1103/PhysRevD.78.025022}{\emph{Phys. Rev. D}
  {\bfseries 78} (2008) 025022},
  [\href{https://arxiv.org/abs/0804.4190}{{\ttfamily 0804.4190}}].

\bibitem{Smilga:2017arl}
A.~Smilga, \emph{{Classical and quantum dynamics of higher-derivative
  systems}}, \href{https://doi.org/10.1142/S0217751X17300253}{\emph{Int. J.
  Mod. Phys. A} {\bfseries 32} (2017) 1730025},
  [\href{https://arxiv.org/abs/1710.11538}{{\ttfamily 1710.11538}}].

\bibitem{Pais:1950za}
A.~Pais and G.~E. Uhlenbeck, \emph{{On Field theories with nonlocalized
  action}}, \href{https://doi.org/10.1103/PhysRev.79.145}{\emph{Phys. Rev.}
  {\bfseries 79} (1950) 145--165}.

\bibitem{Boyle:2021jaz}
L.~Boyle and N.~Turok, \emph{{Cancelling the vacuum energy and Weyl anomaly in
  the standard model with dimension-zero scalar fields}},
  \href{https://arxiv.org/abs/2110.06258}{{\ttfamily 2110.06258}}.

\bibitem{Maldacena:2011mk}
J.~Maldacena, \emph{{Einstein Gravity from Conformal Gravity}},
  \href{https://arxiv.org/abs/1105.5632}{{\ttfamily 1105.5632}}.

\bibitem{Smilga:2004cy}
A.~V. Smilga, \emph{{Benign versus malicious ghosts in higher-derivative
  theories}},
  \href{https://doi.org/10.1016/j.nuclphysb.2004.10.037}{\emph{Nucl. Phys. B}
  {\bfseries 706} (2005) 598--614},
  [\href{https://arxiv.org/abs/hep-th/0407231}{{\ttfamily hep-th/0407231}}].

\bibitem{Safari:2021ocb}
M.~Safari, A.~Stergiou, G.~P. Vacca and O.~Zanusso, \emph{{Scale and conformal
  invariance in higher derivative shift symmetric theories}},
  \href{https://doi.org/10.1007/JHEP02(2022)034}{\emph{JHEP} {\bfseries 02}
  (2022) 034}, [\href{https://arxiv.org/abs/2112.01084}{{\ttfamily
  2112.01084}}].

\bibitem{Levy:2018bdc}
T.~Levy and Y.~Oz, \emph{{Liouville Conformal Field Theories in Higher
  Dimensions}}, \href{https://doi.org/10.1007/JHEP06(2018)119}{\emph{JHEP}
  {\bfseries 06} (2018) 119},
  [\href{https://arxiv.org/abs/1804.02283}{{\ttfamily 1804.02283}}].

\bibitem{Bzowski:2013sza}
A.~Bzowski, P.~McFadden and K.~Skenderis, \emph{{Implications of conformal
  invariance in momentum space}},
  \href{https://doi.org/10.1007/JHEP03(2014)111}{\emph{JHEP} {\bfseries 03}
  (2014) 111}, [\href{https://arxiv.org/abs/1304.7760}{{\ttfamily 1304.7760}}].

\bibitem{Bianchi:2021snj}
L.~Bianchi, A.~Chalabi, V.~Proch\'azka, B.~Robinson and J.~Sisti,
  \emph{{Monodromy defects in free field theories}},
  \href{https://doi.org/10.1007/JHEP08(2021)013}{\emph{JHEP} {\bfseries 08}
  (2021) 013}, [\href{https://arxiv.org/abs/2104.01220}{{\ttfamily
  2104.01220}}].

\bibitem{Klebanov:2011gs}
I.~R. Klebanov, S.~S. Pufu and B.~R. Safdi, \emph{{F-Theorem without
  Supersymmetry}}, \href{https://doi.org/10.1007/JHEP10(2011)038}{\emph{JHEP}
  {\bfseries 10} (2011) 038},
  [\href{https://arxiv.org/abs/1105.4598}{{\ttfamily 1105.4598}}].

\end{thebibliography}\endgroup

\end{document}